\documentclass[aps,article,twocolumn,superscriptaddress,nofootinbib]{revtex4-2}

\usepackage[utf8]{inputenc}
\usepackage[fleqn]{amsmath}
\usepackage{amssymb}
\usepackage{graphicx}
\usepackage{ulem,xcolor}
\usepackage{epstopdf, epsfig}
\usepackage{graphicx}
\usepackage{mathtools}
\usepackage{bm}

\newcommand{\Ltfbar}[1]{\overline{L}^{{#1}}_{\rm{TF}}}
\newcommand{\Lcol}{L_{\rm{sh}}}
\newcommand{\Tcol}{T_{\rm{sh}}}

\newcommand{\Wcs}{W}
\newcommand{\Wbw}{\overline{W}}

\newcommand{\mica}[1]{{\color{black} #1}}

\newcommand{\tys}[1]{{\color{black} #1}}
\newcommand{\ttys}[1]{{\color{black} #1}}
\newcommand{\marta}[1]{{\color{black} #1}}

\newcommand{\authormail}[2]{#1\thanks{#2}}

\begin{document}

\title{\tys{Properties of Pair Plasmas Emerging from Electromagnetic Showers in Matter}}

\author{M. Pouyez}
\email{pouyezmattys@gmail.com}
\affiliation{LULI, Sorbonne Université, CNRS, CEA, École Polytechnique, Institut Polytechnique de Paris, F-75255 Paris, France}
\author{G. Nicotera}
\affiliation{LULI, CNRS, CEA, Sorbonne Universit\'{e}, École Polytechnique, Institut Polytechnique de Paris, F-91128 Palaiseau, France}
\author{M. Galbiati}
\authormail{}{marta.galbiati@polytechnique.edu}
\affiliation{LULI, CNRS, CEA, Sorbonne Universit\'{e}, École Polytechnique, Institut Polytechnique de Paris, F-91128 Palaiseau, France}
\author{T. Grismayer}
\affiliation{GoLP/Instituto de Plasmas e Fusão Nuclear, Instituto Superior Técnico, Universidade de Lisboa, 1049-001 Lisboa, Portugal}
\author{L. Lancia}
\affiliation{LULI, CNRS, CEA, Sorbonne Universit\'{e}, École Polytechnique, Institut Polytechnique de Paris, F-91128 Palaiseau, France}
\author{C. Riconda}
\affiliation{LULI, Sorbonne Université, CNRS, CEA, École Polytechnique, Institut Polytechnique de Paris, F-75255 Paris, France}
\author{M. Grech}
\affiliation{LULI, CNRS, CEA, Sorbonne Universit\'{e}, École Polytechnique, Institut Polytechnique de Paris, F-91128 Palaiseau, France}

\begin{abstract}

Electromagnetic showers from high-energy electron beams interacting with a target are a promising path to creating pair plasmas in the laboratory. Here, we solve analytically the kinetic equations describing this process. Two regimes are defined by the ratio of the target thickness $L$ to the \tys{shower} length $\Lcol$, which depends on the electron energy and target composition. For thin targets ($L < \Lcol$), we derive explicit expressions for the spectra of produced photons and pairs, as well as the number of pairs. For thick targets ($L > \Lcol$), we obtain the total pair number and photon spectrum. Analytical results agree well with Geant4 simulations, and it is found that significant pair escape requires $L < \Lcol$. The divergence, density and characteristic dimensions of the escaping pair jets are derived, and a criterion for pair plasma formation is obtained. While current laser wakefield beams are not well adapted, multi-petawatt lasers may provide new electron or photon sources suitable for laboratory pair plasma production.


\end{abstract}

\maketitle

\section{Introduction}

When a high-energy electron or positron interacts with the intense electric field of a nucleus, it has a probability to radiate its energy via the Bremsstrahlung process. The emitted photon, also interacting with the nuclear field \cite{bethe1934stopping}, can in turn transform into an electron-positron pair via the Bethe-Heitler process \cite{bethe1934stopping}. This interplay of photon emission and pair creation initiates a cascade of secondary particles, also known as an electromagnetic shower (EMS). 

\marta{EMS in matter are extensively studied in particle physics \cite{fabjan2003calorimetry}. In astrophysics, they are relevant to studies of cosmic-ray–induced air showers \cite{kampert2012extensive,bertolotti2013electromagnetic,watson2019highest} and their detection \cite{hillas1985cerenkov,aharonian2024high}.} With the emergence of ultra-high intensity lasers, and even more since the advent of petawatt-class laser systems \cite{generation_bahk_2004,vulcan_hernandezgomez_2010,apollon_papadopoulos_2016,ELI,ZEUS,nam2018performance,petawatt_danson_2019,technology_bromage_2019,khazanov2023exawatt} capable of producing high-energy electron~\cite{arefiev2016beyond,kim2017stable,gonsalves2019petawatt,jirka2020scaling,kim2021multi,miao2022multi,poder2024multi,babjak2024direct} and photon~\cite{Sarri2014_NonLinearThomson,albert2016applications,Gong2018_GeVFlash,Gu2018_AttosecondGamma, huang2019highly,formenti2022modeling, galbiati2023numerical,Matheron2024_SelfAlignedCompton,galbiati2025numerical} beams together with extremely intense electromagnetic (EM) fields, EMS have also gained the attention of the laser-plasma community~\cite{reiss1971production,shearer1973pair,burke1997positron,blackburn2017scaling,lobet2017generation,mercuri2021impact,salgado2021towards,golub2022nonlinear,pouyez2024multiplicity,qu2024creating,elsner2025entering}. While these later works focused on EMS developing in strong EM fields rather than matter, EMS in matter\marta{, exploiting laser-driven sources, nuclear reactors \cite{Hugenschmidt2012}, or conventional accelerators,} are still considered a promising path toward the generation of quasi-neutral electron-positron pair plasmas in the laboratory~\cite{liang1998pair,Gahn2000,nakashima2002numerical,relativistic_chen_2009,myatt2009optimizing,sarri2013table,generation_sarri_2015,xu2016ultrashort,HChen2023,Arrowsmith_2024,noh2024charge}\marta{, as demonstrated by the recent achievements at CERN \cite{Arrowsmith_2024}}. Producing such plasmas would unlock the experimental investigation of various processes - from plasma instabilities to particle acceleration - 
of utmost importance for extreme plasma astrophysics~\cite{Lobet2015PRL,Uzdensky2019,Stoneking2020}.

The theoretical study of EMS in matter began in the 1930's, motivated by the discovery of the cosmic ray showers~\marta{\cite{blackett1932photography,rossi1933eigenschaften,pfotzer1936dreifachkoinzidenzen,Auger1939,auger1939grandes}}. The mathematical description of such showers was developed almost simultaneously by Bhabha and Heitler~\cite{bhabha1937passage} and by Carlson and Oppenheimer \cite{carlson1937multiplicative} in 1937. Further descriptions were established at the beginning of the 1940's~\cite{landau1938cascade,snyder1938transition,tamm1939soft} and today, the book of Rossi and Greisen~\cite{rossi1941cosmic} remains one of the most detailed and clear mathematical analyses of the EMS evolution. The angular structure of the EMS was first addressed by Molière \cite{moliere1946cosmic}, who provided a key estimate for the angular spread of shower particles at a given thickness. Further analytical analyses \cite{roberg1949angular,eyges1951angular,green1952spread,green1954core,Greisen1956,kamata1958lateral} have extended this work by solving the full three-dimensional diffusion equation for different shower depths (length traversed by the primary particle). 

However, all these advanced analytical tools rely on several approximations for the shower depth, particle energies and rate definitions. As a result, they fall short of providing an analytical framework for describing the production of an electron-positron pair plasma from the collision of an electron beam with matter. In particular, a good way to estimate the density of the jet of pairs emerging from such a collision is yet to be found.

In this work, we address this problem. To do so, we rely on the analytical framework developed in \cite{pouyez2024kinetic} to describe EMS in strong EM fields, which we adapt to EMS in matter. This allows us to derive explicit expressions for the number of pairs produced per incident particle ($N_\pm/N_0$), henceforth referred to as the shower multiplicity, as well as for the produced pair and photon spectra. These expressions are thoroughly benchmarked against Monte-Carlo simulations, and in particular Geant4 simulations \cite{agostinelli2003geant4}. Additionally, these simulations allow us to characterise the divergence of the pairs escaping the target. Together, these studies provide us with simple estimates for the characteristic density and size of the pair jets emerging from the target, and thus allow us to identify the conditions under which a pair plasma could be produced.

Throughout the paper, we used SI units and standard notations for physical constants:
$c$ denotes the speed of light in vacuum, $m$ the electron mass, $e$  the elementary charge, $\varepsilon_0$ the permittivity of vacuum and $\hbar$ the reduced Planck constant. The fine structure constant is denoted by $\alpha = e^2/(4\pi\varepsilon_0\hbar c)$, and $r_e=e^2/(4\pi\varepsilon_0m_ec^2)$ denotes the classical radius of the electron.

\section{Number of pairs}\label{sec:NumberOfPairs}

In this section, we aim to determine the number of pairs generated by an initial population of electrons with energy $\gamma_0 mc^2$ as it interacts with a mono-atomic target. 

Our starting point is the cascade equations written for successive generations of leptons and photons~\cite{pouyez2024multiplicity}: a lepton of generation $n$ creates photons of generation $n$, which in turn produce new pairs of generation $n+1$. The temporal evolution of the energy distributions of each generation ($n$) of electrons ($\scriptstyle{-}$), positrons ($\scriptstyle{+}$), and photons ($\gamma$) read\footnote{Under the approximation that all particles propagate at the speed of light.}:
\begin{eqnarray}
    \nonumber \partial_t f_\pm^{(n)}(\gamma,t) &=& \int_0^{\infty}\!\! d\gamma_\gamma\, w(\gamma+\gamma_\gamma,\gamma_\gamma)\,f_\pm^{(n)}(\gamma+\gamma_\gamma,t) \nonumber\\
    &-& W(\gamma)\,f_\pm^{(n)}(\gamma,t) \nonumber\\
    &+&\int_0^{\infty}\!\! d\gamma_\gamma\, \overline{w}(\gamma_\gamma,\gamma)\,f_\gamma^{(n-1)}(\gamma_\gamma,t)\,, \label{eq:fpn} \\
    \nonumber \partial_t f_\gamma^{(n)}(\gamma_\gamma,t) &=& \int_1^{\infty}\!\! d\gamma\, w(\gamma,\gamma_\gamma)\,f_-^{(n)}(\gamma,t)\\
    &+&\int_1^{\infty}\!\! d\gamma\, w(\gamma,\gamma_\gamma)\,f_+^{(n)}(\gamma,t) \nonumber\\
    &-& \Wbw(\gamma_\gamma)f_\gamma^{(n)}(\gamma_\gamma,t) \, , \label{eq:fgn}
\end{eqnarray}
\noindent where $(\gamma-1)\,mc^2$ is the lepton kinetic energy, $\gamma_\gamma\,mc^2$ is the photon energy, and $w(\gamma,\gamma_\gamma)$ and $\overline{w}(\gamma_\gamma,\gamma)$ are the energy differential rates of photon emission (Bremsstrahlung process) and pair production (Bethe-Heitler process), respectively. We have also introduced $\Wcs(\gamma)=\int_0^{\infty}\!d\gamma_\gamma\,w(\gamma,\gamma_\gamma)$ and 
$\Wbw(\gamma_\gamma)=\int_1^\infty \!d\gamma\,\overline{w}(\gamma_\gamma,\gamma)$. 

\marta{The kinetic equations \eqref{eq:fpn} and \eqref{eq:fgn} account only for the Bremsstrahlung and Bethe–Heitler processes. We neglect here other processes such as collisions, ionisation, hadronic processes, and linear Breit-Wheeler, as discussed in detail in the {\it Supplemental Material}~\cite{suppMat}.}

Multiple definitions of the Bremsstrahlung and Bethe–Heitler cross-sections exist in the literature~\cite{koch1959bremsstrahlung, motz1969pair}. In this work, we adopt the formulations used in \cite{martinez2019high} and the corresponding differential rates $w$ and $\overline{w}$ are recalled in the {\it Supplemental Material}~\cite{suppMat}. These rates are different from the ones used in Geant4, which rely on interpolated data from Seltzer and Berger \cite{seltzer1985bremsstrahlung, seltzer1986bremsstrahlung}. As will be shown later in this work, this difference does not impact the overall study as it mainly leads to a systematic overestimate (by a factor $\sim 1.5$) of the total Bremsstrahlung cross-section. Furthermore, using the cross-sections from \cite{martinez2019high} rather than the Seltzer and Berger data \cite{seltzer1985bremsstrahlung,seltzer1986bremsstrahlung} has two advantages: it provides us with an analytical model for the cross-sections, and it can be applied to both neutral and ionised targets.

\subsection{\tys{Shower} time}

To solve the kinetic Eqs. \eqref{eq:fpn} and \eqref{eq:fgn}, we follow the methodology developed in~\cite{pouyez2024kinetic} and consider the \tys{shower} time $\Tcol$ (equivalently \tys{shower} length $\Lcol=c\Tcol$). \tys{This  reference time is introduced to distinguish two regimes of shower evolution. It is defined, from the kinetic equation for $f_-^{(0)}$, as the time required for the average energy to decrease from $\gamma_0 mc^2$ to $mc^2$} \footnote{\tys{The integration is stopped at $mc^2$ since, at this stage, other processes have been neglected. If ionisation is also taken into account, it is more appropriate to stop the integration at $\mathcal{E}_c(Z)$ [Eq. \eqref{Eq:fit:Ec}], the characteristic energy at which ionisation dominates over radiation. However, this correction does not affect the asymptotic expression and will be neglected in the definition of the shower time.}}:
\begin{eqnarray}
    \Tcol&=& \int_1^{\gamma_0}\frac{d\gamma}{\int_0^\infty  d\gamma_\gamma \gamma_\gamma w(\gamma,\gamma_\gamma)} \\
    &&\xrightarrow{\gamma_0 \gg 1} \frac{\ln(\gamma_0)}{K(n_i,Z,Z^*,T)} \label{Eq:radiativetime}
\end{eqnarray}
where $K$, which has the dimension of a frequency, is a function of the target atomic density $n_i$, atomic number $Z$, ionisation degree $Z^*$, and temperature $T$. Its full expression is provided in the {\it Supplemental Material}~\cite{suppMat}. In the following, we restrict our study to neutral targets \tys{and always refer to the asymptotic definition of the shower time, i.e., Eq.\eqref{Eq:radiativetime}.  The function $K$ is now a} function of $n_i$ and $Z$ only:
\begin{eqnarray}
K(n_i,Z)=4\,Z^2\,n_i\,r_e^2 c\, \alpha\,\left[\frac{4}{3}I(0)+\frac{13}{9}-\frac{4}{3}f_C(Z)\right]
\end{eqnarray}
with $I(0)=\frac{1}{2}\left[\ln\left(1+\Ltfbar{2}\right)-\Ltfbar{2}/(1+\Ltfbar{2})\right]
$ a function of the Thomas-Fermi length normalised to the Compton radius $\Ltfbar{}=Z^{-1/3}\,\alpha^{-1}$, 
and $f_C$ the Coulomb correction defined as:
\begin{eqnarray}\label{eq:ur:fc}
   f_C(Z)=\frac{\alpha^2Z^2}{1+\alpha^2Z^2}\sum_{n=1}^\infty (-\alpha^2Z^2)^n[\zeta(2n+1)-1]\,.
\end{eqnarray}
\tys{In most cases, the Coulomb correction is well represented considering only the first five terms of the previous sum.}

In this framework, the \tys{shower} time refers to the time over which an electron loses all of its kinetic energy through Bremsstrahlung radiation. This differs from the radiation time commonly used in the literature, which corresponds to the time over which an electron reduces its energy by a factor $1/e$ through radiation. However, with that definition, an electron continues to contribute to the shower development after the radiation time. In contrast, \tys{the shower time defined in} Eq. \eqref{Eq:radiativetime} delimits the moment when the electron ceases to participate in the photon emission. \tys{Note that the shower time and the radiation time differ by a factor $\sim \ln(\gamma_0)$}.

\begin{figure*}[htbp]
    \centering
    \includegraphics[width=\textwidth]{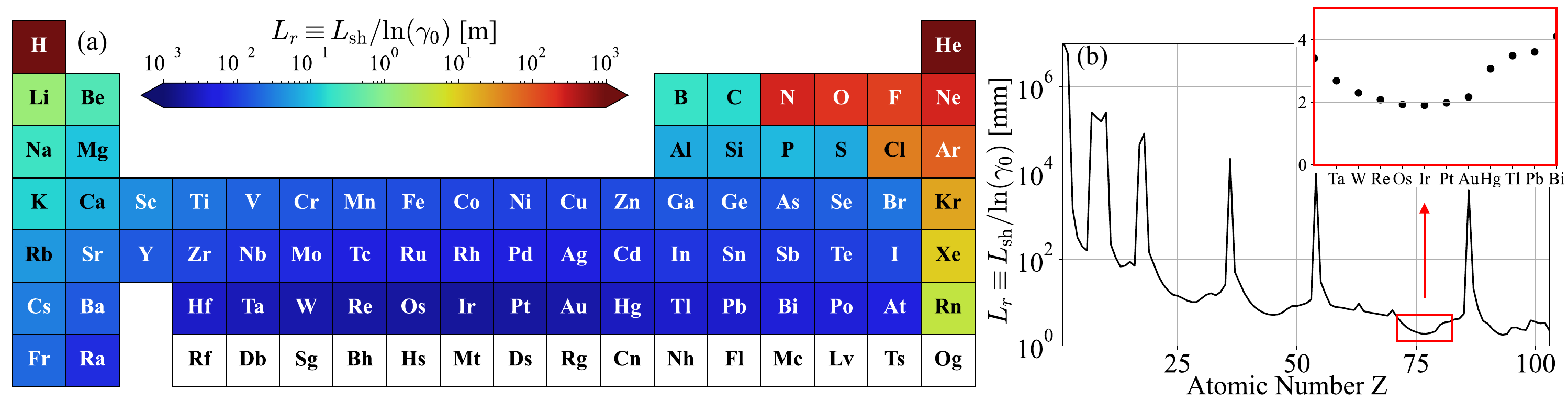} 
    \caption{Radiation length $L_r\equiv\Lcol/\ln(\gamma_0)$ for relevant elements of the periodic table in a neutral state and at standard conditions of temperature and pressure (a) and as a function of the atomic number (b). \mica{The inset in panel (b) corresponds to the radiation length of the element between Hafnium and Bismuth. }
    }
    \label{fig:fig1}
\end{figure*}

In the following, it is also convenient to express distance in units of \tys{shower} length $\Lcol=c\Tcol$. We have reported in Fig. \ref{fig:fig1} the value of \tys{the radiation length}\footnote{\tys{In the literature, the radiation length reads $1/[4Z^2\,n_i\,r_e^2\ln(189Z^{-1/3})]$, but here we expressed it following our definition of the Bremsstrahlung cross section: $L_r=c/K(n_i,Z)$.}} $L_r \equiv \Lcol/\ln(\gamma_0)=c/K(n_i,Z)$ for different relevant elements of the periodic table in a neutral state and at standard conditions of temperature and pressure.

\subsection{Short-time solution}\label{Number:short_time}

At a time $t\ll \Tcol$, the incident electrons have not yet cooled down, and we can neglect the radiative operator in the leptons' dynamics [first two terms of Eq. \eqref{eq:fpn}]. Furthermore, $\forall \gamma_\gamma$ $\overline{W}(\gamma_\gamma)\,t \ll 1$ we Taylor expand\footnote{Using the asymptotic definition of the rates, this approximation breaks down for $t/\Tcol \gtrsim 2/\ln(\gamma_0-1)$. Nevertheless, this approximation still holds for a wide range of times and energies (Example: for $\gamma_0=10^5$, it is a good approximation until $t/\Tcol \lesssim 1/6$).} the photon pair creation probability $1-\exp{[-\overline{W}(\gamma_\gamma)\,t]}$. Solving the kinetic Eqs. \eqref{eq:fpn} and \eqref{eq:fgn} under these two assumptions, we obtain: 

\begin{eqnarray}
    f_\gamma^{(n-1)}(\gamma_\gamma,t)&\simeq& \frac{2^{n-1}}{(2n-1)!}G^{(n-1)}(\gamma_\gamma) \left(\frac{t}{\Tcol}\right)^{2n-1} ,\label{eq:sh:sol:fgn} \\
    f_\pm^{(n)}(\gamma,t)&\simeq& \frac{2^{n-1}}{(2n)!}L^{(n)}(\gamma)\left(\frac{t}{\Tcol}\right)^{2n}, \label{eq:sh:sol:fpn}
\end{eqnarray}
with
\begin{eqnarray}
    G^{(n)}(\gamma_\gamma)&=& \int_1^{\infty}\!\! d\gamma \, \Tcol \, w(\gamma,\gamma_\gamma)L^{(n)}(\gamma) ,\label{eq:sh:Gn} \\
    L^{(n)}(\gamma)&=& \int_0^{\infty}\!\! d\gamma_\gamma  \, \Tcol \, \overline{w}(\gamma_\gamma,\gamma)G^{(n-1)}(\gamma_\gamma), \label{eq:sh:Ln}\\
    L^{(0)}(\gamma)&=&f_-^{(0)}(\gamma,t=0).
\end{eqnarray}
\tys{Similarly to the shower developing in a pure magnetic field \cite{pouyez2024kinetic}, in matter, the} $nth$ generation of pairs grows as $(t/\Tcol)^{2n}$. In the limit $t\ll \Tcol$, the first generation thus dominates the total pairs. Therefore, the integration of $f_\pm^{(1)}$ over energy leads to a good estimate for the number of produced pairs. For $N_0$ incident electrons at $\gamma_0mc^2$, the shower multiplicity $N_\pm/N_0$ reads:
\begin{eqnarray}\label{eq:shorttime:Npm}
    N_\pm(t)&&/N_0=\left(\frac{t}{\Tcol}\right)^2\frac{\Tcol^2}{2}\int_0^{\gamma_0-1} d\gamma_\gamma \, \overline{W}(\gamma_\gamma)w(\gamma_0,\gamma_\gamma)\,\,\,\,\,\,\,\,\nonumber\\
    &&\,\,\,\,\xrightarrow{\gamma_0 \gg 1}\frac{1}{2} \left(\frac{t}{\Tcol}\right)^2 \frac{R(n_i,Z)}{K(n_i,Z)} \ln(\gamma_0)^2\ln(c_1 \gamma_0)
    \, .\label{eq:st:N1exact}
\end{eqnarray}
where $c_1=0.016$ and where the functions $R$ and $K$ are given in \cite{suppMat}. Interestingly, the only dependence on the target properties is contained in the ratio $R/K$ and the \tys{shower} time. For a non-ionised material, the ratio $R/K$ is a function of the atomic number $Z$ only and decreases very slowly: from $0.570$ to $0.568$ for $Z \in [10,100]$. \tys{In agreement with \cite{heitler1984quantum} and with the recent experiments of Kim et al. \cite{kim2024electron}, we have proved that the number of produced pairs is independent of the target element when its thickness is expressed in units of the shower length (similarly in units of the radiation length).}

\subsection{Long-time solution}\label{Number:Long_time}

The number of pairs can also be derived in the long-time-scale regime using the methodology presented in \cite{pouyez2024kinetic}. The mathematical derivation is provided in the {\it Supplementary Material} \cite{suppMat}. Here, we outline the main approximations of the calculation.

The first assumption is that all leptons have already radiated their energy into photons. By computing the total photon spectrum emitted by a single lepton, we establish a relation between the spectrum of all emitted photons of generation $n$ and the spectrum of the pairs of generation $n$ at the moment of their creation.

Second, we assume that over such a time scale, the probability of the pair production process is close to a Heaviside function evaluated at $\gamma_\gamma-\gamma_c$, where \marta{$\gamma_\gamma m c^2$} is the energy of the parent photon and \marta{$\gamma_c m c^2$} is a threshold energy. This assumption means that only the photons created with an energy larger than \marta{$\gamma_c m c^2$} have produced pairs. The critical energy  \marta{$\gamma_c m c^2$} is chosen as the photon energy below which other processes dominate over pair production. It was extracted from the NIST database \cite{NistWeb} as the energy for which the photon scattering rate becomes equal to the pair production rate. For neutral materials, a fit provides:
\begin{eqnarray}\label{Eq:fit:gammac}
    \gamma_c(Z)=\frac{237.6}{Z}+\frac{27.5}{Z^{1/3}}.
\end{eqnarray}

Finally, we assume that, upon pair production, a photon transfers half of its energy to each resulting lepton.

The first approximation provides us with the total photon spectrum emitted by a given population of leptons. The second approximation identifies the photons that participate in the pair production, and the third approximation defines how these photons split their energy into the new pairs. Using these approximations, we obtain a recursive relation between the number of pairs $N_\pm^{(n)}$ and the total photon spectrum $F_\gamma^{(n)}$ of generation $(n)$. For $n\leq \ln(\gamma_0/\gamma_c)/\ln(2)$ the total photon spectrum is:
\setlength{\mathindent}{5pt}
\begin{align}
    F_\gamma^{(n)}(\gamma_\gamma)=\frac{2^n}{(2n+1)!}\frac{1}{\gamma_\gamma}\ln\left(\frac{\gamma_0}{2^n \gamma_\gamma}\right)^{2n+1}\Theta(\gamma_0 -2^n \gamma_\gamma)\label{eq:lt:sol:Fn}
\end{align}
while the number of pairs of generation $(n)$ reads:
\begin{eqnarray}
    N_\pm^{(n)}/N_0&=&\frac{2^{n-1}}{(2n)!}\ln\left(\frac{\gamma_0}{2^{n-1}\gamma_c}\right)^{2n} \Theta(\gamma_0-2^{n-1}\gamma_c).\,\,\,\,\, \label{eq:lt:sol:Nn}
\end{eqnarray}
From the Heaviside function $\Theta$ in equation \eqref{eq:lt:sol:Nn} we obtain that the maximal number of generations is given by $n=\ln(\gamma_0/\gamma_c)/\ln(2)+1$. Furthermore, the final number of pairs is obtained by summing this equation over all generations. In the limit $\gamma_0\rightarrow \infty$, taking the continuous limit for the sum and using a saddle-point approximation for the integral, we finally obtain:
\begin{eqnarray}\label{eq:longtime:Npm}
    N_\pm/N_0 =\frac{1}{2+\ln(2)}\frac{\gamma_0}{\gamma_c(Z)}
\end{eqnarray}
where the only material dependence is now contained in the photon threshold energy $\gamma_c(Z)$.

\tys{From this analysis, it is also interesting to estimate the energy lost in the sub-threshold photons (incapable of pair production). Since all leptons have completely cooled down, we estimate their energy as their rest mass energy and the energy fraction of converted pairs as $2mc^2N_\pm/(\gamma_0mc^2 N_0)$. It follows from Eq. \ref{eq:longtime:Npm} that this ratio is independent of $\gamma_0$.  We can then calculate the fraction of the incident energy that remains in low-energy photons as $
1-2/(2+\ln(2))/\gamma_c(Z)$. When only Bremsstrahlung and Bethe-Heitler processes are considered, the natural threshold corresponds to an energy of $\gamma_cmc^2=2mc^2$, meaning that $\sim 63 \%$ of the initial energy is lost to sub-threshold photons.}

\begin{figure*}[htbp]
    \centering
    \includegraphics[width=\textwidth]{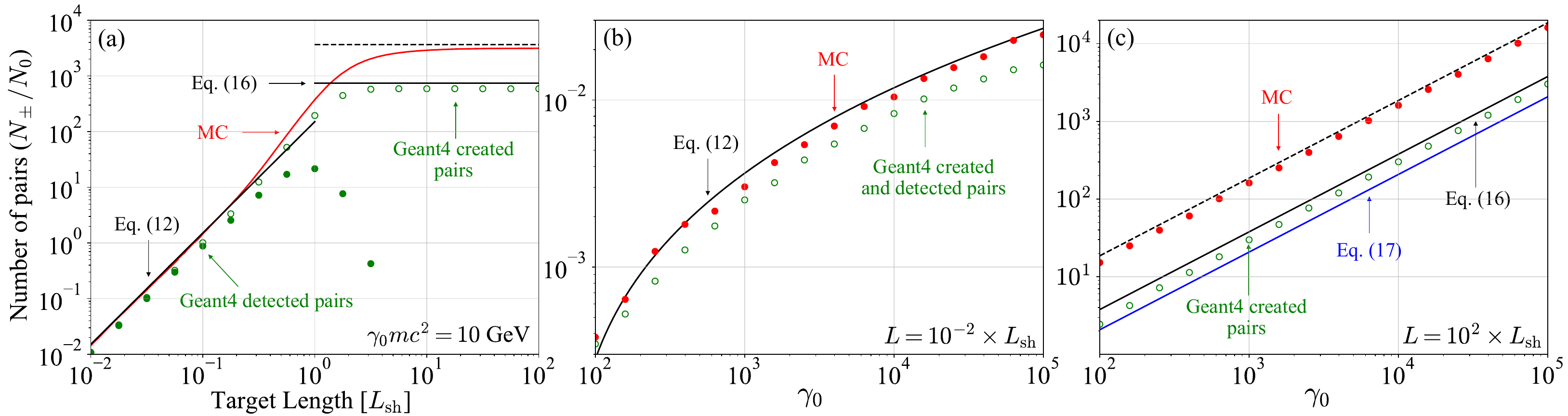}
    \caption{Number of produced pairs in the collision of electrons with Tantalum $Z=73$. In panel (a) as a function of the target thickness $L$ for $10$ GeV incident electrons. In panels (b) and (c) as a function of the incident electron energy $\gamma_0$ for $L/\Lcol=10^{-2}$ and $L/\Lcol=10^2$, respectively. The black curves represent the solutions from this work, given by Eqs. \eqref{eq:shorttime:Npm} and \eqref{eq:longtime:Npm}, while the blue line corresponds to Heitler's model, Eq. \eqref{eq:Heitler}. Dashed black lines indicate the solution of Eq.\eqref{eq:longtime:Npm} with $\gamma_c=2$. The red line in panel (a) and red dots in panels (b) and (c) show the results from the \mica{Monte Carlo (MC)} simulations. Green circles denote the total number of pairs, and green dots represent the number of outgoing pairs extracted from Geant4 simulations.
    }
    \label{fig:fig2}
\end{figure*}

\subsection{Comparison with simulations}\label{sec:NumberOfPairs:comparison}

In this section, we compare analytical predictions of pair production with numerical results obtained from both Geant4 simulations and numerical integration of Eqs. \eqref{eq:fpn} and \eqref{eq:fgn}. The numerical integration is carried out using an in-house Monte Carlo (MC) code that includes only Bremsstrahlung and Bethe–Heitler processes, with cross-sections defined in \cite{martinez2019high}. We refer to these results as MC simulations throughout the text. In contrast, the Geant4 simulations incorporate broader physical processes: for photons, the pair production, the Compton scattering, and the photoelectric processes, while for leptons, the Bremsstrahlung, the ionisation, the multiple Coulomb scattering, and the annihilation processes. \marta{In addition, they include matter suppression effects \cite{klein1999suppression} such as the Landau-Pomeranchuk-Migdal \cite{Migdal1956} and Ter-Mikaelian \cite{Terikaelian1972} effects.} \tys{In what follows, all the quantities from Geant4 simulations are extracted considering positrons only. However, since the absorption is not symmetric for electrons and positrons below $10$ MeV and since  ionisation can contribute to generating new electrons, we may expect a slight difference in the number of escaping positrons and electrons.} 

In Figure \ref{fig:fig2}, we present the total number of produced pairs by the collision of electrons with a neutral target of tantalum under standard temperature and pressure conditions. Panel (a) represents the number of pairs as a function of the target thickness $L/\Lcol$ for $10$ GeV incident electrons, while panels (b) and (c) show the dependence on incident electron energy $\gamma_0$ for thin ($L/\Lcol=10^{-2}$) and thick ($L/\Lcol=10^2$) targets, respectively. 

The number of pairs derived for short-time  Eq.~\eqref{eq:shorttime:Npm} and long-time Eq.~\eqref{eq:longtime:Npm} interactions are represented in solid black lines (where we use the relation $L/\Lcol=t/\Tcol$). The total number of pairs calculated using Geant4 simulations is shown as green circles, while the number of pairs collected on a detector right after the target is presented in green dots. The solutions of the EMS using the MC simulations are shown as a red line in panel (a), and as red dots in panels (b) and (c). The maximum number of pairs predicted by Eq.~\eqref{eq:longtime:Npm}, using a threshold value of $\gamma_c=2$ (natural threshold for Bethe-Heitler process) is indicated by the black dashed line.

Excellent agreement between theory, Geant4 simulations, and numerical solutions of Eqs.  \eqref{eq:fpn} and \eqref{eq:fgn} are found in both asymptotic regimes. In the case of a thin target, the discrepancy between the Geant4 results and the theoretical or MC simulations arises from the differences in the cross-section definitions. \tys{In this regime, the particles are still very energetic and the ionisation process taken into account in Geant4 is negligible. However, the suppression effects \cite{Migdal1956,Terikaelian1972,klein1999suppression} and thus the different cross sections considered in the Geant4 simulation make the number of produced pairs smaller compared to our estimate. This causes Eq.~\eqref{eq:st:N1exact} to overestimate the Geant4 result by a factor of about $1.5$}. The decrease in the number of outgoing pairs (green dots) beyond the \tys{shower} length indicates that leptons get trapped within the material. 
Indeed, after travelling a distance $L \sim \Lcol$ through the target, most of the leptons have cooled down and processes such as ionisation and collisions become dominant, which confine the pairs inside the material. As a result, to maximise the number of outgoing pairs and, consequently, the plasma density, it is optimal to use a target with a thickness smaller than $\Lcol$. For such a small target thickness, an excellent agreement between Geant4 simulations and Eq.~\eqref{eq:shorttime:Npm} is found. \tys{For thick targets, we find a very good agreement between Geant4 and the theoretical prediction Eq.~\eqref{eq:longtime:Npm} with $\gamma_c$ from Eq.~\eqref{Eq:fit:gammac}. The analytical solution remains $\sim 1.25$ times larger than the Geant4 results.  Similarly, the MC simulations and Eq.~\eqref{eq:longtime:Npm} using $\gamma_c=2$ (horizontal dashed line) are in good agreement (analytical solution $\sim 1.15$ times larger than MC simulation). Note that a direct comparison between MC and Geant4 is not meaningful since the two codes account for different physical processes. In Geant4, additional mechanisms, primarily ionisation, limit photon conversion into pairs. In contrast, the MC simulations include only Bremsstrahlung and Bethe–Heitler processes, so that all photons with energy above $2mc^2$ produce pairs. Consequently, the influence of the different processes on the total number of pairs is quantified by a factor of $\sim \gamma_c(Z)/2$ (about 5 for a Tantalum target). }

Although \tys{the thick target} solution has no practical importance for pair plasma production, it is a fundamental aspect of the shower development, as well as important to determine the total energy deposition.

Finally, we note that, although Fig.~\ref{fig:fig2} focuses on a Tantalum target, the short-time predictions are universal across different target materials, as long as the ratio $L/\Lcol$ is similar. In contrast, the long-time regime becomes material-dependent through the critical energy $\gamma_c$.

\begin{figure}
    \centering
    \includegraphics[width=0.9\linewidth]{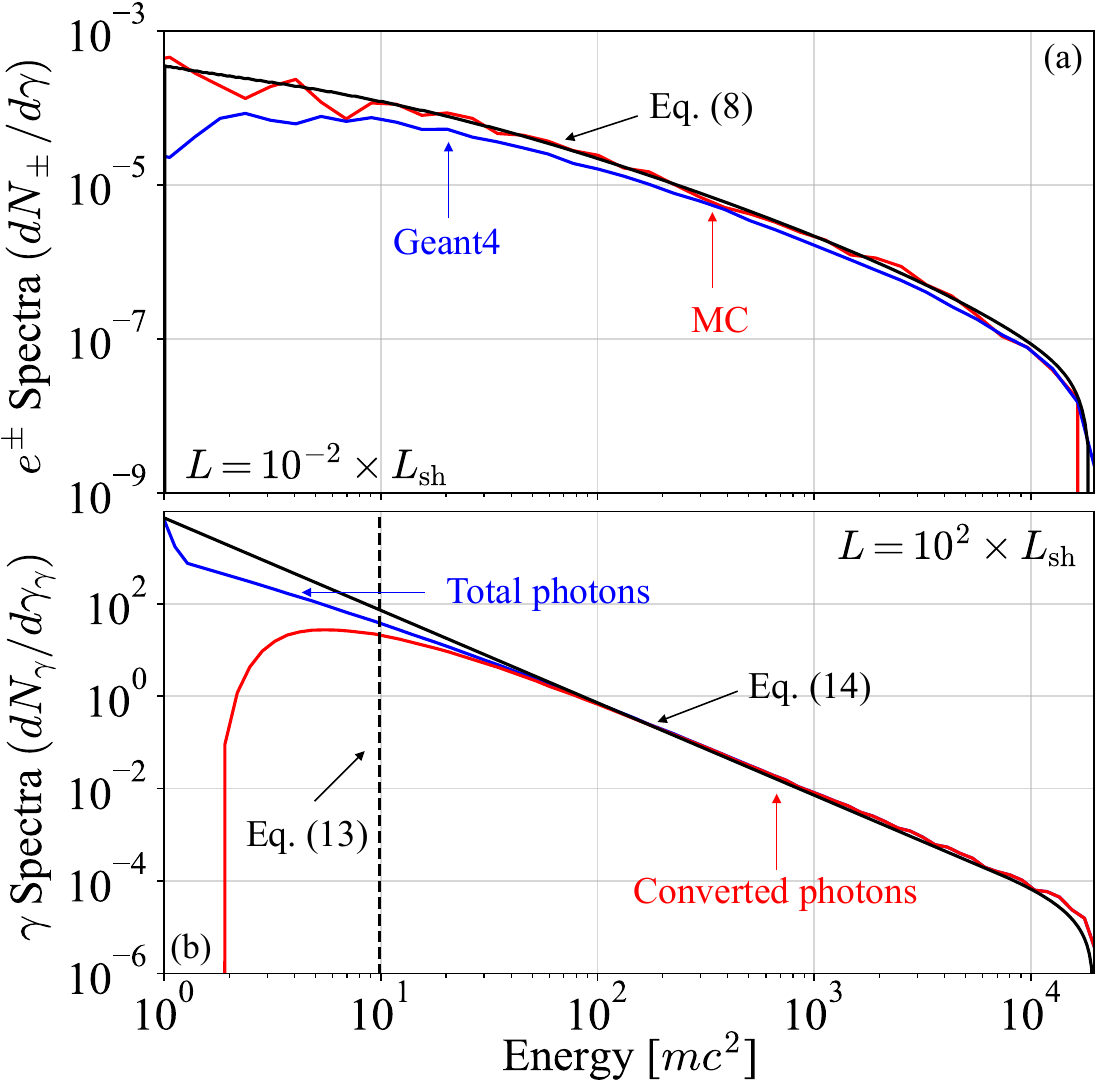}
    \caption{\mica{Particle spectra for thin and thick target}. In panel (a), the spectra of electron-positron pairs and in panel (b) the spectra of photons resulting from the collision of $10$ GeV electrons with a Tantalum target of thickness $L=10^{-2}\Lcol$ and $L=10^2\Lcol$, respectively. In panel (a), the black line stands for Eq. \eqref{eq:sh:sol:fpn} (n=1), while the blue and red curves are obtained from Geant4 and MC simulations, respectively. In panel (b), the black line represents the total photon spectrum obtained by summing Eq. \eqref{eq:lt:sol:Fn} over each generation, and the black dotted line corresponds to Eq. \eqref{Eq:fit:gammac}. The blue and red curve, extracted from Geant4 simulations, represents, respectively, the total photon spectrum and the spectrum of all the photons that have generated pairs.
    }
    \label{fig:fig3}
\end{figure}

Let us now turn to Fig.~\ref{fig:fig3} that provides the spectrum of electron-positron pairs (a) and photons emerging (b) from the collision of $10$ GeV electrons with a Tantalum target of thickness $L=10^{-2}\Lcol$ and $L=10^2\Lcol$, respectively. 
In Fig.~\ref{fig:fig3}(a), the analytical pair spectrum given by Eq.~\eqref{eq:sh:sol:fpn}, considering only $n=1$ (in black), is compared to Geant4 (in blue) and MC (in red) simulations and is found in excellent agreement. The slight differences are explained by the discrepancies in the cross-section, as mentioned before. Figure \ref{fig:fig3}~(b) shows the spectrum of all generated photons during the interaction with a thick target. The solid black curve, computed by summing over all generations Eq.~\eqref{eq:lt:sol:Fn}, and the blue curve, obtained from Geant4 simulations\footnote{MC simulation results are not included here, as they do not account for all significant processes in this regime.}, correspond to the total photon spectrum emitted during the interaction. The red curve represents the Geant4 spectrum of photons that have generated a pair, while the vertical black dashed line indicates the threshold energy $\gamma_c$. Our analysis predicted a sharp threshold in the photon spectrum responsible for pair production near $\gamma_c$. However, because the photon scattering probability does not behave like a Heaviside step function at $\gamma_c$, this sharp cutoff does not appear in the Geant4 simulations. Nevertheless, by slightly overestimating the contributions from photons just above $\gamma_c$ and neglecting those just below it, integrating the total photon spectrum from $\gamma_c$ to infinity still yields an accurate estimate of the number of pairs as demonstrated by Fig. \ref{fig:fig2}.

\subsection{Comparison with previous solutions}\label{sec:NumberOfPairs:previousSol}

\subsubsection{Solutions with Mellin transform}

In 1938, Landau and Rumer \cite{landau1938cascade} solved the diffusion equations, or equivalently the kinetics equations, by applying the Mellin transform with respect to energy for the distribution functions. By considering the rates of Bremsstrahlung and Bethe-Heitler processes in the asymptotic limit $1\ll \gamma,\gamma_\gamma \ll \gamma_0$, they compute the distribution of pairs and photons in Mellin space. The solution takes the form of a sum of two exponential terms, one of which they neglect to compute the inverse Mellin transform. This last approximation is only valid for \textit{a time not too small} \cite{rossi1941cosmic}, meaning their solution can not accurately describe the system at very short time scales. 

Moreover, since the number of pairs is $N_\pm(t)=\int_1^{\gamma_0} d\gamma f_\pm(\gamma,t)$ and the pair spectrum is derived in the limit $\gamma\gg 1$, the total number of pairs can not be obtained. However, the number of pairs with an energy greater than some threshold $\gamma$ is well represented by $\int_\gamma^{\gamma_0} d\gamma' f_\pm(\gamma',t)$ in the limit $1\ll \gamma \ll \gamma_0$. For the long-time scale regime, the leptons have already cooled down and their energy approaches $mc^2$. It follows that their solution no longer captures the pair population in the long-time scale regime.

Thus, all the solutions and their approximate form obtained in this framework \cite{landau1938cascade,tamm1939soft,snyder1938transition,rossi1941cosmic,Greisen1956} are representative of the EMS evolution only in the intermediate time regime $t\sim \Tcol$, while our solutions complete the description for the asymptotic regimes $t\ll \Tcol$ and $t\gg \Tcol$.

\subsubsection{Heitler's model}

One of the simplest and most useful tools for the description of EMS is the toy model introduced by Heitler \cite{heitler1984quantum}. It is now commonly used in a large physics community to estimate the final number of pairs emerging from EMS: in astrophysics for magnetic showers \cite{akhiezer1994kinetic}, in high-energy particle physics for hadronic showers \cite{matthews2005heitler}, and in the strong-field community where a generalisation to arbitrary splitting ratios has been proposed recently~\cite{selivanov2024final}. In this heuristic model, the electron (or positron) radiates a photon after travelling a length $d$ corresponding to the length for which a lepton loses half of its energy through radiation. Similarly, after travelling $d$, the photon generates a pair, giving half of its energy to each of the outgoing leptons. These processes stop when leptons reach a critical energy noted $\mathcal{E}_c$. Heitler shows that the final number of pairs then reads\footnote{In the second edition of his book \cite{heitler1984quantum} page $234$, Heitler introduces a supplementary factor $1/3$ to take the critical energy into account correctly. He corrected this in the third edition, page $388$, introducing instead a factor $1/\kappa$, with $\kappa$ of order unity. In this work, we use the formula given in the third edition with $\kappa=1$ and divided by $2$ to account only for the number of pairs and not for the total number of leptons as he did. }:
\begin{eqnarray}\label{eq:Heitler}
    N_\pm/N_0=\frac{\gamma_0mc^2}{3\mathcal{E}_c}.
\end{eqnarray}
\noindent This asymptotic solution closely aligns with the one obtained in this work,  Eq.~\eqref{eq:longtime:Npm}. Both exhibit the same energy dependence, $N_\pm \propto \gamma_0$, but they differ in their essence and in the definition of the critical energy. 
In this work, we prove that if only the two QED channels are considered, i.e. Bremsstrahlung and Bethe-Heitler, the factor in front of $\gamma_0/\gamma_c$ should tend towards $(2+\ln(2))^{-1}\simeq 0.37$ when $\gamma_0 \rightarrow \infty$.  By contrast, Heitler’s model yields a slightly lower prefactor of $1/3$. However, by choosing a heuristic rate, the photon emitted always receives the same fraction of the lepton energy, and low-energy photons can be emitted only by low-energy leptons. In our framework, low-energy photons are continuously emitted and follow the full Bremsstrahlung spectrum. 
Furthermore, Heitler defines the critical energy using a condition on the electron processes, while we use a condition on the photon instead. The critical energy of the Heitler model is indeed defined as the value for which the energy lost by radiation is of the same order of magnitude as the energy lost by ionisation. It reads \cite{heitler1984quantum,fabjan2003calorimetry}:
\begin{eqnarray}\label{Eq:fit:Ec}
    \mathcal{E}_c(Z)/mc^2=\frac{1193.7}{Z+1.24},
\end{eqnarray}
while the critical energy of this work [Eq. \eqref{Eq:fit:gammac}] corresponds to the value for which the photon scattering rate equals the pair production rate. \tys{Overall, these differences cause the Heitler model to underestimate the number of pairs by roughly a factor of two compared to the result obtained in this work [Eq.~\eqref{eq:longtime:Npm}]. This underestimate is visible in Fig.~\ref{fig:fig2} (c), where the Heitler solution Eq.~\eqref{eq:Heitler} is shown as blue solid lines.}

\section{Divergence of produced pairs}\label{sec:angle}

An analytical expression for the number of electron-positron pairs produced during short and long interactions has been established. We now turn to the divergence angle of the outgoing pairs. \tys{Having established that the thin target regime is optimal for maximising the number of pairs, in Secs. \ref{sec:angle} and \ref{sec:density} we focus solely on this regime.} 

\begin{figure*}[htbp]
    \centering
    \includegraphics[width=\textwidth]{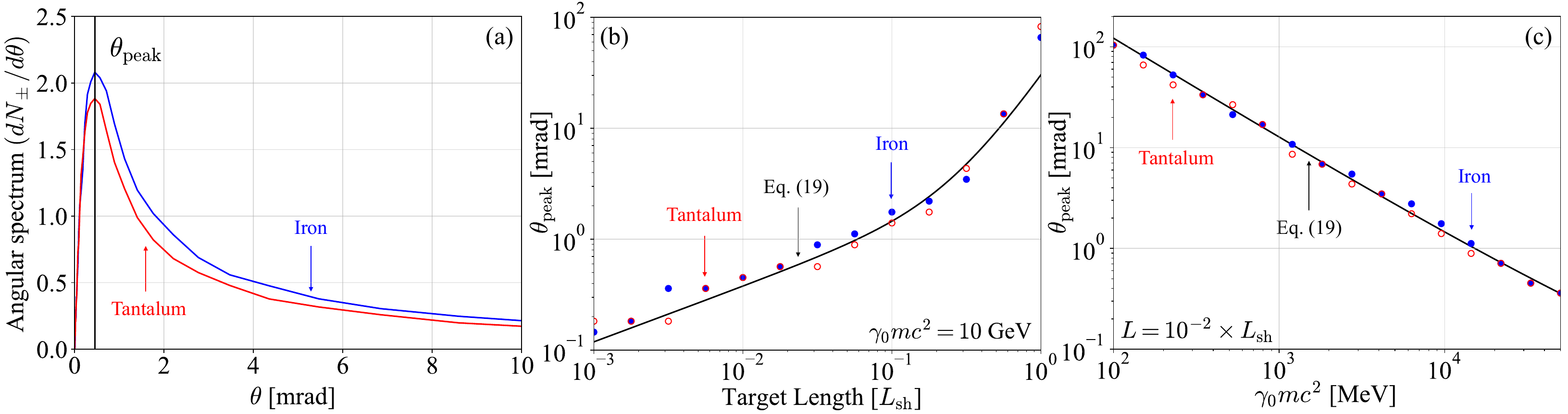}
    \caption{\mica{Angular description of the pair jet.} In panel (a), the angular distribution of the outgoing pairs $dN_\pm/d\theta$ for $10$ GeV incident electrons colliding with a target of length $L= 10^{-2}\Lcol$.
    In panels $(b)$ and $(c)$ the angle $\theta_{\rm{peak}}$ as a function of the target length (for $10$ GeV incident electron) and as a function of initial electron energy (for $L=10^{-2}\Lcol$). Results for Tantalum and Iron converters are shown in red and blue, respectively. The black curve stands for Eq. \eqref{eq:Angle:theta}.
    }
    \label{fig:fig4}
\end{figure*}

As outlined in the introduction, several advanced analytical studies on the angular distribution of shower particles were conducted in the 1950s \cite{moliere1946cosmic,roberg1949angular,eyges1951angular,green1952spread,green1954core,Greisen1956,kamata1958lateral}. By solving the three-dimensional diffusion equation for EMS considering the angular differential cross section for Bremsstrahlung, Bethe-Heitler, ionisation and multiple Coulomb scattering processes, they were able to estimate the radial distribution of the outgoing pairs. The Kamata and Nishimura model \cite{kamata1958lateral} remains the most widely adopted, offering a simple and practical approximation for the radial distribution of shower particles.

However, this solution relies on several simplifying assumptions that lead to notable inaccuracies, as confirmed by later numerical studies \cite{capdevielle2005relation,apel2006comparison}. These limitations are thoroughly reviewed in \cite{dey2016slope} and were partially addressed in Sec. \ref{sec:NumberOfPairs:previousSol}. 

Furthermore, one of the main contributions to the angular spread of the outgoing pairs is the Multiple Coulomb Scattering (MCS) angles, which have already been studied in \cite{rossi1941cosmic,moliere1948theorie,bethe1953moliere,scott1963theory,highland1975some,lynch1991approximations}. In these works, authors have estimated the root mean square (RMS) angle resulting from MCS. Under the assumption of a Gaussian angular distribution, Rossi and Greisen \cite{rossi1941cosmic} have shown that the RMS angle scales as $\propto \gamma_0^{-1}\sqrt{L}$, and demonstrated that when $L$ is expressed in units of the radiation length ($L_r\equiv \Lcol/\ln(\gamma_0)$), it becomes independent of the material properties. However, the accumulation of MCS contributes to a large tail in the angular distribution function \cite{lynch1991approximations}, making the Gaussian approximation unreliable and the RMS angle a poor descriptor of the angular spread. \tys{Because of this tail we choose to characterise the angle at which the distribution function is maximised instead of the RMS angle.}

As emphasised by Capdevielle and Gawin \cite{capdevielle1982radial}, only Monte Carlo simulations can accurately reproduce the radial profile of shower particles in our regime of interest and previous models could not provide a reliable estimate of the divergence angle of the outgoing pairs. Nonetheless, the scaling laws derived from these studies will guide our upcoming parametric analysis.

We performed $16\times16\times 17$ Geant4 simulations\footnote{The physics list used here is the same as the one defined in Sec. \ref{sec:NumberOfPairs:comparison}.} for $16$ different atomic numbers (ranging from $6$ to $82$), $16$ different incident energy (logarithmically spaced from $100$ MeV to $50$ GeV) and $17$ different target lengths (logarithmically spaced from $L/\Lcol=10^{-3}$ to $L/\Lcol=10$). 
For each simulation, we recorded $p_\perp$ and $p_\parallel$, the perpendicular and parallel components to the incident electrons' direction of the outgoing positrons. We choose to characterise only the properties of the positrons to exclude the contribution from the incident electrons. \tys{Due to the asymmetry of the positrons and electron absorption and due to the generation of new electrons via ionisation, the final distribution angle of electron and positron can differ. For simplicity, we neglect this aspect.} From $p_\perp$ and $p_\parallel$, we reconstruct the angular distribution of the outgoing pairs $dN_\pm/d\theta$ where $\theta=\arctan(p_\perp/p_\parallel)$.

\tys{In Fig. \ref{fig:fig4}(a), we present the angular distribution $dN_\pm/d\theta$ of the outgoing pairs resulting from the collision of $10$ GeV electrons with Tantalum (red) and Iron (blue) targets of length $L=10^{-2}\Lcol$. The position of the maxima, denoted $\theta_{\rm{peak}}$, is highlighted with the black line. 

The $16\times16\times17$ Geant4 simulations, performed over the parameter ranges $Z \in [6,82]$, $\gamma_0 mc^2 \in [0.1,50]$ GeV and $L/\Lcol \in [10^{-3},1]$, provide the following fit for value of the angle $\theta_{\rm{peak}}$:} 
\begin{eqnarray}\label{eq:Angle:theta}
    \theta_{\rm{peak}}(\gamma_0,L)= \frac{c_2}{\gamma_0}\sqrt{\frac{L}{\Lcol}}+\frac{c_3}{\sqrt{\gamma_0}}\left(\frac{L}{\Lcol}\right)^{2} 
\end{eqnarray}
where $c_2\simeq 73.6$ and $c_3\simeq 3.7$ are fitting parameters. The position of $\theta_{\rm{peak}}$ for both materials is also displayed with a black straight line. For short interactions, similar dependencies as the ones discussed in \cite{rossi1941cosmic,moliere1948theorie,highland1975some,lynch1991approximations} for the RMS MCS angle are found: $\theta_{\rm{peak}}$ grows with $\sqrt{L/\Lcol}$, decrease as $\gamma_0^{-1}$ and is independent of the material when $L$ is expressed in units of the \tys{shower} length. \tys{Thus, the first term of Eq. \eqref{eq:Angle:theta} represents mainly the angle accumulated through Multiple Coulomb scattering. More details on the link between the RMS angle from MCS and $\theta_{\rm{peak}}$ are provided in the \textit{Supplementary Material} \cite{suppMat}. The last term in Eq. \eqref{eq:Angle:theta} is due to  successive generations contributing more and more to the shower.} 

To highlight the different dependencies and assess the validity of the fit, we also show in panels $(b)$ and $(c)$ the angle $\theta_{\rm{peak}}$ as a function of the target length (for $10$ GeV incident electron) and as a function of initial electron energy (for $L=10^{-2}\Lcol$) in the cases of Tantalum (red) and Iron (blue) converters. As depicted in this figure, the fitted function given by Eq. \eqref{eq:Angle:theta} accurately describes $\theta_{\rm{peak}}$ for $L<\Lcol$, which is the regime of interest for pair plasma production. \tys{It also validates one important feature of the angular aperture of the outgoing pairs: the momentum angle $\theta_{\rm{peak}}$ is independent of the material when its length is expressed in units of the shower length. }

Here, we chose to focus solely on $\theta_{\rm{peak}}$ because, as it will be shown in the next section, it provides a reliable estimate for the outgoing pair density.

\section{Toward pair plasma in the laboratory}\label{sec:density}

Having determined both the number of electron-positron pairs and their divergence angle as functions of the incident electron energy and target parameters, we are now able to estimate the pair density. This estimate will ultimately yield a straightforward criterion on the initial beam and target parameters for the achievement of the plasma state.

To obtain a pair plasma, the outgoing jet of leptons should first satisfy the quasi-neutrality $N_+ \sim N_-$, and secondly, to ensure collective behaviour, its size should exceed the plasma skin depth \cite{HChen2023}. Let us further note here that the produced pair jets are relativistically hot, in the sense that their characteristic temperature greatly exceeds their rest energy. As a result, the Debye length is of the same order as the skin-depth (they only differ by a factor $\sqrt{3}$, of order 1). It follows that the characteristic number of particles in the Debye sphere, also known as the plasma parameter, will be, in general, very large.

\mica{In what follows, we will first derive the escaping jet density in the laboratory frame. We will then derive the typical sizes and skin-depth of the escaping jet in its center-of-mass frame. Finally, a simple criterion for achieving the plasma state will be provided.}

\subsection{Outgoing pair density (in the lab frame)}

We adopt polar coordinates defining $(x_\perp, x_\parallel)$ as the transverse and longitudinal positions of the positrons (or equivalently of the generated electrons). We introduce the spatial angle $\theta_x = \arctan(x_\perp / x_\parallel)$ to describe their trajectory. Assuming the positrons are ultra-relativistic with $v_\perp \ll v_\parallel \simeq c$ and $\theta_{\rm {peak}} \ll 1$, we have $dx_\parallel/dt = c$ and $dx_\perp/dt \simeq c \, \theta_{\rm{peak}}$. Integrating the equation of motion, we obtain for $x_\parallel \leq L$:
\begin{eqnarray}
x_\perp =x_\parallel\left[\frac{2}{3}\frac{c_2}{\gamma_0}\sqrt{\frac{x_\parallel}{\Lcol}}+\frac{1}{3}\frac{c_3}{\sqrt{\gamma_0}}\left(\frac{x_\parallel}{\Lcol}\right)^2\right],
\end{eqnarray}
and for $x_\parallel \geq L$:,
\begin{eqnarray}
    x_\perp=x_\parallel&\bigg[&\frac{c_2}{\gamma_0}\sqrt{\frac{L}{\Lcol}}\left(1-\frac{1}{3}\frac{L}{x_\parallel}\right) \\
    &\,&\,\,+\frac{c_3}{\sqrt{\gamma_0}}\left(\frac{L}{\Lcol}\right)^2\left(1-\frac{2}{3}\frac{L}{x_\parallel}\right)\bigg]. \nonumber
\end{eqnarray}
We model the incoming $N_0$ electrons as a cylindrical bunch of length $L_0$ and radius $R_0$. We neglect the initial divergence of the beam as the dominant contribution arises from interactions within the target\footnote{As shown below, see also Fig.~\ref{fig:fig4}, the typical angular divergence of the escaping pairs is of the order of several 10s of mrad when typical LWFA beams have mrad divergence.}. Since all the particles are assumed to travel at a velocity $v_\parallel \simeq c$, the longitudinal size of the emerging pair beam remains approximately $L_0$ immediately after exiting the target.\footnote{As the beam propagates, the longitudinal size of the beam increases due to the broad energy spectra as shown in Fig. \ref{fig:fig3}~(a). Placing ourselves just after the target, we can neglect this effect.} Considering the electron-positron beam as a cylinder, the radius of the beam is estimated as $R_\pm\simeq R_0+x_\perp$ and the plasma density finally reads $n_\pm\simeq N_\pm/(\pi L_0 R_\pm^2)$. Just after the target, we have:
\begin{eqnarray}\label{eq:density:density}
n_\pm=\frac{c_4 N_0 \left(L/\Lcol\right)^2\ln(\gamma_0)^2\ln(c_1\gamma_0)}{2\pi L_0\left[R_0+\Lcol g(\gamma_0,L/\Lcol)\right]^2}
\end{eqnarray}
where $c_4=R(n_i,Z)/K(n_i,Z)\simeq0.569$ for a neutral target and
\begin{eqnarray}
    g(\gamma_0,L/\Lcol)=\frac{2}{3}\frac{c_2}{\gamma_0}\left(\frac{L}{\Lcol}\right)^{3/2}+\frac{1}{3}\frac{c_3}{\sqrt{\gamma_0}}\left(\frac{L}{\Lcol}\right)^3.
\end{eqnarray}
The optimisation of the pair density is now studied to predict the maximal density reachable in the laboratory.

Equation \eqref{eq:density:density} shows that the pair density depends independently on the \tys{shower} length $\Lcol$ and the normalised target thickness $L/\Lcol$. Since $\Lcol$ is material-specific, the density is inherently dependent on the target composition. For a fixed target length, the density always decreases with increasing $\Lcol$. \tys{This effect arises from the volume expansion of the resulting beam, which grows with the shower length of the material. Therefore, the optimal material for pair-plasma production is the one that minimises the shower length (equivalently, radiation length $L_r$).} As shown in the insert of Fig. \ref{fig:fig1}(b), the optimal material is Iridium ($Z=77$) and its neighbours (Osmium and Platinum) with a \tys{shower} length of $\Lcol = 1.89 \ln(\gamma_0)$ mm. For comparison with commonly used materials \cite{sarri2013table,generation_sarri_2015,xu2016ultrashort,Arrowsmith_2024,noh2024charge}, Lead ($Z=82$) and Tantalum ($Z=73)$ have a \tys{shower} length of $\Lcol = 3.60 \ln(\gamma_0)$ mm and $\Lcol = 2.68\ln(\gamma_0)$ mm respectively. As a result, the maximum density achievable with Lead or Tantalum is approximately $2.5$ and $1.5$ times smaller than the one obtained with a target of Iridium. This conclusion also holds for the most accessible material presented above: Platinum.

The interplay between pair production and volume expansion is also explicitly reflected in Eq. \eqref{eq:density:density}. Initially, the density increases as $(L/\Lcol)^2$, driven by the growth of the pair multiplicity, but at a certain point, it decreases as $(L/\Lcol)^{-1}$ due to the expansion of the volume. By differentiating Eq. \eqref{eq:density:density} with respect to the target length $L/\Lcol$, we find that the maximum of the density is reached at:
\begin{eqnarray}\label{eq:density:Lmax}
\frac{L_{\rm{opt}}}{\Lcol} = \left( \frac{3}{c_2} \gamma_0 \frac{R_0}{\Lcol} \right)^{2/3},
\end{eqnarray}
and is given by:
\begin{eqnarray}\label{eq:density:max_density}
\max(n_\pm) = c_5 n_0 \gamma_0^{4/3} \ln(\gamma_0)^2 \ln(c_1 \gamma_0) \left( \frac{R_0}{\Lcol} \right)^{4/3},
\end{eqnarray}
where $c_5 = 3^{-2/3} c_4 c_2^{-4/3} / 2 \simeq 4.43 \times 10^{-4}$, and $n_0 = N_0 / (\pi L_0 R_0^2)$ is the initial electron density. Furthermore, at this optimal length, the beam radius is $R_\pm=3R_0$. \tys{Note that equations \eqref{eq:density:Lmax} and \eqref{eq:density:max_density} are valid\footnote{Typically, $R_0\sim 1 \mu$m and $\Lcol\sim \ln(\gamma_0)$ mm, thus $L_{\rm{opt}}/\Lcol<1$ holds for $\gamma_0mc^2 \lesssim 150$ GeV.} only for $L_{\rm{opt}}/\Lcol<1$.}

Having derived the maximal density reachable, we are now in a position to provide the condition for the pair plasma production as a function of the initial parameters only.

\subsection{Conditions for pair plasma production}

As we are looking for thin targets, both the incident electrons and the generated electron-positron pairs escape the material. Consequently, the outgoing beam consists of $N_\pm$ positrons and $N_0 + N_\pm$ electrons. As a result, the quasi-neutrality, necessary for obtaining a pair plasma, can be satisfied only for the high multiplicity regime: $N_\pm \gg N_0$. In this regime, the pair multiplicity is accurately captured by Eq.~\eqref{eq:shorttime:Npm}. Since the target length is constrained by the \tys{shower} length, achieving this condition relies primarily on maximising the incident electron energy.\footnote{At the maximum density given by Eq. \eqref{eq:density:Lmax}, for $R_0 \sim 1,\mu\text{m}$ and $\Lcol \sim \ln(\gamma_0)$ mm, Eq. \eqref{eq:shorttime:Npm} yields $N_\pm > 2N_0$ for $\gamma_0 mc^2 \gtrsim 5$ GeV}

\mica{The second condition for achieving a pair plasma is given by comparing the produced pair jet's skin depth to its smallest characteristic size. To do so, we will compute all quantities in the centre-of-mass frame of the escaping jet. This frame moves with respect to the laboratory frame, in the direction of the incident electron beam, at a velocity~\cite{suppMat}:
\begin{eqnarray}
    V_b = \frac{c^2{\vert\bf P}_{\rm tot}\vert}{\mathcal{E}_{\rm tot}} \simeq 1 - \frac{\pi}{2}\,\theta_{\rm peak}(\gamma_0,L_{\rm opt})\,,
\end{eqnarray}
with ${\bf P}_{\rm tot}$ and $\mathcal{E}_{\rm tot}$ the total momentum and energy computed in the lab frame, respectively. The corresponding Lorentz boost is then given by:
\begin{eqnarray}\label{eq:LorentzBoost:asymtptotic}
    \Gamma_b = \frac{1}{\sqrt{1-V_b^2/c^2}} \simeq \frac{1}{\sqrt{\pi\,\theta_{\rm peak}(\gamma_0,L_{\rm opt})}}\,,
\end{eqnarray}
with the approximated form being valid only for small enough $\theta_{\rm peak}$  ($\Gamma_b > 1$). When computed at the optimal length, $\Gamma_b \simeq 0.30\,[\gamma_0^{2}\ln^{1/3}\gamma_0\,(5 {\rm \mu m}/R_0)]^{1/6}$ scales as $0.37\,\gamma_0^{0.35}\,(5 {\rm \mu m}/R_0)^{1/6}$.
For typical parameters of interest for this study, $\Gamma_b$ will be in the range $1$ to $50$.

In this frame and as discussed in the \textit{Supplementary Material} \cite{suppMat}, the plasma skin depth reads:
\begin{eqnarray}\label{eq:skindepth}
    \delta^R = \sqrt{\frac{\varepsilon_0 m c^2 h_0^R}{ 2 e^2 n_\pm^R}}\,.
\end{eqnarray}
\noindent where $n_{\pm}^R = n_{\pm}/\Gamma_b$ is the density in the rest frame. The enthalpy of the escaping pair jets in the centre-of-mass frame is denoted as $h_0^R$ and, as discussed in the {\it Supplementary Material}~\cite{suppMat}, is well approximated by:
\begin{eqnarray}
    h_0^R \simeq \frac{\langle\gamma_\pm\rangle}{\Gamma_b}\,.
\end{eqnarray}
\noindent with $\langle\gamma_{\pm}\rangle$ the average Lorentz factor of the escaping pairs computed in the laboratory frame. It is estimated considering the distribution function of the first generation of pairs Eq. \eqref{eq:sh:sol:fpn}, as:
\begin{eqnarray}\label{eq:average_energy}
    \langle \gamma_\pm \rangle &&= \frac{\int_0^{\gamma_0}d\gamma_\gamma \Wbw(\gamma_\gamma)w(\gamma_0,\gamma_\gamma)\gamma_\gamma/2}{\int_0^{\gamma_0}d\gamma_\gamma \Wbw(\gamma_\gamma)w(\gamma_0,\gamma_\gamma)} \sim 0.25\gamma_0^{0.89}\,,
\end{eqnarray}
the numerical values in the last expression are fitting parameters.}


\begin{figure*}[htpb]
    \centering
    \includegraphics[width=\linewidth]{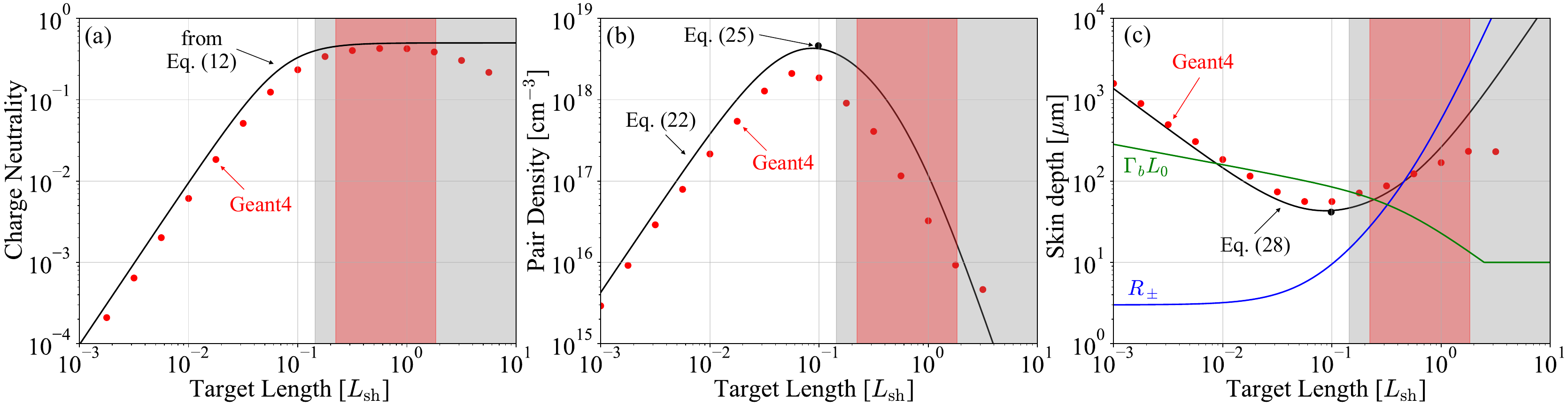}
    \caption{\mica{Neutrality, density and skin depth of pair jet generated from a typical LWFA-seeded electron.} \tys{Panels show (a) charge neutrality $N_+/(N_++N_-)$, (b) positron density at the target rear, and (c) skin depth at the target rear, as functions of the target thickness $L/\Lcol$. Red points are Geant4 results, while black lines are theoretical estimates. In (a), the black line corresponds to $N_\pm/(1+2N_\pm)$ with $N_\pm$ from Eq. \eqref{eq:st:N1exact}. In (b), the red points report the maximal density from Geant4, while the black line follows Eq. \eqref{eq:density:density}, and the black dot represents the estimated maximal density from Eqs. \eqref{eq:density:max_density}, \eqref{eq:density:Lmax}. In (c), the red points correspond to Eq. \eqref{eq:skindepth} using Geant4 density and average energy. The black line is computed by Eq. \eqref{eq:skindepth} using  Eqs. \eqref{eq:density:density} and \eqref{eq:average_energy}, while to obtain the black dot Eq. \eqref{eq:density:max_density} was used. Green and blue lines mark the estimates of the longitudinal ($L_\pm \sim \Gamma_b L_0$) and transverse ($R_\pm$) sizes of the outgoing beam. Red (Geant4) and grey (from Eq. \eqref{eq:shorttime:Npm}) regions denote the quasi-neutrality $N_+/(N_++N_-)>0.4$. The results are obtained considering a cylindrical electron beam with $eN_0 = 2$ nC, $R_0 = 3$ $\mu$m, $L_0 = 10$ $\mu$m, and $\gamma_0 mc^2 = 3$ GeV colliding with Tantalum.}}
    \label{fig:fig5}
\end{figure*}

\mica{
Achieving a plasma state that could sustain collective effect requires the skin-depth [Eq.~\eqref{eq:skindepth}] to be much smaller than the smallest characteristic size of the escaping pair jet $\ell_\pm^*$. Although this state is not necessarily achieved at the maximal density, in what follows\footnote{All starred quantities are expressed in the centre-of-mass frame and for optimal target lenght $L_{\rm{opt}}$.}, we express all quantities for the optimal target length $L_{\rm{opt}}$ Eq.~\eqref{eq:density:Lmax}. Thus, $ \ell_\pm^*= {\rm min}\left(L_\pm^*\simeq\Gamma_b\,L_0,R_\pm^*\simeq3\,R_0\right)$, and the plasma state condition can be casted in the form\footnote{\mica{The numerical value is formally given by $c_6/(c_5 8\pi)\simeq 22.4$ but here we simplify it to $20$}}:
\begin{eqnarray}\label{eq:density:formalconditionplasma}
r_e \ell^{*2}_\pm n_0\left(\frac{R_0\ln(\gamma_0)}{\Lcol}\right)^{4/3}s(\gamma_0)\gg20\,,
\end{eqnarray}
with 
$s(\gamma_0)=\gamma_0^{4/3-0.89}\ln(\gamma_0)^{2/3}\ln(c_1\gamma_0)$ which is ${\simeq 1.63\,\gamma_0^{0.72}}$  for $\gamma_0\in[10^3,10^5]$. 
}

\mica{
It is interesting to rewrite the previous condition considering a Platinum target, identified earlier as optimal for an experiment. Two practical forms can then be derived, whether one considers a pancake-like or a cigar-like pair jet (in the centre-of-mass frame).
}

\mica{
For a pancake-like (disk-shaped) pair jet $(L_\pm^* \simeq \Gamma_b L_0) < (R_\pm^* \simeq 3 R_0)$, $\ell_\pm^*=\Gamma_b L_0$, and condition Eq.~\eqref{eq:density:formalconditionplasma} can be rewritten (valid for $L_{\rm{opt}}/\Lcol<1$) as:
\begin{eqnarray}\label{eq:density:conditionplasma}
    \left[\frac{\mathcal{E}_0}{10\,\rm{GeV}}\right]^{1.42}\left[\frac{eN_0}{1\,\rm{nC}}\right]\left[\frac{L_0}{1\,\mu\rm{m}}\right] \left[\frac{R_0}{5\,\mu\rm{m}}\right]^{-1} \gg 1 \,.\,\,\,\,\,\,\,\,\,\,
\end{eqnarray}
While this condition may look, at first sight, favourable, it is important to note that it is valid only for seeding an electron beam with a pancake-like (disk) shape: $L_0 < 3 R_0/\Gamma_b$ with $\Gamma_b \in [1-50]$. Such beams are not typical for either conventional accelerator or laser wakefield technology. Furthermore, even though compression techniques may be envisioned to decrease $L_0$, $R_0$ cannot be increased too much to ensure that the optimal length $L_{\rm opt}$ [Eq.\eqref{eq:density:max_density}] remains below $L_{\rm sh}$. Hence, the former condition, Eq.~\eqref{eq:density:conditionplasma}, is of limited practical interest.
}

\mica{
Much more relevant to current electron beam technologies is the case of a prolate (cigar-shaped) pair jet $(L_\pm^* \simeq \Gamma_b L_0) > (R_\pm^* \simeq 3 R_0)$. The limiting size is then given by $\ell_\pm^* = 3 R_0$, and condition Eq.~\eqref{eq:density:formalconditionplasma} can be rewritten (valid for $L_{\rm{opt}}/\Lcol<1$):
\begin{eqnarray}\label{eq:density:conditionplasma2}
    \left[\frac{\mathcal{E}_0}{10\,\rm{GeV}}\right]^{0.72}\left[\frac{eN_0}{10\,\rm{nC}}\right]\left[\frac{L_0}{20\,\mu\rm{m}}\right]^{-1}\left[\frac{R_0}{5\,\mu\rm{m}}\right]^{4/3} \gg 1 .\,\,\,\,\,\,\,\,\,\,
\end{eqnarray}
}

\noindent \mica{This equation shows an exacting condition in particularly on the bunch energy and charge.
Since the two practical conditions Eqs. \eqref{eq:density:conditionplasma} and \eqref{eq:density:conditionplasma2} [valid respectively for a pancake-like ($\Gamma_b L_0<3R_0$) and cigar-like ($\Gamma_b L_0>3R_0$) beam] exhibit opposite scalings with radius and longitudinal size, the aspect-ratio $\Gamma_b L_0=3 R_0$ offers the optimal beam shape for reaching collective effects.
}

\subsection{Application to shower seeded by LWFA electron beam}

To illustrate and validate the previous analysis, we now apply our results to a Laser Wakefield Accelerated (LWFA) electron beam colliding against a Tantalum target. \tys{The parameters used in the following example are inspired from the LWFA beam described in \cite{lobet2017generation}. For additional examples, similar simulations have been reported by Song et al. \cite{song2023characterization}.} The beam is modelled as a cylinder of charge $eN_0 = 2$ nC, radius $R_0 = 3$ $\mu$m, length $L_0 = 10~\mu$m, and energy $\gamma_0 mc^2 = 3$ GeV.

\tys{In Fig. \ref{fig:fig5}(a), we show the charge neutrality $N_+/(N_-+N_+)$ of the escaping particles as a function of the normalised target thickness $L/\Lcol$. Red dots are extracted from Geant4 simulations, while the black line is theoretical and computed by considering that all the incident electrons escape from the target. Thus, we have estimate $N_-=N_++N_0$ and $N_+$ from Eq. \eqref{eq:st:N1exact}. The red and grey shaded areas represent regions where $N_-/(N_++N_-)$ exceed $0.4$, respectively extracted from Geant4 simulations and using Eq.~\eqref{eq:shorttime:Npm}. These regions indicate where the outgoing bunch approaches quasi-neutrality. The right boundary of the Geant4 region reflects that, for sufficiently thick targets, the produced pairs are unable to escape the material.}

Additionally, in Fig. \ref{fig:fig5}(b), we present the pair density at the rear of the target as a function of the normalised target thickness $L/\Lcol$. Red dots are extracted from Geant4 simulations while the black line corresponds to Eq.~\eqref{eq:density:density}. As discussed earlier, our model slightly overestimates the simulated density but still shows excellent agreement in both trend and magnitude.

In panel (c), we show the corresponding skin depth as a function of the target length. The red points are extracted from Geant4 results, while the black line is calculated from Eq. \eqref{eq:skindepth} using Eqs. \eqref{eq:density:density} and \eqref{eq:average_energy}. As shown by the blue and green lines representing\tys{, respectively, the longitudinal $L_0$ and transversal $R_\pm$ size of the outgoing beam, the minimal size} remains at least $10$ times smaller than its skin depth for all target lengths. As a result, this typical LWFA electron beam can not produce a pair plasma when colliding with a neutral target.

\tys{This conclusion also applies to the pioneering experiment of Sarri et al. \cite{generation_sarri_2015}, where an electron beam with a maximal energy of $\gamma_0 mc^2 \sim 600$ MeV and charge $eN_0 = 0.3$ nC was collided with a Lead target. Assuming all the electrons at the maximal energy and the beam to be distributed within a cylinder of radius $R_0 \sim 5$ $\mu$m and length $L_0 \sim 5 $ $\mu$m, the number of produced pairs predicted by Eq.~\eqref{eq:shorttime:Npm} is consistent with their measurements.
However, while the maximum pair density in ref. \cite{generation_sarri_2015} is obtained for a target thickness of $\sim 1$ cm, our analysis indicates that the maximum occurs instead for $L \sim 1$ mm in agreement with Song et al. \cite{song2023characterization}. Furthermore, Eq.~\eqref{eq:density:density} yields a density of $\sim 7 \times 10^{15}$ cm$^{-3}$ for a 1 cm target, which is about two orders of magnitude lower than the value reported in \cite{generation_sarri_2015}.
We attribute this discrepancy to differences in the angular distribution of the outgoing beam, as also discussed \cite{williams2020comment} in reference to a similar experiment \cite{sarri2013table}. For a Lead target of $2.5$ cm, our estimate gives a peak angle of $\sim 170$ mrad, whereas their reported divergence is $5-20$ mrad. As a consequence, for a $2.5$ cm target, we obtain an outgoing beam radius of $1.9$ mm, in contrast to their reported value of $\sim 200$ $\mu$m. Under the same parameters, this corresponds to a plasma skin depth of $1.4$ mm, roughly $100$ times larger than the minimal characteristic beam size (the longitudinal dimension). Despite this, we argue that electron–positron filamentation could still be observable in this configuration, since the beam radius and the skin depth remain of comparable magnitude.}

\tys{During the referral process of this work, the recent study~\cite{ludwig2025laser} was brought to our attention, which discusses - via particle-in-cell simulations - the possibility to reach electron energies of 100 GeV using a single stage, all optical LWFA with guiding channel. In particular, one simulation reported the creation of a $100$ GeV electron beam, with radius $R_0 \sim 84~{\rm \mu m}$,  longitudinal size $L_0 \sim 64 {\rm \mu m}$, and a charge $eN_0 \sim 5~{\rm nC}$ about 1 GeV. Even though our condition for pair plasma production [Eqs.~\eqref{eq:density:conditionplasma} and~\eqref{eq:density:conditionplasma2}] is not strictly applicable here as $L_{\rm opt} \sim L_{\rm sh}$, we obtain that the left-hand-side of Eq.~\eqref{eq:density:formalconditionplasma}\footnote{We have used Eq.~\eqref{eq:density:formalconditionplasma} with $l_\pm^{*2}=64$ $\mu$m since $\Gamma_b$ from Eq. \eqref{eq:LorentzBoost:asymtptotic} is smaller than one and thus Eq. \eqref{eq:density:conditionplasma} is not applicable.} is of order~2 times the right-hand-side. Hence, in this 
scenario envisioned for forthcoming experiments at multi-PW laser facilities,  abundant electron-positron pair production is expected, but the plasma condition remains marginal.} 

The main limitation stems from the initial beam charge available from LWFA sources, which is well below the threshold defined by Eqs.~\eqref{eq:density:conditionplasma} and \eqref{eq:density:conditionplasma2} as well as by their limited dimensions. Since condition Eqs.~\eqref{eq:density:conditionplasma} and \eqref{eq:density:conditionplasma2} and the quasi-neutrality condition [Eq.~\eqref{eq:shorttime:Npm} with $N_\pm \gg N_0$] hold for arbitrary electron source, alternative sources providing higher charge, energy or \tys{smaller} spatial dimension could be envisioned, such as electron beams from direct laser acceleration in plasma channels~\cite{babjak2024direct} or electron beams from solid density targets~\cite{ShenPRX2021,MariniPRR2023}. In addition, while our analysis focuses on neutral targets, one could also consider employing ionised materials in which the Bremsstrahlung and Bethe-Heitler cross sections could be increased~\cite{martinez2019high} \tys{and the divergence angle reduced}. Last but not least, the approach presented in this paper can easily be generalised \tys{in order to consider photon-seeded sources such as the ones generated by nuclear reactor \cite{Hugenschmidt2012}, hadronic cascade \cite{Arrowsmith_2024}, non-linear Compton Scattering \cite{Matheron2024_SelfAlignedCompton} or Bremsstrahlung source \cite{noh2024charge}.} It is also worth noticing that the use of high-energy photon sources instead of electron beams can be an adequate solution since the quasi-neutrality is always satisfied in this geometry.\\

\section{Conclusion}

The optimal conditions for creating electron-positron plasmas from electron-seeded EMS in matter are investigated. Building on our previous work on EMS in strong-fields~\cite{pouyez2024kinetic}, we derive explicit expressions for the number of pairs generated from a solid target irradiated by a relativistic electron beam. These expressions, obtained considering both thin and thick targets, are in very good agreement with Geant4 simulations. These also provide us with a simple scaling for the angular divergence of the pairs escaping the target.

\tys{Explicit} expressions are then derived for the density and characteristic size of the escaping pair jet, and a simple criterion for pair plasma production is given as a function of the driving electron beam charge, energy and characteristic size (before it enters the target). Thus, our study identifies the optimal electron beam and target properties for pair plasma production in the laboratory.

We have applied our findings to the case of LWFA electron beams and demonstrated that, with current LWFA technology, the plasma state cannot be achieved. Our results are general and can be extended to other types of electron beams that may be better suited for pair plasma production. Further studies using the same methodology on ionised targets or photon sources are promising, since we believe that it could significantly increase the density of the outgoing pairs.

\tys{Our findings can be relevant to different areas of physics beyond the strong-field community. Specifically, they are of interest to the particle physics community, offering a predictive framework for pair production with potential applications in calorimetry and detector design. They are also valuable to the astrophysics community, as producing pair plasmas in the laboratory will help understanding processes relevant to the most extreme environments in the Universe.}

\section*{Acknowledgments}

The authors thank Sebastian Meuren, Arseny Mironov and Marija Vranic for fruitful discussions. T.G. was supported by FCT (Portugal) Grant No. CEECIND/04050/2021. Financial support by the ANR (g4QED project,
Grant No. ANR-23-CE30-0011) is acknowledged. G.N. was supported by EUR PLASMAScience (France) founded by ANR Grant No. ANE-18-EURE-0014. We acknowledge the referees for their thorough reports that allowed us to improve the manuscript.


\begin{thebibliography}{108}%
\makeatletter
\providecommand \@ifxundefined [1]{%
 \@ifx{#1\undefined}
}%
\providecommand \@ifnum [1]{%
 \ifnum #1\expandafter \@firstoftwo
 \else \expandafter \@secondoftwo
 \fi
}%
\providecommand \@ifx [1]{%
 \ifx #1\expandafter \@firstoftwo
 \else \expandafter \@secondoftwo
 \fi
}%
\providecommand \natexlab [1]{#1}%
\providecommand \enquote  [1]{``#1''}%
\providecommand \bibnamefont  [1]{#1}%
\providecommand \bibfnamefont [1]{#1}%
\providecommand \citenamefont [1]{#1}%
\providecommand \href@noop [0]{\@secondoftwo}%
\providecommand \href [0]{\begingroup \@sanitize@url \@href}%
\providecommand \@href[1]{\@@startlink{#1}\@@href}%
\providecommand \@@href[1]{\endgroup#1\@@endlink}%
\providecommand \@sanitize@url [0]{\catcode `\\12\catcode `\$12\catcode `\&12\catcode `\#12\catcode `\^12\catcode `\_12\catcode `\%12\relax}%
\providecommand \@@startlink[1]{}%
\providecommand \@@endlink[0]{}%
\providecommand \url  [0]{\begingroup\@sanitize@url \@url }%
\providecommand \@url [1]{\endgroup\@href {#1}{\urlprefix }}%
\providecommand \urlprefix  [0]{URL }%
\providecommand \Eprint [0]{\href }%
\providecommand \doibase [0]{https://doi.org/}%
\providecommand \selectlanguage [0]{\@gobble}%
\providecommand \bibinfo  [0]{\@secondoftwo}%
\providecommand \bibfield  [0]{\@secondoftwo}%
\providecommand \translation [1]{[#1]}%
\providecommand \BibitemOpen [0]{}%
\providecommand \bibitemStop [0]{}%
\providecommand \bibitemNoStop [0]{.\EOS\space}%
\providecommand \EOS [0]{\spacefactor3000\relax}%
\providecommand \BibitemShut  [1]{\csname bibitem#1\endcsname}%
\let\auto@bib@innerbib\@empty
\bibitem [{\citenamefont {Bethe}\ and\ \citenamefont {Heitler}(1934)}]{bethe1934stopping}%
  \BibitemOpen
  \bibfield  {author} {\bibinfo {author} {\bibfnamefont {H.}~\bibnamefont {Bethe}}\ and\ \bibinfo {author} {\bibfnamefont {W.}~\bibnamefont {Heitler}},\ }\bibfield  {title} {\bibinfo {title} {On the stopping of fast particles and on the creation of positive electrons},\ }\href@noop {} {\bibfield  {journal} {\bibinfo  {journal} {Proceedings of the Royal Society of London. Series A, Containing Papers of a Mathematical and Physical Character}\ }\textbf {\bibinfo {volume} {146}},\ \bibinfo {pages} {83} (\bibinfo {year} {1934})}\BibitemShut {NoStop}%
\bibitem [{\citenamefont {Fabjan}\ and\ \citenamefont {Gianotti}(2003)}]{fabjan2003calorimetry}%
  \BibitemOpen
  \bibfield  {author} {\bibinfo {author} {\bibfnamefont {C.~W.}\ \bibnamefont {Fabjan}}\ and\ \bibinfo {author} {\bibfnamefont {F.}~\bibnamefont {Gianotti}},\ }\bibfield  {title} {\bibinfo {title} {Calorimetry for particle physics},\ }\href@noop {} {\bibfield  {journal} {\bibinfo  {journal} {Reviews of Modern Physics}\ }\textbf {\bibinfo {volume} {75}},\ \bibinfo {pages} {1243} (\bibinfo {year} {2003})}\BibitemShut {NoStop}%
\bibitem [{\citenamefont {Kampert}\ and\ \citenamefont {Watson}(2012)}]{kampert2012extensive}%
  \BibitemOpen
  \bibfield  {author} {\bibinfo {author} {\bibfnamefont {K.-H.}\ \bibnamefont {Kampert}}\ and\ \bibinfo {author} {\bibfnamefont {A.~A.}\ \bibnamefont {Watson}},\ }\bibfield  {title} {\bibinfo {title} {Extensive air showers and ultra high-energy cosmic rays: a historical review},\ }\href@noop {} {\bibfield  {journal} {\bibinfo  {journal} {The European Physical Journal H}\ }\textbf {\bibinfo {volume} {37}},\ \bibinfo {pages} {359} (\bibinfo {year} {2012})}\BibitemShut {NoStop}%
\bibitem [{\citenamefont {Bertolotti}\ and\ \citenamefont {Bertolotti}(2013)}]{bertolotti2013electromagnetic}%
  \BibitemOpen
  \bibfield  {author} {\bibinfo {author} {\bibfnamefont {M.}~\bibnamefont {Bertolotti}}\ and\ \bibinfo {author} {\bibfnamefont {M.}~\bibnamefont {Bertolotti}},\ }\bibfield  {title} {\bibinfo {title} {Electromagnetic showers},\ }\href@noop {} {\bibfield  {journal} {\bibinfo  {journal} {Celestial Messengers: Cosmic Rays: The Story of a Scientific Adventure}\ ,\ \bibinfo {pages} {127}} (\bibinfo {year} {2013})}\BibitemShut {NoStop}%
\bibitem [{\citenamefont {Watson}(2019)}]{watson2019highest}%
  \BibitemOpen
  \bibfield  {author} {\bibinfo {author} {\bibfnamefont {A.}~\bibnamefont {Watson}},\ }\bibfield  {title} {\bibinfo {title} {The highest-energy cosmic-rays--the past, the present and the future},\ }in\ \href@noop {} {\emph {\bibinfo {booktitle} {EPJ Web of Conferences}}},\ Vol.\ \bibinfo {volume} {210}\ (\bibinfo {organization} {EDP Sciences},\ \bibinfo {year} {2019})\ p.\ \bibinfo {pages} {00001}\BibitemShut {NoStop}%
\bibitem [{\citenamefont {Hillas}(1985)}]{hillas1985cerenkov}%
  \BibitemOpen
  \bibfield  {author} {\bibinfo {author} {\bibfnamefont {A.~M.}\ \bibnamefont {Hillas}},\ }\bibfield  {title} {\bibinfo {title} {Cerenkov light images of eas produced by primary gamma},\ }in\ \href@noop {} {\emph {\bibinfo {booktitle} {19th Intern. Cosmic Ray Conf-Vol. 3}}},\ \bibinfo {series and number} {\bibinfo {number} {OG-9.5-3}}\ (\bibinfo {year} {1985})\BibitemShut {NoStop}%
\bibitem [{\citenamefont {Aharonian}\ \emph {et~al.}(2024)\citenamefont {Aharonian}, \citenamefont {Benkhali}, \citenamefont {Aschersleben}, \citenamefont {Ashkar}, \citenamefont {Backes}, \citenamefont {Martins}, \citenamefont {Batzofin}, \citenamefont {Becherini}, \citenamefont {Berge}, \citenamefont {Bernl{\"o}hr} \emph {et~al.}}]{aharonian2024high}%
  \BibitemOpen
  \bibfield  {author} {\bibinfo {author} {\bibfnamefont {F.}~\bibnamefont {Aharonian}}, \bibinfo {author} {\bibfnamefont {F.~A.}\ \bibnamefont {Benkhali}}, \bibinfo {author} {\bibfnamefont {J.}~\bibnamefont {Aschersleben}}, \bibinfo {author} {\bibfnamefont {H.}~\bibnamefont {Ashkar}}, \bibinfo {author} {\bibfnamefont {M.}~\bibnamefont {Backes}}, \bibinfo {author} {\bibfnamefont {V.~B.}\ \bibnamefont {Martins}}, \bibinfo {author} {\bibfnamefont {R.}~\bibnamefont {Batzofin}}, \bibinfo {author} {\bibfnamefont {Y.}~\bibnamefont {Becherini}}, \bibinfo {author} {\bibfnamefont {D.}~\bibnamefont {Berge}}, \bibinfo {author} {\bibfnamefont {K.}~\bibnamefont {Bernl{\"o}hr}}, \emph {et~al.},\ }\bibfield  {title} {\bibinfo {title} {High-statistics measurement of the cosmic-ray electron spectrum with hess},\ }\href@noop {} {\bibfield  {journal} {\bibinfo  {journal} {Physical Review Letters}\ }\textbf {\bibinfo {volume} {133}},\ \bibinfo {pages} {221001} (\bibinfo {year} {2024})}\BibitemShut {NoStop}%
\bibitem [{\citenamefont {Bahk}\ \emph {et~al.}(2004)\citenamefont {Bahk}, \citenamefont {Rousseau}, \citenamefont {Planchon}, \citenamefont {Chvykov}, \citenamefont {Kalintchenko}, \citenamefont {Maksimchuk}, \citenamefont {Mourou},\ and\ \citenamefont {Yanovsky}}]{generation_bahk_2004}%
  \BibitemOpen
  \bibfield  {author} {\bibinfo {author} {\bibfnamefont {S.-W.}\ \bibnamefont {Bahk}}, \bibinfo {author} {\bibfnamefont {P.}~\bibnamefont {Rousseau}}, \bibinfo {author} {\bibfnamefont {T.~A.}\ \bibnamefont {Planchon}}, \bibinfo {author} {\bibfnamefont {V.}~\bibnamefont {Chvykov}}, \bibinfo {author} {\bibfnamefont {G.}~\bibnamefont {Kalintchenko}}, \bibinfo {author} {\bibfnamefont {A.}~\bibnamefont {Maksimchuk}}, \bibinfo {author} {\bibfnamefont {G.}~\bibnamefont {Mourou}},\ and\ \bibinfo {author} {\bibfnamefont {V.}~\bibnamefont {Yanovsky}},\ }\bibfield  {title} {\bibinfo {title} {Generation and characterization of the highest laser intensities (10(22) w/cm2).},\ }\bibfield  {journal} {\bibinfo  {journal} {Optics Letters}\ }\href {https://doi.org/10.1364/OL.29.002837} {10.1364/OL.29.002837} (\bibinfo {year} {2004})\BibitemShut {NoStop}%
\bibitem [{\citenamefont {Hernandez-Gomez}\ \emph {et~al.}(2010)\citenamefont {Hernandez-Gomez}, \citenamefont {Blake}, \citenamefont {Chekhlov}, \citenamefont {Clarke}, \citenamefont {Dunne}, \citenamefont {Galimberti}, \citenamefont {Hancock}, \citenamefont {Heathcote}, \citenamefont {Holligan}, \citenamefont {Lyachev}, \citenamefont {Matousek}, \citenamefont {Musgrave}, \citenamefont {Neely}, \citenamefont {Norreys}, \citenamefont {Ross}, \citenamefont {Tang}, \citenamefont {Winstone}, \citenamefont {Wyborn},\ and\ \citenamefont {Collier}}]{vulcan_hernandezgomez_2010}%
  \BibitemOpen
  \bibfield  {author} {\bibinfo {author} {\bibfnamefont {C.}~\bibnamefont {Hernandez-Gomez}}, \bibinfo {author} {\bibfnamefont {S.}~\bibnamefont {Blake}}, \bibinfo {author} {\bibfnamefont {O.}~\bibnamefont {Chekhlov}}, \bibinfo {author} {\bibfnamefont {R.}~\bibnamefont {Clarke}}, \bibinfo {author} {\bibfnamefont {A.}~\bibnamefont {Dunne}}, \bibinfo {author} {\bibfnamefont {M.}~\bibnamefont {Galimberti}}, \bibinfo {author} {\bibfnamefont {S.}~\bibnamefont {Hancock}}, \bibinfo {author} {\bibfnamefont {R.}~\bibnamefont {Heathcote}}, \bibinfo {author} {\bibfnamefont {P.}~\bibnamefont {Holligan}}, \bibinfo {author} {\bibfnamefont {A.}~\bibnamefont {Lyachev}}, \bibinfo {author} {\bibfnamefont {P.}~\bibnamefont {Matousek}}, \bibinfo {author} {\bibfnamefont {I.}~\bibnamefont {Musgrave}}, \bibinfo {author} {\bibfnamefont {D.}~\bibnamefont {Neely}}, \bibinfo {author} {\bibfnamefont {P.}~\bibnamefont {Norreys}}, \bibinfo {author} {\bibfnamefont {I.}~\bibnamefont {Ross}}, \bibinfo {author} {\bibfnamefont
  {Y.}~\bibnamefont {Tang}}, \bibinfo {author} {\bibfnamefont {T.}~\bibnamefont {Winstone}}, \bibinfo {author} {\bibfnamefont {B.}~\bibnamefont {Wyborn}},\ and\ \bibinfo {author} {\bibfnamefont {J.}~\bibnamefont {Collier}},\ }\bibfield  {title} {\bibinfo {title} {The vulcan 10 pw project}\ }\href {https://doi.org/10.1088/1742-6596/244/3/032006} {10.1088/1742-6596/244/3/032006} (\bibinfo {year} {2010})\BibitemShut {NoStop}%
\bibitem [{\citenamefont {Papadopoulos}\ \emph {et~al.}(2016)\citenamefont {Papadopoulos}, \citenamefont {Zou}, \citenamefont {Blanc}, \citenamefont {Chériaux}, \citenamefont {Georges}, \citenamefont {Druon}, \citenamefont {Mennerat}, \citenamefont {Ramirez}, \citenamefont {Martin}, \citenamefont {Fréneaux}, \citenamefont {Beluze}, \citenamefont {Lebas}, \citenamefont {Monot}, \citenamefont {Mathieu},\ and\ \citenamefont {Audebert}}]{apollon_papadopoulos_2016}%
  \BibitemOpen
  \bibfield  {author} {\bibinfo {author} {\bibfnamefont {D.}~\bibnamefont {Papadopoulos}}, \bibinfo {author} {\bibfnamefont {J.}~\bibnamefont {Zou}}, \bibinfo {author} {\bibfnamefont {C.~L.}\ \bibnamefont {Blanc}}, \bibinfo {author} {\bibfnamefont {G.}~\bibnamefont {Chériaux}}, \bibinfo {author} {\bibfnamefont {P.}~\bibnamefont {Georges}}, \bibinfo {author} {\bibfnamefont {F.}~\bibnamefont {Druon}}, \bibinfo {author} {\bibfnamefont {G.}~\bibnamefont {Mennerat}}, \bibinfo {author} {\bibfnamefont {P.}~\bibnamefont {Ramirez}}, \bibinfo {author} {\bibfnamefont {L.}~\bibnamefont {Martin}}, \bibinfo {author} {\bibfnamefont {A.}~\bibnamefont {Fréneaux}}, \bibinfo {author} {\bibfnamefont {A.}~\bibnamefont {Beluze}}, \bibinfo {author} {\bibfnamefont {N.}~\bibnamefont {Lebas}}, \bibinfo {author} {\bibfnamefont {P.}~\bibnamefont {Monot}}, \bibinfo {author} {\bibfnamefont {F.}~\bibnamefont {Mathieu}},\ and\ \bibinfo {author} {\bibfnamefont {P.}~\bibnamefont {Audebert}},\ }\bibfield  {title} {\bibinfo {title} {The
  apollon 10 pw laser: experimental and theoretical investigation of the temporal characteristics},\ }\bibfield  {journal} {\bibinfo  {journal} {High Power Laser Science and Engineering}\ }\href {https://doi.org/10.1017/HPL.2016.34} {10.1017/HPL.2016.34} (\bibinfo {year} {2016})\BibitemShut {NoStop}%
\bibitem [{ELI()}]{ELI}%
  \BibitemOpen
  \href@noop {} {}\bibinfo {note} {{\it Extreme light infrastructure} (ELI), \url{https://eli-laser.eu}}\BibitemShut {NoStop}%
\bibitem [{ZEU()}]{ZEUS}%
  \BibitemOpen
  \href@noop {} {}\bibinfo {note} {{\it Zettawatt-equivalent ultrashort pulse laser system } (ZEUS), \url{https://zeus.engin.umich.edu}}\BibitemShut {NoStop}%
\bibitem [{\citenamefont {Nam}\ \emph {et~al.}(2018)\citenamefont {Nam}, \citenamefont {Sung}, \citenamefont {Lee}, \citenamefont {Youn},\ and\ \citenamefont {Lee}}]{nam2018performance}%
  \BibitemOpen
  \bibfield  {author} {\bibinfo {author} {\bibfnamefont {C.~H.}\ \bibnamefont {Nam}}, \bibinfo {author} {\bibfnamefont {J.~H.}\ \bibnamefont {Sung}}, \bibinfo {author} {\bibfnamefont {H.~W.}\ \bibnamefont {Lee}}, \bibinfo {author} {\bibfnamefont {J.~W.}\ \bibnamefont {Youn}},\ and\ \bibinfo {author} {\bibfnamefont {S.~K.}\ \bibnamefont {Lee}},\ }\bibfield  {title} {\bibinfo {title} {Performance of the 20 fs, 4 pw ti: sapphire laser at corels},\ }in\ \href@noop {} {\emph {\bibinfo {booktitle} {CLEO: Science and Innovations}}}\ (\bibinfo {organization} {Optica Publishing Group},\ \bibinfo {year} {2018})\ pp.\ \bibinfo {pages} {STu4O--3}\BibitemShut {NoStop}%
\bibitem [{\citenamefont {Danson}\ \emph {et~al.}(2019)\citenamefont {Danson}, \citenamefont {Haefner}, \citenamefont {Bromage}, \citenamefont {Butcher}, \citenamefont {Chanteloup}, \citenamefont {Chowdhury}, \citenamefont {Galvanauskas}, \citenamefont {Gizzi}, \citenamefont {Hein}, \citenamefont {Hillier}, \citenamefont {Hopps}, \citenamefont {Kato}, \citenamefont {Khazanov}, \citenamefont {Kodama}, \citenamefont {Korn}, \citenamefont {Li}, \citenamefont {Li}, \citenamefont {Limpert}, \citenamefont {Ma}, \citenamefont {Nam}, \citenamefont {Neely}, \citenamefont {Papadopoulos}, \citenamefont {Penman}, \citenamefont {Qian}, \citenamefont {Rocca}, \citenamefont {Shaykin}, \citenamefont {Siders}, \citenamefont {Spindloe}, \citenamefont {Szatmári}, \citenamefont {Trines}, \citenamefont {Zhu}, \citenamefont {Zhu},\ and\ \citenamefont {Zuegel}}]{petawatt_danson_2019}%
  \BibitemOpen
  \bibfield  {author} {\bibinfo {author} {\bibfnamefont {C.}~\bibnamefont {Danson}}, \bibinfo {author} {\bibfnamefont {C.}~\bibnamefont {Haefner}}, \bibinfo {author} {\bibfnamefont {J.}~\bibnamefont {Bromage}}, \bibinfo {author} {\bibfnamefont {T.}~\bibnamefont {Butcher}}, \bibinfo {author} {\bibfnamefont {J.}~\bibnamefont {Chanteloup}}, \bibinfo {author} {\bibfnamefont {E.}~\bibnamefont {Chowdhury}}, \bibinfo {author} {\bibfnamefont {A.}~\bibnamefont {Galvanauskas}}, \bibinfo {author} {\bibfnamefont {L.}~\bibnamefont {Gizzi}}, \bibinfo {author} {\bibfnamefont {J.}~\bibnamefont {Hein}}, \bibinfo {author} {\bibfnamefont {D.}~\bibnamefont {Hillier}}, \bibinfo {author} {\bibfnamefont {N.}~\bibnamefont {Hopps}}, \bibinfo {author} {\bibfnamefont {Y.}~\bibnamefont {Kato}}, \bibinfo {author} {\bibfnamefont {E.}~\bibnamefont {Khazanov}}, \bibinfo {author} {\bibfnamefont {R.}~\bibnamefont {Kodama}}, \bibinfo {author} {\bibfnamefont {G.}~\bibnamefont {Korn}}, \bibinfo {author} {\bibfnamefont {R.}~\bibnamefont {Li}},
  \bibinfo {author} {\bibfnamefont {Y.}~\bibnamefont {Li}}, \bibinfo {author} {\bibfnamefont {J.}~\bibnamefont {Limpert}}, \bibinfo {author} {\bibfnamefont {J.}~\bibnamefont {Ma}}, \bibinfo {author} {\bibfnamefont {C.}~\bibnamefont {Nam}}, \bibinfo {author} {\bibfnamefont {D.}~\bibnamefont {Neely}}, \bibinfo {author} {\bibfnamefont {D.}~\bibnamefont {Papadopoulos}}, \bibinfo {author} {\bibfnamefont {R.}~\bibnamefont {Penman}}, \bibinfo {author} {\bibfnamefont {L.}~\bibnamefont {Qian}}, \bibinfo {author} {\bibfnamefont {J.}~\bibnamefont {Rocca}}, \bibinfo {author} {\bibfnamefont {A.}~\bibnamefont {Shaykin}}, \bibinfo {author} {\bibfnamefont {C.}~\bibnamefont {Siders}}, \bibinfo {author} {\bibfnamefont {C.}~\bibnamefont {Spindloe}}, \bibinfo {author} {\bibfnamefont {S.}~\bibnamefont {Szatmári}}, \bibinfo {author} {\bibfnamefont {R.}~\bibnamefont {Trines}}, \bibinfo {author} {\bibfnamefont {J.}~\bibnamefont {Zhu}}, \bibinfo {author} {\bibfnamefont {P.}~\bibnamefont {Zhu}},\ and\ \bibinfo {author} {\bibfnamefont
  {J.}~\bibnamefont {Zuegel}},\ }\bibfield  {title} {\bibinfo {title} {Petawatt and exawatt class lasers worldwide},\ }\bibfield  {journal} {\bibinfo  {journal} {High Power Laser Science and Engineering}\ }\href {https://doi.org/10.1017/HPL.2019.36} {10.1017/HPL.2019.36} (\bibinfo {year} {2019})\BibitemShut {NoStop}%
\bibitem [{\citenamefont {Bromage}\ \emph {et~al.}(2019)\citenamefont {Bromage}, \citenamefont {Bahk}, \citenamefont {Begishev}, \citenamefont {Dorrer}, \citenamefont {Guardalben}, \citenamefont {Hoffman}, \citenamefont {Oliver}, \citenamefont {Roides}, \citenamefont {Schiesser}, \citenamefont {Iii}, \citenamefont {Spilatro}, \citenamefont {Webb}, \citenamefont {Weiner},\ and\ \citenamefont {Zuegel}}]{technology_bromage_2019}%
  \BibitemOpen
  \bibfield  {author} {\bibinfo {author} {\bibfnamefont {J.}~\bibnamefont {Bromage}}, \bibinfo {author} {\bibfnamefont {S.}~\bibnamefont {Bahk}}, \bibinfo {author} {\bibfnamefont {I.}~\bibnamefont {Begishev}}, \bibinfo {author} {\bibfnamefont {C.}~\bibnamefont {Dorrer}}, \bibinfo {author} {\bibfnamefont {M.}~\bibnamefont {Guardalben}}, \bibinfo {author} {\bibfnamefont {B.}~\bibnamefont {Hoffman}}, \bibinfo {author} {\bibfnamefont {J.}~\bibnamefont {Oliver}}, \bibinfo {author} {\bibfnamefont {R.}~\bibnamefont {Roides}}, \bibinfo {author} {\bibfnamefont {E.}~\bibnamefont {Schiesser}}, \bibinfo {author} {\bibfnamefont {M.~J.~S.}\ \bibnamefont {Iii}}, \bibinfo {author} {\bibfnamefont {M.}~\bibnamefont {Spilatro}}, \bibinfo {author} {\bibfnamefont {B.}~\bibnamefont {Webb}}, \bibinfo {author} {\bibfnamefont {D.}~\bibnamefont {Weiner}},\ and\ \bibinfo {author} {\bibfnamefont {J.}~\bibnamefont {Zuegel}},\ }\bibfield  {title} {\bibinfo {title} {Technology development for ultraintense all-opcpa systems},\ }\bibfield
  {journal} {\bibinfo  {journal} {High Power Laser Science and Engineering}\ }\href {https://doi.org/10.1017/HPL.2018.64} {10.1017/HPL.2018.64} (\bibinfo {year} {2019})\BibitemShut {NoStop}%
\bibitem [{\citenamefont {Khazanov}\ \emph {et~al.}(2023)\citenamefont {Khazanov}, \citenamefont {Shaykin}, \citenamefont {Kostyukov}, \citenamefont {Ginzburg}, \citenamefont {Mukhin}, \citenamefont {Yakovlev}, \citenamefont {Soloviev}, \citenamefont {Kuznetsov}, \citenamefont {Mironov}, \citenamefont {Korzhimanov} \emph {et~al.}}]{khazanov2023exawatt}%
  \BibitemOpen
  \bibfield  {author} {\bibinfo {author} {\bibfnamefont {E.}~\bibnamefont {Khazanov}}, \bibinfo {author} {\bibfnamefont {A.}~\bibnamefont {Shaykin}}, \bibinfo {author} {\bibfnamefont {I.}~\bibnamefont {Kostyukov}}, \bibinfo {author} {\bibfnamefont {V.}~\bibnamefont {Ginzburg}}, \bibinfo {author} {\bibfnamefont {I.}~\bibnamefont {Mukhin}}, \bibinfo {author} {\bibfnamefont {I.}~\bibnamefont {Yakovlev}}, \bibinfo {author} {\bibfnamefont {A.}~\bibnamefont {Soloviev}}, \bibinfo {author} {\bibfnamefont {I.}~\bibnamefont {Kuznetsov}}, \bibinfo {author} {\bibfnamefont {S.}~\bibnamefont {Mironov}}, \bibinfo {author} {\bibfnamefont {A.}~\bibnamefont {Korzhimanov}}, \emph {et~al.},\ }\bibfield  {title} {\bibinfo {title} {exawatt center for extreme light studies},\ }\href@noop {} {\bibfield  {journal} {\bibinfo  {journal} {High Power Laser Science and Engineering}\ }\textbf {\bibinfo {volume} {11}},\ \bibinfo {pages} {e78} (\bibinfo {year} {2023})}\BibitemShut {NoStop}%
\bibitem [{\citenamefont {Arefiev}\ \emph {et~al.}(2016)\citenamefont {Arefiev}, \citenamefont {Khudik}, \citenamefont {Robinson}, \citenamefont {Shvets}, \citenamefont {Willingale},\ and\ \citenamefont {Schollmeier}}]{arefiev2016beyond}%
  \BibitemOpen
  \bibfield  {author} {\bibinfo {author} {\bibfnamefont {A.}~\bibnamefont {Arefiev}}, \bibinfo {author} {\bibfnamefont {V.}~\bibnamefont {Khudik}}, \bibinfo {author} {\bibfnamefont {A.}~\bibnamefont {Robinson}}, \bibinfo {author} {\bibfnamefont {G.}~\bibnamefont {Shvets}}, \bibinfo {author} {\bibfnamefont {L.}~\bibnamefont {Willingale}},\ and\ \bibinfo {author} {\bibfnamefont {M.}~\bibnamefont {Schollmeier}},\ }\bibfield  {title} {\bibinfo {title} {Beyond the ponderomotive limit: Direct laser acceleration of relativistic electrons in sub-critical plasmas},\ }\href@noop {} {\bibfield  {journal} {\bibinfo  {journal} {Physics of Plasmas}\ }\textbf {\bibinfo {volume} {23}} (\bibinfo {year} {2016})}\BibitemShut {NoStop}%
\bibitem [{\citenamefont {Kim}\ \emph {et~al.}(2017)\citenamefont {Kim}, \citenamefont {Pathak}, \citenamefont {Hong~Pae}, \citenamefont {Lifschitz}, \citenamefont {Sylla}, \citenamefont {Shin}, \citenamefont {Hojbota}, \citenamefont {Lee}, \citenamefont {Sung}, \citenamefont {Lee} \emph {et~al.}}]{kim2017stable}%
  \BibitemOpen
  \bibfield  {author} {\bibinfo {author} {\bibfnamefont {H.~T.}\ \bibnamefont {Kim}}, \bibinfo {author} {\bibfnamefont {V.}~\bibnamefont {Pathak}}, \bibinfo {author} {\bibfnamefont {K.}~\bibnamefont {Hong~Pae}}, \bibinfo {author} {\bibfnamefont {A.}~\bibnamefont {Lifschitz}}, \bibinfo {author} {\bibfnamefont {F.}~\bibnamefont {Sylla}}, \bibinfo {author} {\bibfnamefont {J.~H.}\ \bibnamefont {Shin}}, \bibinfo {author} {\bibfnamefont {C.}~\bibnamefont {Hojbota}}, \bibinfo {author} {\bibfnamefont {S.~K.}\ \bibnamefont {Lee}}, \bibinfo {author} {\bibfnamefont {J.~H.}\ \bibnamefont {Sung}}, \bibinfo {author} {\bibfnamefont {H.~W.}\ \bibnamefont {Lee}}, \emph {et~al.},\ }\bibfield  {title} {\bibinfo {title} {Stable multi-gev electron accelerator driven by waveform-controlled pw laser pulses},\ }\href@noop {} {\bibfield  {journal} {\bibinfo  {journal} {Scientific reports}\ }\textbf {\bibinfo {volume} {7}},\ \bibinfo {pages} {10203} (\bibinfo {year} {2017})}\BibitemShut {NoStop}%
\bibitem [{\citenamefont {Gonsalves}\ \emph {et~al.}(2019)\citenamefont {Gonsalves}, \citenamefont {Nakamura}, \citenamefont {Daniels}, \citenamefont {Benedetti}, \citenamefont {Pieronek}, \citenamefont {De~Raadt}, \citenamefont {Steinke}, \citenamefont {Bin}, \citenamefont {Bulanov}, \citenamefont {Van~Tilborg} \emph {et~al.}}]{gonsalves2019petawatt}%
  \BibitemOpen
  \bibfield  {author} {\bibinfo {author} {\bibfnamefont {A.}~\bibnamefont {Gonsalves}}, \bibinfo {author} {\bibfnamefont {K.}~\bibnamefont {Nakamura}}, \bibinfo {author} {\bibfnamefont {J.}~\bibnamefont {Daniels}}, \bibinfo {author} {\bibfnamefont {C.}~\bibnamefont {Benedetti}}, \bibinfo {author} {\bibfnamefont {C.}~\bibnamefont {Pieronek}}, \bibinfo {author} {\bibfnamefont {T.}~\bibnamefont {De~Raadt}}, \bibinfo {author} {\bibfnamefont {S.}~\bibnamefont {Steinke}}, \bibinfo {author} {\bibfnamefont {J.}~\bibnamefont {Bin}}, \bibinfo {author} {\bibfnamefont {S.}~\bibnamefont {Bulanov}}, \bibinfo {author} {\bibfnamefont {J.}~\bibnamefont {Van~Tilborg}}, \emph {et~al.},\ }\bibfield  {title} {\bibinfo {title} {Petawatt laser guiding and electron beam acceleration to 8 gev in a laser-heated capillary discharge waveguide},\ }\href@noop {} {\bibfield  {journal} {\bibinfo  {journal} {Physical review letters}\ }\textbf {\bibinfo {volume} {122}},\ \bibinfo {pages} {084801} (\bibinfo {year} {2019})}\BibitemShut
  {NoStop}%
\bibitem [{\citenamefont {Jirka}\ \emph {et~al.}(2020)\citenamefont {Jirka}, \citenamefont {Vranic}, \citenamefont {Grismayer},\ and\ \citenamefont {Silva}}]{jirka2020scaling}%
  \BibitemOpen
  \bibfield  {author} {\bibinfo {author} {\bibfnamefont {M.}~\bibnamefont {Jirka}}, \bibinfo {author} {\bibfnamefont {M.}~\bibnamefont {Vranic}}, \bibinfo {author} {\bibfnamefont {T.}~\bibnamefont {Grismayer}},\ and\ \bibinfo {author} {\bibfnamefont {L.}~\bibnamefont {Silva}},\ }\bibfield  {title} {\bibinfo {title} {Scaling laws for direct laser acceleration in a radiation-reaction dominated regime},\ }\href@noop {} {\bibfield  {journal} {\bibinfo  {journal} {New Journal of Physics}\ }\textbf {\bibinfo {volume} {22}},\ \bibinfo {pages} {083058} (\bibinfo {year} {2020})}\BibitemShut {NoStop}%
\bibitem [{\citenamefont {Kim}\ \emph {et~al.}(2021)\citenamefont {Kim}, \citenamefont {Pathak}, \citenamefont {Hojbota}, \citenamefont {Mirzaie}, \citenamefont {Pae}, \citenamefont {Kim}, \citenamefont {Yoon}, \citenamefont {Sung},\ and\ \citenamefont {Lee}}]{kim2021multi}%
  \BibitemOpen
  \bibfield  {author} {\bibinfo {author} {\bibfnamefont {H.~T.}\ \bibnamefont {Kim}}, \bibinfo {author} {\bibfnamefont {V.~B.}\ \bibnamefont {Pathak}}, \bibinfo {author} {\bibfnamefont {C.~I.}\ \bibnamefont {Hojbota}}, \bibinfo {author} {\bibfnamefont {M.}~\bibnamefont {Mirzaie}}, \bibinfo {author} {\bibfnamefont {K.~H.}\ \bibnamefont {Pae}}, \bibinfo {author} {\bibfnamefont {C.~M.}\ \bibnamefont {Kim}}, \bibinfo {author} {\bibfnamefont {J.~W.}\ \bibnamefont {Yoon}}, \bibinfo {author} {\bibfnamefont {J.~H.}\ \bibnamefont {Sung}},\ and\ \bibinfo {author} {\bibfnamefont {S.~K.}\ \bibnamefont {Lee}},\ }\bibfield  {title} {\bibinfo {title} {Multi-gev laser wakefield electron acceleration with pw lasers},\ }\href@noop {} {\bibfield  {journal} {\bibinfo  {journal} {Applied Sciences}\ }\textbf {\bibinfo {volume} {11}},\ \bibinfo {pages} {5831} (\bibinfo {year} {2021})}\BibitemShut {NoStop}%
\bibitem [{\citenamefont {Miao}\ \emph {et~al.}(2022)\citenamefont {Miao}, \citenamefont {Shrock}, \citenamefont {Feder}, \citenamefont {Hollinger}, \citenamefont {Morrison}, \citenamefont {Nedbailo}, \citenamefont {Picksley}, \citenamefont {Song}, \citenamefont {Wang}, \citenamefont {Rocca} \emph {et~al.}}]{miao2022multi}%
  \BibitemOpen
  \bibfield  {author} {\bibinfo {author} {\bibfnamefont {B.}~\bibnamefont {Miao}}, \bibinfo {author} {\bibfnamefont {J.}~\bibnamefont {Shrock}}, \bibinfo {author} {\bibfnamefont {L.}~\bibnamefont {Feder}}, \bibinfo {author} {\bibfnamefont {R.}~\bibnamefont {Hollinger}}, \bibinfo {author} {\bibfnamefont {J.}~\bibnamefont {Morrison}}, \bibinfo {author} {\bibfnamefont {R.}~\bibnamefont {Nedbailo}}, \bibinfo {author} {\bibfnamefont {A.}~\bibnamefont {Picksley}}, \bibinfo {author} {\bibfnamefont {H.}~\bibnamefont {Song}}, \bibinfo {author} {\bibfnamefont {S.}~\bibnamefont {Wang}}, \bibinfo {author} {\bibfnamefont {J.}~\bibnamefont {Rocca}}, \emph {et~al.},\ }\bibfield  {title} {\bibinfo {title} {Multi-gev electron bunches from an all-optical laser wakefield accelerator},\ }\href@noop {} {\bibfield  {journal} {\bibinfo  {journal} {Physical Review X}\ }\textbf {\bibinfo {volume} {12}},\ \bibinfo {pages} {031038} (\bibinfo {year} {2022})}\BibitemShut {NoStop}%
\bibitem [{\citenamefont {Poder}\ \emph {et~al.}(2024)\citenamefont {Poder}, \citenamefont {Wood}, \citenamefont {Lopes}, \citenamefont {Cole}, \citenamefont {Alatabi}, \citenamefont {Backhouse}, \citenamefont {Foster}, \citenamefont {Hughes}, \citenamefont {Kamperidis}, \citenamefont {Kononenko} \emph {et~al.}}]{poder2024multi}%
  \BibitemOpen
  \bibfield  {author} {\bibinfo {author} {\bibfnamefont {K.}~\bibnamefont {Poder}}, \bibinfo {author} {\bibfnamefont {J.}~\bibnamefont {Wood}}, \bibinfo {author} {\bibfnamefont {N.}~\bibnamefont {Lopes}}, \bibinfo {author} {\bibfnamefont {J.}~\bibnamefont {Cole}}, \bibinfo {author} {\bibfnamefont {S.}~\bibnamefont {Alatabi}}, \bibinfo {author} {\bibfnamefont {M.}~\bibnamefont {Backhouse}}, \bibinfo {author} {\bibfnamefont {P.}~\bibnamefont {Foster}}, \bibinfo {author} {\bibfnamefont {A.}~\bibnamefont {Hughes}}, \bibinfo {author} {\bibfnamefont {C.}~\bibnamefont {Kamperidis}}, \bibinfo {author} {\bibfnamefont {O.}~\bibnamefont {Kononenko}}, \emph {et~al.},\ }\bibfield  {title} {\bibinfo {title} {Multi-gev electron acceleration in wakefields strongly driven by oversized laser spots},\ }\href@noop {} {\bibfield  {journal} {\bibinfo  {journal} {Physical review letters}\ }\textbf {\bibinfo {volume} {132}},\ \bibinfo {pages} {195001} (\bibinfo {year} {2024})}\BibitemShut {NoStop}%
\bibitem [{\citenamefont {Babjak}\ \emph {et~al.}(2024)\citenamefont {Babjak}, \citenamefont {Willingale}, \citenamefont {Arefiev},\ and\ \citenamefont {Vranic}}]{babjak2024direct}%
  \BibitemOpen
  \bibfield  {author} {\bibinfo {author} {\bibfnamefont {R.}~\bibnamefont {Babjak}}, \bibinfo {author} {\bibfnamefont {L.}~\bibnamefont {Willingale}}, \bibinfo {author} {\bibfnamefont {A.}~\bibnamefont {Arefiev}},\ and\ \bibinfo {author} {\bibfnamefont {M.}~\bibnamefont {Vranic}},\ }\bibfield  {title} {\bibinfo {title} {Direct laser acceleration in underdense plasmas with multi-pw lasers: a path to high-charge, gev-class electron bunches},\ }\href@noop {} {\bibfield  {journal} {\bibinfo  {journal} {Physical Review Letters}\ }\textbf {\bibinfo {volume} {132}},\ \bibinfo {pages} {125001} (\bibinfo {year} {2024})}\BibitemShut {NoStop}%
\bibitem [{\citenamefont {Sarri}\ \emph {et~al.}(2014)\citenamefont {Sarri}, \citenamefont {Corvan}, \citenamefont {Schumaker}, \citenamefont {Cole}, \citenamefont {Di~Piazza}, \citenamefont {Ahmed}, \citenamefont {Harvey}, \citenamefont {Keitel}, \citenamefont {Krushelnick}, \citenamefont {Mangles}, \citenamefont {Najmudin}, \citenamefont {Symes}, \citenamefont {Thomas}, \citenamefont {Yeung}, \citenamefont {Zhao},\ and\ \citenamefont {Zepf}}]{Sarri2014_NonLinearThomson}%
  \BibitemOpen
  \bibfield  {author} {\bibinfo {author} {\bibfnamefont {G.}~\bibnamefont {Sarri}}, \bibinfo {author} {\bibfnamefont {D.~J.}\ \bibnamefont {Corvan}}, \bibinfo {author} {\bibfnamefont {W.}~\bibnamefont {Schumaker}}, \bibinfo {author} {\bibfnamefont {J.~M.}\ \bibnamefont {Cole}}, \bibinfo {author} {\bibfnamefont {A.}~\bibnamefont {Di~Piazza}}, \bibinfo {author} {\bibfnamefont {H.}~\bibnamefont {Ahmed}}, \bibinfo {author} {\bibfnamefont {C.}~\bibnamefont {Harvey}}, \bibinfo {author} {\bibfnamefont {C.~H.}\ \bibnamefont {Keitel}}, \bibinfo {author} {\bibfnamefont {K.}~\bibnamefont {Krushelnick}}, \bibinfo {author} {\bibfnamefont {S.~P.~D.}\ \bibnamefont {Mangles}}, \bibinfo {author} {\bibfnamefont {Z.}~\bibnamefont {Najmudin}}, \bibinfo {author} {\bibfnamefont {D.}~\bibnamefont {Symes}}, \bibinfo {author} {\bibfnamefont {A.~G.~R.}\ \bibnamefont {Thomas}}, \bibinfo {author} {\bibfnamefont {M.}~\bibnamefont {Yeung}}, \bibinfo {author} {\bibfnamefont {Z.}~\bibnamefont {Zhao}},\ and\ \bibinfo {author} {\bibfnamefont
  {M.}~\bibnamefont {Zepf}},\ }\bibfield  {title} {\bibinfo {title} {Ultrahigh brilliance multi-mev $\gamma$-ray beams from nonlinear relativistic thomson scattering},\ }\href {https://doi.org/10.1103/PhysRevLett.113.224801} {\bibfield  {journal} {\bibinfo  {journal} {Physical Review Letters}\ }\textbf {\bibinfo {volume} {113}},\ \bibinfo {pages} {224801} (\bibinfo {year} {2014})}\BibitemShut {NoStop}%
\bibitem [{\citenamefont {Albert}\ and\ \citenamefont {Thomas}(2016)}]{albert2016applications}%
  \BibitemOpen
  \bibfield  {author} {\bibinfo {author} {\bibfnamefont {F.}~\bibnamefont {Albert}}\ and\ \bibinfo {author} {\bibfnamefont {A.~G.}\ \bibnamefont {Thomas}},\ }\bibfield  {title} {\bibinfo {title} {Applications of laser wakefield accelerator-based light sources},\ }\href@noop {} {\bibfield  {journal} {\bibinfo  {journal} {Plasma Physics and Controlled Fusion}\ }\textbf {\bibinfo {volume} {58}},\ \bibinfo {pages} {103001} (\bibinfo {year} {2016})}\BibitemShut {NoStop}%
\bibitem [{\citenamefont {Gong}\ \emph {et~al.}(2018)\citenamefont {Gong}, \citenamefont {Hu}, \citenamefont {Lu}, \citenamefont {Yu}, \citenamefont {Wang}, \citenamefont {Fu}, \citenamefont {Liu},\ and\ \citenamefont {Yan}}]{Gong2018_GeVFlash}%
  \BibitemOpen
  \bibfield  {author} {\bibinfo {author} {\bibfnamefont {Z.}~\bibnamefont {Gong}}, \bibinfo {author} {\bibfnamefont {R.}~\bibnamefont {Hu}}, \bibinfo {author} {\bibfnamefont {H.}~\bibnamefont {Lu}}, \bibinfo {author} {\bibfnamefont {J.}~\bibnamefont {Yu}}, \bibinfo {author} {\bibfnamefont {D.}~\bibnamefont {Wang}}, \bibinfo {author} {\bibfnamefont {E.-G.}\ \bibnamefont {Fu}}, \bibinfo {author} {\bibfnamefont {J.}~\bibnamefont {Liu}},\ and\ \bibinfo {author} {\bibfnamefont {X.}~\bibnamefont {Yan}},\ }\bibfield  {title} {\bibinfo {title} {Brilliant gev gamma-ray flash from inverse compton scattering in the qed regime},\ }\href {https://doi.org/10.1088/1361-6587/aaa9b1} {\bibfield  {journal} {\bibinfo  {journal} {Plasma Physics and Controlled Fusion}\ }\textbf {\bibinfo {volume} {60}},\ \bibinfo {pages} {044004} (\bibinfo {year} {2018})}\BibitemShut {NoStop}%
\bibitem [{\citenamefont {Gu}\ \emph {et~al.}(2018)\citenamefont {Gu}, \citenamefont {Klimo}, \citenamefont {Bulanov},\ and\ \citenamefont {Weber}}]{Gu2018_AttosecondGamma}%
  \BibitemOpen
  \bibfield  {author} {\bibinfo {author} {\bibfnamefont {Y.-J.}\ \bibnamefont {Gu}}, \bibinfo {author} {\bibfnamefont {O.}~\bibnamefont {Klimo}}, \bibinfo {author} {\bibfnamefont {S.~V.}\ \bibnamefont {Bulanov}},\ and\ \bibinfo {author} {\bibfnamefont {S.}~\bibnamefont {Weber}},\ }\bibfield  {title} {\bibinfo {title} {Brilliant gamma-ray beam and electron–positron pair production by enhanced attosecond pulses},\ }\href {https://doi.org/10.1038/s42005-018-0095-3} {\bibfield  {journal} {\bibinfo  {journal} {Communications Physics}\ }\textbf {\bibinfo {volume} {1}},\ \bibinfo {pages} {93} (\bibinfo {year} {2018})}\BibitemShut {NoStop}%
\bibitem [{\citenamefont {Huang}\ \emph {et~al.}(2019)\citenamefont {Huang}, \citenamefont {Kim}, \citenamefont {Zhou}, \citenamefont {Cho}, \citenamefont {Nakajima}, \citenamefont {Ryu}, \citenamefont {Ruan},\ and\ \citenamefont {Nam}}]{huang2019highly}%
  \BibitemOpen
  \bibfield  {author} {\bibinfo {author} {\bibfnamefont {T.}~\bibnamefont {Huang}}, \bibinfo {author} {\bibfnamefont {C.~M.}\ \bibnamefont {Kim}}, \bibinfo {author} {\bibfnamefont {C.}~\bibnamefont {Zhou}}, \bibinfo {author} {\bibfnamefont {M.~H.}\ \bibnamefont {Cho}}, \bibinfo {author} {\bibfnamefont {K.}~\bibnamefont {Nakajima}}, \bibinfo {author} {\bibfnamefont {C.~M.}\ \bibnamefont {Ryu}}, \bibinfo {author} {\bibfnamefont {S.}~\bibnamefont {Ruan}},\ and\ \bibinfo {author} {\bibfnamefont {C.~H.}\ \bibnamefont {Nam}},\ }\bibfield  {title} {\bibinfo {title} {Highly efficient laser-driven compton gamma-ray source},\ }\href@noop {} {\bibfield  {journal} {\bibinfo  {journal} {New Journal of Physics}\ }\textbf {\bibinfo {volume} {21}},\ \bibinfo {pages} {013008} (\bibinfo {year} {2019})}\BibitemShut {NoStop}%
\bibitem [{\citenamefont {Formenti}\ \emph {et~al.}(2022)\citenamefont {Formenti}, \citenamefont {Galbiati},\ and\ \citenamefont {Passoni}}]{formenti2022modeling}%
  \BibitemOpen
  \bibfield  {author} {\bibinfo {author} {\bibfnamefont {A.}~\bibnamefont {Formenti}}, \bibinfo {author} {\bibfnamefont {M.}~\bibnamefont {Galbiati}},\ and\ \bibinfo {author} {\bibfnamefont {M.}~\bibnamefont {Passoni}},\ }\bibfield  {title} {\bibinfo {title} {Modeling and simulations of ultra-intense laser-driven bremsstrahlung with double-layer targets},\ }\href {https://doi.org/10.1088/1361-6587/ac4fce} {\bibfield  {journal} {\bibinfo  {journal} {Plasma Physics and Controlled Fusion}\ }\textbf {\bibinfo {volume} {64}},\ \bibinfo {pages} {044009} (\bibinfo {year} {2022})}\BibitemShut {NoStop}%
\bibitem [{\citenamefont {Galbiati}\ \emph {et~al.}(2023)\citenamefont {Galbiati}, \citenamefont {Formenti}, \citenamefont {Grech},\ and\ \citenamefont {Passoni}}]{galbiati2023numerical}%
  \BibitemOpen
  \bibfield  {author} {\bibinfo {author} {\bibfnamefont {M.}~\bibnamefont {Galbiati}}, \bibinfo {author} {\bibfnamefont {A.}~\bibnamefont {Formenti}}, \bibinfo {author} {\bibfnamefont {M.}~\bibnamefont {Grech}},\ and\ \bibinfo {author} {\bibfnamefont {M.}~\bibnamefont {Passoni}},\ }\bibfield  {title} {\bibinfo {title} {Numerical investigation of non-linear inverse compton scattering in double-layer targets},\ }\bibfield  {journal} {\bibinfo  {journal} {Frontiers in Physics}\ }\textbf {\bibinfo {volume} {11}},\ \href {https://doi.org/10.3389/fphy.2023.1117543} {10.3389/fphy.2023.1117543} (\bibinfo {year} {2023})\BibitemShut {NoStop}%
\bibitem [{\citenamefont {Matheron}\ \emph {et~al.}(2024)\citenamefont {Matheron}, \citenamefont {Marquès}, \citenamefont {Lelasseux}, \citenamefont {Shou}, \citenamefont {Andriyash}, \citenamefont {Phung}, \citenamefont {Ayoul}, \citenamefont {Beluze}, \citenamefont {Dăncuş}, \citenamefont {Dorchies}, \citenamefont {D'Souza}, \citenamefont {Dumergue}, \citenamefont {Frotin}, \citenamefont {Gautier}, \citenamefont {Gobert}, \citenamefont {Gugiu}, \citenamefont {Krishnamurthy}, \citenamefont {Kargapolov}, \citenamefont {Kroupp}, \citenamefont {Lancia}, \citenamefont {Lazăr}, \citenamefont {Leblanc}, \citenamefont {Lo}, \citenamefont {Mataja}, \citenamefont {Mathieu}, \citenamefont {Papadopoulos}, \citenamefont {San Miguel~Claveria}, \citenamefont {Ta~Phuoc}, \citenamefont {Talposi}, \citenamefont {Tata}, \citenamefont {Ur}, \citenamefont {Ursescu}, \citenamefont {Văsescu}, \citenamefont {Doria}, \citenamefont {Malka}, \citenamefont {Ghenuche},\ and\ \citenamefont
  {Corde}}]{Matheron2024_SelfAlignedCompton}%
  \BibitemOpen
  \bibfield  {author} {\bibinfo {author} {\bibfnamefont {A.}~\bibnamefont {Matheron}}, \bibinfo {author} {\bibfnamefont {J.-R.}\ \bibnamefont {Marquès}}, \bibinfo {author} {\bibfnamefont {V.}~\bibnamefont {Lelasseux}}, \bibinfo {author} {\bibfnamefont {Y.}~\bibnamefont {Shou}}, \bibinfo {author} {\bibfnamefont {I.~A.}\ \bibnamefont {Andriyash}}, \bibinfo {author} {\bibfnamefont {V.~L.~J.}\ \bibnamefont {Phung}}, \bibinfo {author} {\bibfnamefont {Y.}~\bibnamefont {Ayoul}}, \bibinfo {author} {\bibfnamefont {A.}~\bibnamefont {Beluze}}, \bibinfo {author} {\bibfnamefont {I.}~\bibnamefont {Dăncuş}}, \bibinfo {author} {\bibfnamefont {F.}~\bibnamefont {Dorchies}}, \bibinfo {author} {\bibfnamefont {F.}~\bibnamefont {D'Souza}}, \bibinfo {author} {\bibfnamefont {M.}~\bibnamefont {Dumergue}}, \bibinfo {author} {\bibfnamefont {M.}~\bibnamefont {Frotin}}, \bibinfo {author} {\bibfnamefont {J.}~\bibnamefont {Gautier}}, \bibinfo {author} {\bibfnamefont {F.}~\bibnamefont {Gobert}}, \bibinfo {author} {\bibfnamefont
  {M.}~\bibnamefont {Gugiu}}, \bibinfo {author} {\bibfnamefont {S.}~\bibnamefont {Krishnamurthy}}, \bibinfo {author} {\bibfnamefont {I.}~\bibnamefont {Kargapolov}}, \bibinfo {author} {\bibfnamefont {E.}~\bibnamefont {Kroupp}}, \bibinfo {author} {\bibfnamefont {L.}~\bibnamefont {Lancia}}, \bibinfo {author} {\bibfnamefont {A.}~\bibnamefont {Lazăr}}, \bibinfo {author} {\bibfnamefont {A.}~\bibnamefont {Leblanc}}, \bibinfo {author} {\bibfnamefont {M.}~\bibnamefont {Lo}}, \bibinfo {author} {\bibfnamefont {D.}~\bibnamefont {Mataja}}, \bibinfo {author} {\bibfnamefont {F.}~\bibnamefont {Mathieu}}, \bibinfo {author} {\bibfnamefont {D.}~\bibnamefont {Papadopoulos}}, \bibinfo {author} {\bibfnamefont {P.}~\bibnamefont {San Miguel~Claveria}}, \bibinfo {author} {\bibfnamefont {K.}~\bibnamefont {Ta~Phuoc}}, \bibinfo {author} {\bibfnamefont {A.-M.}\ \bibnamefont {Talposi}}, \bibinfo {author} {\bibfnamefont {S.}~\bibnamefont {Tata}}, \bibinfo {author} {\bibfnamefont {C.~A.}\ \bibnamefont {Ur}}, \bibinfo {author}
  {\bibfnamefont {D.}~\bibnamefont {Ursescu}}, \bibinfo {author} {\bibfnamefont {L.}~\bibnamefont {Văsescu}}, \bibinfo {author} {\bibfnamefont {D.}~\bibnamefont {Doria}}, \bibinfo {author} {\bibfnamefont {V.}~\bibnamefont {Malka}}, \bibinfo {author} {\bibfnamefont {P.}~\bibnamefont {Ghenuche}},\ and\ \bibinfo {author} {\bibfnamefont {S.}~\bibnamefont {Corde}},\ }\bibfield  {title} {\bibinfo {title} {Compton photons at the gev scale from self-aligned collisions with a plasma mirror},\ }\href {https://arxiv.org/abs/2412.19337} {\bibfield  {journal} {\bibinfo  {journal} {arXiv preprint arXiv:2412.19337}\ } (\bibinfo {year} {2024})}\BibitemShut {NoStop}%
\bibitem [{\citenamefont {Galbiati}\ \emph {et~al.}(2025)\citenamefont {Galbiati}, \citenamefont {Ambrogioni}, \citenamefont {Monaco}, \citenamefont {De~Magistris}, \citenamefont {Orecchia}, \citenamefont {Mirani}, \citenamefont {Maffini},\ and\ \citenamefont {Passoni}}]{galbiati2025numerical}%
  \BibitemOpen
  \bibfield  {author} {\bibinfo {author} {\bibfnamefont {M.}~\bibnamefont {Galbiati}}, \bibinfo {author} {\bibfnamefont {K.}~\bibnamefont {Ambrogioni}}, \bibinfo {author} {\bibfnamefont {L.~F.~C.}\ \bibnamefont {Monaco}}, \bibinfo {author} {\bibfnamefont {M.~S.~G.}\ \bibnamefont {De~Magistris}}, \bibinfo {author} {\bibfnamefont {D.}~\bibnamefont {Orecchia}}, \bibinfo {author} {\bibfnamefont {F.}~\bibnamefont {Mirani}}, \bibinfo {author} {\bibfnamefont {A.}~\bibnamefont {Maffini}},\ and\ \bibinfo {author} {\bibfnamefont {M.}~\bibnamefont {Passoni}},\ }\bibfield  {title} {\bibinfo {title} {Numerical proof-of-concept of a photon, proton, and positron laser-driven source with nanostructured targets},\ }\href@noop {} {\bibfield  {journal} {\bibinfo  {journal} {arXiv preprint arXiv:2503.21630}\ } (\bibinfo {year} {2025})}\BibitemShut {NoStop}%
\bibitem [{\citenamefont {Reiss}(1971)}]{reiss1971production}%
  \BibitemOpen
  \bibfield  {author} {\bibinfo {author} {\bibfnamefont {H.~R.}\ \bibnamefont {Reiss}},\ }\bibfield  {title} {\bibinfo {title} {Production of electron pairs from a zero-mass state},\ }\href@noop {} {\bibfield  {journal} {\bibinfo  {journal} {Physical Review Letters}\ }\textbf {\bibinfo {volume} {26}},\ \bibinfo {pages} {1072} (\bibinfo {year} {1971})}\BibitemShut {NoStop}%
\bibitem [{\citenamefont {Shearer}\ \emph {et~al.}(1973)\citenamefont {Shearer}, \citenamefont {Garrison}, \citenamefont {Wong},\ and\ \citenamefont {Swain}}]{shearer1973pair}%
  \BibitemOpen
  \bibfield  {author} {\bibinfo {author} {\bibfnamefont {J.}~\bibnamefont {Shearer}}, \bibinfo {author} {\bibfnamefont {J.}~\bibnamefont {Garrison}}, \bibinfo {author} {\bibfnamefont {J.}~\bibnamefont {Wong}},\ and\ \bibinfo {author} {\bibfnamefont {J.}~\bibnamefont {Swain}},\ }\bibfield  {title} {\bibinfo {title} {Pair production by relativistic electrons from an intense laser focus},\ }\href@noop {} {\bibfield  {journal} {\bibinfo  {journal} {Physical Review A}\ }\textbf {\bibinfo {volume} {8}},\ \bibinfo {pages} {1582} (\bibinfo {year} {1973})}\BibitemShut {NoStop}%
\bibitem [{\citenamefont {Burke}\ \emph {et~al.}(1997)\citenamefont {Burke}, \citenamefont {Field}, \citenamefont {Horton-Smith}, \citenamefont {Spencer}, \citenamefont {Walz}, \citenamefont {Berridge}, \citenamefont {Bugg}, \citenamefont {Shmakov}, \citenamefont {Weidemann}, \citenamefont {Bula} \emph {et~al.}}]{burke1997positron}%
  \BibitemOpen
  \bibfield  {author} {\bibinfo {author} {\bibfnamefont {D.}~\bibnamefont {Burke}}, \bibinfo {author} {\bibfnamefont {R.}~\bibnamefont {Field}}, \bibinfo {author} {\bibfnamefont {G.}~\bibnamefont {Horton-Smith}}, \bibinfo {author} {\bibfnamefont {J.}~\bibnamefont {Spencer}}, \bibinfo {author} {\bibfnamefont {D.}~\bibnamefont {Walz}}, \bibinfo {author} {\bibfnamefont {S.}~\bibnamefont {Berridge}}, \bibinfo {author} {\bibfnamefont {W.}~\bibnamefont {Bugg}}, \bibinfo {author} {\bibfnamefont {K.}~\bibnamefont {Shmakov}}, \bibinfo {author} {\bibfnamefont {A.}~\bibnamefont {Weidemann}}, \bibinfo {author} {\bibfnamefont {C.}~\bibnamefont {Bula}}, \emph {et~al.},\ }\bibfield  {title} {\bibinfo {title} {Positron production in multiphoton light-by-light scattering},\ }\href@noop {} {\bibfield  {journal} {\bibinfo  {journal} {Physical Review Letters}\ }\textbf {\bibinfo {volume} {79}},\ \bibinfo {pages} {1626} (\bibinfo {year} {1997})}\BibitemShut {NoStop}%
\bibitem [{\citenamefont {Blackburn}\ \emph {et~al.}(2017)\citenamefont {Blackburn}, \citenamefont {Ilderton}, \citenamefont {Murphy},\ and\ \citenamefont {Marklund}}]{blackburn2017scaling}%
  \BibitemOpen
  \bibfield  {author} {\bibinfo {author} {\bibfnamefont {T.}~\bibnamefont {Blackburn}}, \bibinfo {author} {\bibfnamefont {A.}~\bibnamefont {Ilderton}}, \bibinfo {author} {\bibfnamefont {C.}~\bibnamefont {Murphy}},\ and\ \bibinfo {author} {\bibfnamefont {M.}~\bibnamefont {Marklund}},\ }\bibfield  {title} {\bibinfo {title} {Scaling laws for positron production in laser--electron-beam collisions},\ }\href@noop {} {\bibfield  {journal} {\bibinfo  {journal} {Physical Review A}\ }\textbf {\bibinfo {volume} {96}},\ \bibinfo {pages} {022128} (\bibinfo {year} {2017})}\BibitemShut {NoStop}%
\bibitem [{\citenamefont {Lobet}\ \emph {et~al.}(2017)\citenamefont {Lobet}, \citenamefont {Davoine}, \citenamefont {d’Humi{\`e}res},\ and\ \citenamefont {Gremillet}}]{lobet2017generation}%
  \BibitemOpen
  \bibfield  {author} {\bibinfo {author} {\bibfnamefont {M.}~\bibnamefont {Lobet}}, \bibinfo {author} {\bibfnamefont {X.}~\bibnamefont {Davoine}}, \bibinfo {author} {\bibfnamefont {E.}~\bibnamefont {d’Humi{\`e}res}},\ and\ \bibinfo {author} {\bibfnamefont {L.}~\bibnamefont {Gremillet}},\ }\bibfield  {title} {\bibinfo {title} {Generation of high-energy electron-positron pairs in the collision of a laser-accelerated electron beam with a multipetawatt laser},\ }\href@noop {} {\bibfield  {journal} {\bibinfo  {journal} {Physical Review Accelerators and Beams}\ }\textbf {\bibinfo {volume} {20}},\ \bibinfo {pages} {043401} (\bibinfo {year} {2017})}\BibitemShut {NoStop}%
\bibitem [{\citenamefont {Mercuri-Baron}\ \emph {et~al.}(2021)\citenamefont {Mercuri-Baron}, \citenamefont {Grech}, \citenamefont {Niel}, \citenamefont {Grassi}, \citenamefont {Lobet}, \citenamefont {Di~Piazza},\ and\ \citenamefont {Riconda}}]{mercuri2021impact}%
  \BibitemOpen
  \bibfield  {author} {\bibinfo {author} {\bibfnamefont {A.}~\bibnamefont {Mercuri-Baron}}, \bibinfo {author} {\bibfnamefont {M.}~\bibnamefont {Grech}}, \bibinfo {author} {\bibfnamefont {F.}~\bibnamefont {Niel}}, \bibinfo {author} {\bibfnamefont {A.}~\bibnamefont {Grassi}}, \bibinfo {author} {\bibfnamefont {M.}~\bibnamefont {Lobet}}, \bibinfo {author} {\bibfnamefont {A.}~\bibnamefont {Di~Piazza}},\ and\ \bibinfo {author} {\bibfnamefont {C.}~\bibnamefont {Riconda}},\ }\bibfield  {title} {\bibinfo {title} {Impact of the laser spatio-temporal shape on breit--wheeler pair production},\ }\href@noop {} {\bibfield  {journal} {\bibinfo  {journal} {New Journal of Physics}\ }\textbf {\bibinfo {volume} {23}},\ \bibinfo {pages} {085006} (\bibinfo {year} {2021})}\BibitemShut {NoStop}%
\bibitem [{\citenamefont {Salgado}\ \emph {et~al.}(2021)\citenamefont {Salgado}, \citenamefont {Grafenstein}, \citenamefont {Golub}, \citenamefont {D{\"o}pp}, \citenamefont {Eckey}, \citenamefont {Hollatz}, \citenamefont {M{\"u}ller}, \citenamefont {Seidel}, \citenamefont {Seipt}, \citenamefont {Karsch} \emph {et~al.}}]{salgado2021towards}%
  \BibitemOpen
  \bibfield  {author} {\bibinfo {author} {\bibfnamefont {F.}~\bibnamefont {Salgado}}, \bibinfo {author} {\bibfnamefont {K.}~\bibnamefont {Grafenstein}}, \bibinfo {author} {\bibfnamefont {A.}~\bibnamefont {Golub}}, \bibinfo {author} {\bibfnamefont {A.}~\bibnamefont {D{\"o}pp}}, \bibinfo {author} {\bibfnamefont {A.}~\bibnamefont {Eckey}}, \bibinfo {author} {\bibfnamefont {D.}~\bibnamefont {Hollatz}}, \bibinfo {author} {\bibfnamefont {C.}~\bibnamefont {M{\"u}ller}}, \bibinfo {author} {\bibfnamefont {A.}~\bibnamefont {Seidel}}, \bibinfo {author} {\bibfnamefont {D.}~\bibnamefont {Seipt}}, \bibinfo {author} {\bibfnamefont {S.}~\bibnamefont {Karsch}}, \emph {et~al.},\ }\bibfield  {title} {\bibinfo {title} {Towards pair production in the non-perturbative regime},\ }\href@noop {} {\bibfield  {journal} {\bibinfo  {journal} {New Journal of Physics}\ }\textbf {\bibinfo {volume} {23}},\ \bibinfo {pages} {105002} (\bibinfo {year} {2021})}\BibitemShut {NoStop}%
\bibitem [{\citenamefont {Golub}\ \emph {et~al.}(2022)\citenamefont {Golub}, \citenamefont {Villalba-Ch{\'a}vez},\ and\ \citenamefont {M{\"u}ller}}]{golub2022nonlinear}%
  \BibitemOpen
  \bibfield  {author} {\bibinfo {author} {\bibfnamefont {A.}~\bibnamefont {Golub}}, \bibinfo {author} {\bibfnamefont {S.}~\bibnamefont {Villalba-Ch{\'a}vez}},\ and\ \bibinfo {author} {\bibfnamefont {C.}~\bibnamefont {M{\"u}ller}},\ }\bibfield  {title} {\bibinfo {title} {Nonlinear breit-wheeler pair production in collisions of bremsstrahlung $\gamma$ quanta and a tightly focused laser pulse},\ }\href@noop {} {\bibfield  {journal} {\bibinfo  {journal} {Physical Review D}\ }\textbf {\bibinfo {volume} {105}},\ \bibinfo {pages} {116016} (\bibinfo {year} {2022})}\BibitemShut {NoStop}%
\bibitem [{\citenamefont {Pouyez}\ \emph {et~al.}(2024)\citenamefont {Pouyez}, \citenamefont {Mironov}, \citenamefont {Grismayer}, \citenamefont {Mercuri-Baron}, \citenamefont {Perez}, \citenamefont {Vranic}, \citenamefont {Riconda},\ and\ \citenamefont {Grech}}]{pouyez2024multiplicity}%
  \BibitemOpen
  \bibfield  {author} {\bibinfo {author} {\bibfnamefont {M.}~\bibnamefont {Pouyez}}, \bibinfo {author} {\bibfnamefont {A.}~\bibnamefont {Mironov}}, \bibinfo {author} {\bibfnamefont {T.}~\bibnamefont {Grismayer}}, \bibinfo {author} {\bibfnamefont {A.}~\bibnamefont {Mercuri-Baron}}, \bibinfo {author} {\bibfnamefont {F.}~\bibnamefont {Perez}}, \bibinfo {author} {\bibfnamefont {M.}~\bibnamefont {Vranic}}, \bibinfo {author} {\bibfnamefont {C.}~\bibnamefont {Riconda}},\ and\ \bibinfo {author} {\bibfnamefont {M.}~\bibnamefont {Grech}},\ }\bibfield  {title} {\bibinfo {title} {Multiplicity of electron-and photon-seeded electromagnetic showers at multipetawatt laser facilities},\ }\href@noop {} {\bibfield  {journal} {\bibinfo  {journal} {Physical Review E}\ }\textbf {\bibinfo {volume} {110}},\ \bibinfo {pages} {065208} (\bibinfo {year} {2024})}\BibitemShut {NoStop}%
\bibitem [{\citenamefont {Qu}\ and\ \citenamefont {Fisch}(2024)}]{qu2024creating}%
  \BibitemOpen
  \bibfield  {author} {\bibinfo {author} {\bibfnamefont {K.}~\bibnamefont {Qu}}\ and\ \bibinfo {author} {\bibfnamefont {N.~J.}\ \bibnamefont {Fisch}},\ }\bibfield  {title} {\bibinfo {title} {Creating and detecting observable qed plasmas through beam-driven cascade},\ }\href@noop {} {\bibfield  {journal} {\bibinfo  {journal} {Physics of Plasmas}\ }\textbf {\bibinfo {volume} {31}} (\bibinfo {year} {2024})}\BibitemShut {NoStop}%
\bibitem [{\citenamefont {Elsner}\ \emph {et~al.}(2025)\citenamefont {Elsner}, \citenamefont {Golub}, \citenamefont {Villalba-Ch{\'a}vez},\ and\ \citenamefont {M{\"u}ller}}]{elsner2025entering}%
  \BibitemOpen
  \bibfield  {author} {\bibinfo {author} {\bibfnamefont {I.}~\bibnamefont {Elsner}}, \bibinfo {author} {\bibfnamefont {A.}~\bibnamefont {Golub}}, \bibinfo {author} {\bibfnamefont {S.}~\bibnamefont {Villalba-Ch{\'a}vez}},\ and\ \bibinfo {author} {\bibfnamefont {C.}~\bibnamefont {M{\"u}ller}},\ }\bibfield  {title} {\bibinfo {title} {Entering the overcritical regime of nonlinear breit-wheeler pair production in collisions of bremsstrahlung $\gamma$-rays and superintense, tightly focused laser pulses},\ }\href@noop {} {\bibfield  {journal} {\bibinfo  {journal} {Physical Review D}\ }\textbf {\bibinfo {volume} {111}},\ \bibinfo {pages} {096012} (\bibinfo {year} {2025})}\BibitemShut {NoStop}%
\bibitem [{\citenamefont {Hugenschmidt}\ \emph {et~al.}(2012)\citenamefont {Hugenschmidt}, \citenamefont {Piochacz}, \citenamefont {Reiner},\ and\ \citenamefont {Schreckenbach}}]{Hugenschmidt2012}%
  \BibitemOpen
  \bibfield  {author} {\bibinfo {author} {\bibfnamefont {C.}~\bibnamefont {Hugenschmidt}}, \bibinfo {author} {\bibfnamefont {C.}~\bibnamefont {Piochacz}}, \bibinfo {author} {\bibfnamefont {M.}~\bibnamefont {Reiner}},\ and\ \bibinfo {author} {\bibfnamefont {K.}~\bibnamefont {Schreckenbach}},\ }\bibfield  {title} {\bibinfo {title} {The nepomuc upgrade and advanced positron beam experiments},\ }\href {https://doi.org/10.1088/1367-2630/14/5/055027} {\bibfield  {journal} {\bibinfo  {journal} {New Journal of Physics}\ }\textbf {\bibinfo {volume} {14}},\ \bibinfo {pages} {055027} (\bibinfo {year} {2012})}\BibitemShut {NoStop}%
\bibitem [{\citenamefont {Liang}\ \emph {et~al.}(1998)\citenamefont {Liang}, \citenamefont {Wilks},\ and\ \citenamefont {Tabak}}]{liang1998pair}%
  \BibitemOpen
  \bibfield  {author} {\bibinfo {author} {\bibfnamefont {E.~P.}\ \bibnamefont {Liang}}, \bibinfo {author} {\bibfnamefont {S.~C.}\ \bibnamefont {Wilks}},\ and\ \bibinfo {author} {\bibfnamefont {M.}~\bibnamefont {Tabak}},\ }\bibfield  {title} {\bibinfo {title} {Pair production by ultraintense lasers},\ }\href@noop {} {\bibfield  {journal} {\bibinfo  {journal} {Physical review letters}\ }\textbf {\bibinfo {volume} {81}},\ \bibinfo {pages} {4887} (\bibinfo {year} {1998})}\BibitemShut {NoStop}%
\bibitem [{\citenamefont {Gahn}\ \emph {et~al.}(2000)\citenamefont {Gahn}, \citenamefont {Tsakiris}, \citenamefont {Pretzler}, \citenamefont {Witte}, \citenamefont {Delfin}, \citenamefont {Wahlström},\ and\ \citenamefont {Habs}}]{Gahn2000}%
  \BibitemOpen
  \bibfield  {author} {\bibinfo {author} {\bibfnamefont {C.}~\bibnamefont {Gahn}}, \bibinfo {author} {\bibfnamefont {G.~D.}\ \bibnamefont {Tsakiris}}, \bibinfo {author} {\bibfnamefont {G.}~\bibnamefont {Pretzler}}, \bibinfo {author} {\bibfnamefont {K.~J.}\ \bibnamefont {Witte}}, \bibinfo {author} {\bibfnamefont {C.}~\bibnamefont {Delfin}}, \bibinfo {author} {\bibfnamefont {C.-G.}\ \bibnamefont {Wahlström}},\ and\ \bibinfo {author} {\bibfnamefont {D.}~\bibnamefont {Habs}},\ }\bibfield  {title} {\bibinfo {title} {Generating positrons with femtosecond-laser pulses},\ }\href {https://doi.org/10.1063/1.1319526} {\bibfield  {journal} {\bibinfo  {journal} {Applied Physics Letters}\ }\textbf {\bibinfo {volume} {77}},\ \bibinfo {pages} {2662} (\bibinfo {year} {2000})},\ \Eprint {https://arxiv.org/abs/https://pubs.aip.org/aip/apl/article-pdf/77/17/2662/18552952/2662\_1\_online.pdf} {https://pubs.aip.org/aip/apl/article-pdf/77/17/2662/18552952/2662\_1\_online.pdf} \BibitemShut {NoStop}%
\bibitem [{\citenamefont {Nakashima}\ and\ \citenamefont {Takabe}(2002)}]{nakashima2002numerical}%
  \BibitemOpen
  \bibfield  {author} {\bibinfo {author} {\bibfnamefont {K.}~\bibnamefont {Nakashima}}\ and\ \bibinfo {author} {\bibfnamefont {H.}~\bibnamefont {Takabe}},\ }\bibfield  {title} {\bibinfo {title} {Numerical study of pair creation by ultraintense lasers},\ }\href@noop {} {\bibfield  {journal} {\bibinfo  {journal} {Physics of Plasmas}\ }\textbf {\bibinfo {volume} {9}},\ \bibinfo {pages} {1505} (\bibinfo {year} {2002})}\BibitemShut {NoStop}%
\bibitem [{\citenamefont {Chen}\ \emph {et~al.}(2009)\citenamefont {Chen}, \citenamefont {Wilks}, \citenamefont {Bonlie}, \citenamefont {Liang}, \citenamefont {Myatt}, \citenamefont {Price}, \citenamefont {Meyerhofer},\ and\ \citenamefont {Beiersdorfer}}]{relativistic_chen_2009}%
  \BibitemOpen
  \bibfield  {author} {\bibinfo {author} {\bibfnamefont {H.}~\bibnamefont {Chen}}, \bibinfo {author} {\bibfnamefont {S.}~\bibnamefont {Wilks}}, \bibinfo {author} {\bibfnamefont {J.~D.}\ \bibnamefont {Bonlie}}, \bibinfo {author} {\bibfnamefont {E.}~\bibnamefont {Liang}}, \bibinfo {author} {\bibfnamefont {J.}~\bibnamefont {Myatt}}, \bibinfo {author} {\bibfnamefont {D.}~\bibnamefont {Price}}, \bibinfo {author} {\bibfnamefont {D.~D.}\ \bibnamefont {Meyerhofer}},\ and\ \bibinfo {author} {\bibfnamefont {P.}~\bibnamefont {Beiersdorfer}},\ }\bibfield  {title} {\bibinfo {title} {Relativistic positron creation using ultraintense short pulse lasers.},\ }\bibfield  {journal} {\bibinfo  {journal} {Physical Review Letters}\ }\href {https://doi.org/10.1103/PHYSREVLETT.102.105001} {10.1103/PHYSREVLETT.102.105001} (\bibinfo {year} {2009})\BibitemShut {NoStop}%
\bibitem [{\citenamefont {Myatt}\ \emph {et~al.}(2009)\citenamefont {Myatt}, \citenamefont {Delettrez}, \citenamefont {Maximov}, \citenamefont {Meyerhofer}, \citenamefont {Short}, \citenamefont {Stoeckl},\ and\ \citenamefont {Storm}}]{myatt2009optimizing}%
  \BibitemOpen
  \bibfield  {author} {\bibinfo {author} {\bibfnamefont {J.}~\bibnamefont {Myatt}}, \bibinfo {author} {\bibfnamefont {J.}~\bibnamefont {Delettrez}}, \bibinfo {author} {\bibfnamefont {A.}~\bibnamefont {Maximov}}, \bibinfo {author} {\bibfnamefont {D.}~\bibnamefont {Meyerhofer}}, \bibinfo {author} {\bibfnamefont {R.}~\bibnamefont {Short}}, \bibinfo {author} {\bibfnamefont {C.}~\bibnamefont {Stoeckl}},\ and\ \bibinfo {author} {\bibfnamefont {M.}~\bibnamefont {Storm}},\ }\bibfield  {title} {\bibinfo {title} {Optimizing electron-positron pair production on kilojoule-class high-intensity lasers for the purpose of pair-plasma creation},\ }\href@noop {} {\bibfield  {journal} {\bibinfo  {journal} {Physical Review E—Statistical, Nonlinear, and Soft Matter Physics}\ }\textbf {\bibinfo {volume} {79}},\ \bibinfo {pages} {066409} (\bibinfo {year} {2009})}\BibitemShut {NoStop}%
\bibitem [{\citenamefont {Sarri}\ \emph {et~al.}(2013)\citenamefont {Sarri}, \citenamefont {Schumaker}, \citenamefont {Di~Piazza}, \citenamefont {Vargas}, \citenamefont {Dromey}, \citenamefont {Dieckmann}, \citenamefont {Chvykov}, \citenamefont {Maksimchuk}, \citenamefont {Yanovsky}, \citenamefont {He} \emph {et~al.}}]{sarri2013table}%
  \BibitemOpen
  \bibfield  {author} {\bibinfo {author} {\bibfnamefont {G.}~\bibnamefont {Sarri}}, \bibinfo {author} {\bibfnamefont {W.}~\bibnamefont {Schumaker}}, \bibinfo {author} {\bibfnamefont {A.}~\bibnamefont {Di~Piazza}}, \bibinfo {author} {\bibfnamefont {M.}~\bibnamefont {Vargas}}, \bibinfo {author} {\bibfnamefont {B.}~\bibnamefont {Dromey}}, \bibinfo {author} {\bibfnamefont {M.~E.}\ \bibnamefont {Dieckmann}}, \bibinfo {author} {\bibfnamefont {V.}~\bibnamefont {Chvykov}}, \bibinfo {author} {\bibfnamefont {A.}~\bibnamefont {Maksimchuk}}, \bibinfo {author} {\bibfnamefont {V.}~\bibnamefont {Yanovsky}}, \bibinfo {author} {\bibfnamefont {Z.}~\bibnamefont {He}}, \emph {et~al.},\ }\bibfield  {title} {\bibinfo {title} {Table-top laser-based source of femtosecond, collimated, ultrarelativistic positron beams},\ }\href@noop {} {\bibfield  {journal} {\bibinfo  {journal} {Physical review letters}\ }\textbf {\bibinfo {volume} {110}},\ \bibinfo {pages} {255002} (\bibinfo {year} {2013})}\BibitemShut {NoStop}%
\bibitem [{\citenamefont {Sarri}\ \emph {et~al.}(2015)\citenamefont {Sarri}, \citenamefont {Poder}, \citenamefont {Cole}, \citenamefont {Schumaker}, \citenamefont {Piazza}, \citenamefont {Reville}, \citenamefont {Dzelzainis}, \citenamefont {Doria}, \citenamefont {Gizzi}, \citenamefont {Grittani}, \citenamefont {Kar}, \citenamefont {Keitel}, \citenamefont {Krushelnick}, \citenamefont {Kuschel}, \citenamefont {Mangles}, \citenamefont {Najmudin}, \citenamefont {Shukla}, \citenamefont {Silva}, \citenamefont {Symes}, \citenamefont {Thomas}, \citenamefont {Vargas}, \citenamefont {Vieira},\ and\ \citenamefont {Zepf}}]{generation_sarri_2015}%
  \BibitemOpen
  \bibfield  {author} {\bibinfo {author} {\bibfnamefont {G.}~\bibnamefont {Sarri}}, \bibinfo {author} {\bibfnamefont {K.}~\bibnamefont {Poder}}, \bibinfo {author} {\bibfnamefont {J.}~\bibnamefont {Cole}}, \bibinfo {author} {\bibfnamefont {W.}~\bibnamefont {Schumaker}}, \bibinfo {author} {\bibfnamefont {A.~D.}\ \bibnamefont {Piazza}}, \bibinfo {author} {\bibfnamefont {B.}~\bibnamefont {Reville}}, \bibinfo {author} {\bibfnamefont {T.}~\bibnamefont {Dzelzainis}}, \bibinfo {author} {\bibfnamefont {D.}~\bibnamefont {Doria}}, \bibinfo {author} {\bibfnamefont {L.~A.}\ \bibnamefont {Gizzi}}, \bibinfo {author} {\bibfnamefont {G.}~\bibnamefont {Grittani}}, \bibinfo {author} {\bibfnamefont {S.}~\bibnamefont {Kar}}, \bibinfo {author} {\bibfnamefont {C.~H.}\ \bibnamefont {Keitel}}, \bibinfo {author} {\bibfnamefont {K.}~\bibnamefont {Krushelnick}}, \bibinfo {author} {\bibfnamefont {S.}~\bibnamefont {Kuschel}}, \bibinfo {author} {\bibfnamefont {S.}~\bibnamefont {Mangles}}, \bibinfo {author} {\bibfnamefont {Z.}~\bibnamefont
  {Najmudin}}, \bibinfo {author} {\bibfnamefont {N.}~\bibnamefont {Shukla}}, \bibinfo {author} {\bibfnamefont {L.~O.}\ \bibnamefont {Silva}}, \bibinfo {author} {\bibfnamefont {D.}~\bibnamefont {Symes}}, \bibinfo {author} {\bibfnamefont {A.}~\bibnamefont {Thomas}}, \bibinfo {author} {\bibfnamefont {M.}~\bibnamefont {Vargas}}, \bibinfo {author} {\bibfnamefont {J.}~\bibnamefont {Vieira}},\ and\ \bibinfo {author} {\bibfnamefont {M.}~\bibnamefont {Zepf}},\ }\bibfield  {title} {\bibinfo {title} {Generation of neutral and high-density electron-positron pair plasmas in the laboratory},\ }\bibfield  {journal} {\bibinfo  {journal} {Nature Communications}\ }\href {https://doi.org/10.1038/NCOMMS7747} {10.1038/NCOMMS7747} (\bibinfo {year} {2015})\BibitemShut {NoStop}%
\bibitem [{\citenamefont {Xu}\ \emph {et~al.}(2016)\citenamefont {Xu}, \citenamefont {Shen}, \citenamefont {Xu}, \citenamefont {Li}, \citenamefont {Yu}, \citenamefont {Li}, \citenamefont {Lu}, \citenamefont {Wang}, \citenamefont {Wang}, \citenamefont {Liang} \emph {et~al.}}]{xu2016ultrashort}%
  \BibitemOpen
  \bibfield  {author} {\bibinfo {author} {\bibfnamefont {T.}~\bibnamefont {Xu}}, \bibinfo {author} {\bibfnamefont {B.}~\bibnamefont {Shen}}, \bibinfo {author} {\bibfnamefont {J.}~\bibnamefont {Xu}}, \bibinfo {author} {\bibfnamefont {S.}~\bibnamefont {Li}}, \bibinfo {author} {\bibfnamefont {Y.}~\bibnamefont {Yu}}, \bibinfo {author} {\bibfnamefont {J.}~\bibnamefont {Li}}, \bibinfo {author} {\bibfnamefont {X.}~\bibnamefont {Lu}}, \bibinfo {author} {\bibfnamefont {C.}~\bibnamefont {Wang}}, \bibinfo {author} {\bibfnamefont {X.}~\bibnamefont {Wang}}, \bibinfo {author} {\bibfnamefont {X.}~\bibnamefont {Liang}}, \emph {et~al.},\ }\bibfield  {title} {\bibinfo {title} {Ultrashort megaelectronvolt positron beam generation based on laser-accelerated electrons},\ }\href@noop {} {\bibfield  {journal} {\bibinfo  {journal} {Physics of Plasmas}\ }\textbf {\bibinfo {volume} {23}} (\bibinfo {year} {2016})}\BibitemShut {NoStop}%
\bibitem [{\citenamefont {Chen}\ and\ \citenamefont {Fiuza}(2023)}]{HChen2023}%
  \BibitemOpen
  \bibfield  {author} {\bibinfo {author} {\bibfnamefont {H.}~\bibnamefont {Chen}}\ and\ \bibinfo {author} {\bibfnamefont {F.}~\bibnamefont {Fiuza}},\ }\bibfield  {title} {\bibinfo {title} {{Perspectives on relativistic electron–positron pair plasma experiments of astrophysical relevance using high-power lasers}},\ }\href {https://doi.org/10.1063/5.0134819} {\bibfield  {journal} {\bibinfo  {journal} {Physics of Plasmas}\ }\textbf {\bibinfo {volume} {30}},\ \bibinfo {pages} {020601} (\bibinfo {year} {2023})}\BibitemShut {NoStop}%
\bibitem [{\citenamefont {Arrowsmith}\ \emph {et~al.}(2024)\citenamefont {Arrowsmith}, \citenamefont {Simon}, \citenamefont {Bilbao}, \citenamefont {Bott}, \citenamefont {Burger}, \citenamefont {Chen}, \citenamefont {Cruz}, \citenamefont {Davenne}, \citenamefont {Efthymiopoulos}, \citenamefont {Froula}, \citenamefont {Goillot}, \citenamefont {Gudmundsson}, \citenamefont {Haberberger}, \citenamefont {Halliday}, \citenamefont {Hodge}, \citenamefont {Huffman}, \citenamefont {Iaquinta}, \citenamefont {Miniati}, \citenamefont {Reville}, \citenamefont {Sarkar}, \citenamefont {Schekochihin}, \citenamefont {Silva}, \citenamefont {Simpson}, \citenamefont {Stergiou}, \citenamefont {Trines}, \citenamefont {Vieu}, \citenamefont {Charitonidis}, \citenamefont {Bingham},\ and\ \citenamefont {Gregori}}]{Arrowsmith_2024}%
  \BibitemOpen
  \bibfield  {author} {\bibinfo {author} {\bibfnamefont {C.~D.}\ \bibnamefont {Arrowsmith}}, \bibinfo {author} {\bibfnamefont {P.}~\bibnamefont {Simon}}, \bibinfo {author} {\bibfnamefont {P.~J.}\ \bibnamefont {Bilbao}}, \bibinfo {author} {\bibfnamefont {A.~F.~A.}\ \bibnamefont {Bott}}, \bibinfo {author} {\bibfnamefont {S.}~\bibnamefont {Burger}}, \bibinfo {author} {\bibfnamefont {H.}~\bibnamefont {Chen}}, \bibinfo {author} {\bibfnamefont {F.~D.}\ \bibnamefont {Cruz}}, \bibinfo {author} {\bibfnamefont {T.}~\bibnamefont {Davenne}}, \bibinfo {author} {\bibfnamefont {I.}~\bibnamefont {Efthymiopoulos}}, \bibinfo {author} {\bibfnamefont {D.~H.}\ \bibnamefont {Froula}}, \bibinfo {author} {\bibfnamefont {A.}~\bibnamefont {Goillot}}, \bibinfo {author} {\bibfnamefont {J.~T.}\ \bibnamefont {Gudmundsson}}, \bibinfo {author} {\bibfnamefont {D.}~\bibnamefont {Haberberger}}, \bibinfo {author} {\bibfnamefont {J.~W.~D.}\ \bibnamefont {Halliday}}, \bibinfo {author} {\bibfnamefont {T.}~\bibnamefont {Hodge}}, \bibinfo {author}
  {\bibfnamefont {B.~T.}\ \bibnamefont {Huffman}}, \bibinfo {author} {\bibfnamefont {S.}~\bibnamefont {Iaquinta}}, \bibinfo {author} {\bibfnamefont {F.}~\bibnamefont {Miniati}}, \bibinfo {author} {\bibfnamefont {B.}~\bibnamefont {Reville}}, \bibinfo {author} {\bibfnamefont {S.}~\bibnamefont {Sarkar}}, \bibinfo {author} {\bibfnamefont {A.~A.}\ \bibnamefont {Schekochihin}}, \bibinfo {author} {\bibfnamefont {L.~O.}\ \bibnamefont {Silva}}, \bibinfo {author} {\bibfnamefont {R.}~\bibnamefont {Simpson}}, \bibinfo {author} {\bibfnamefont {V.}~\bibnamefont {Stergiou}}, \bibinfo {author} {\bibfnamefont {R.~M. G.~M.}\ \bibnamefont {Trines}}, \bibinfo {author} {\bibfnamefont {T.}~\bibnamefont {Vieu}}, \bibinfo {author} {\bibfnamefont {N.}~\bibnamefont {Charitonidis}}, \bibinfo {author} {\bibfnamefont {R.}~\bibnamefont {Bingham}},\ and\ \bibinfo {author} {\bibfnamefont {G.}~\bibnamefont {Gregori}},\ }\bibfield  {title} {\bibinfo {title} {Laboratory realization of relativistic pair-plasma beams},\ }\bibfield  {journal}
  {\bibinfo  {journal} {Nature Communications}\ }\textbf {\bibinfo {volume} {15}},\ \href {https://doi.org/10.1038/s41467-024-49346-2} {10.1038/s41467-024-49346-2} (\bibinfo {year} {2024})\BibitemShut {NoStop}%
\bibitem [{\citenamefont {Noh}\ \emph {et~al.}(2024)\citenamefont {Noh}, \citenamefont {Song}, \citenamefont {Mirzaie}, \citenamefont {Hojbota}, \citenamefont {Kim}, \citenamefont {Lee}, \citenamefont {Won}, \citenamefont {Song}, \citenamefont {Song}, \citenamefont {Ryu} \emph {et~al.}}]{noh2024charge}%
  \BibitemOpen
  \bibfield  {author} {\bibinfo {author} {\bibfnamefont {Y.}~\bibnamefont {Noh}}, \bibinfo {author} {\bibfnamefont {J.}~\bibnamefont {Song}}, \bibinfo {author} {\bibfnamefont {M.}~\bibnamefont {Mirzaie}}, \bibinfo {author} {\bibfnamefont {C.~I.}\ \bibnamefont {Hojbota}}, \bibinfo {author} {\bibfnamefont {H.-i.}\ \bibnamefont {Kim}}, \bibinfo {author} {\bibfnamefont {S.}~\bibnamefont {Lee}}, \bibinfo {author} {\bibfnamefont {J.}~\bibnamefont {Won}}, \bibinfo {author} {\bibfnamefont {H.}~\bibnamefont {Song}}, \bibinfo {author} {\bibfnamefont {C.}~\bibnamefont {Song}}, \bibinfo {author} {\bibfnamefont {C.-M.}\ \bibnamefont {Ryu}}, \emph {et~al.},\ }\bibfield  {title} {\bibinfo {title} {Charge-neutral, gev-scale electron-positron pair beams produced using bremsstrahlung gamma rays},\ }\href@noop {} {\bibfield  {journal} {\bibinfo  {journal} {Communications Physics}\ }\textbf {\bibinfo {volume} {7}},\ \bibinfo {pages} {44} (\bibinfo {year} {2024})}\BibitemShut {NoStop}%
\bibitem [{\citenamefont {Lobet}\ \emph {et~al.}(2015)\citenamefont {Lobet}, \citenamefont {Ruyer}, \citenamefont {Debayle}, \citenamefont {d'Humi\`eres}, \citenamefont {Grech}, \citenamefont {Lemoine},\ and\ \citenamefont {Gremillet}}]{Lobet2015PRL}%
  \BibitemOpen
  \bibfield  {author} {\bibinfo {author} {\bibfnamefont {M.}~\bibnamefont {Lobet}}, \bibinfo {author} {\bibfnamefont {C.}~\bibnamefont {Ruyer}}, \bibinfo {author} {\bibfnamefont {A.}~\bibnamefont {Debayle}}, \bibinfo {author} {\bibfnamefont {E.}~\bibnamefont {d'Humi\`eres}}, \bibinfo {author} {\bibfnamefont {M.}~\bibnamefont {Grech}}, \bibinfo {author} {\bibfnamefont {M.}~\bibnamefont {Lemoine}},\ and\ \bibinfo {author} {\bibfnamefont {L.}~\bibnamefont {Gremillet}},\ }\bibfield  {title} {\bibinfo {title} {Ultrafast synchrotron-enhanced thermalization of laser-driven colliding pair plasmas},\ }\href {https://doi.org/10.1103/PhysRevLett.115.215003} {\bibfield  {journal} {\bibinfo  {journal} {Phys. Rev. Lett.}\ }\textbf {\bibinfo {volume} {115}},\ \bibinfo {pages} {215003} (\bibinfo {year} {2015})}\BibitemShut {NoStop}%
\bibitem [{\citenamefont {Uzdensky}\ \emph {et~al.}(2019)\citenamefont {Uzdensky}, \citenamefont {Begelman}, \citenamefont {Beloborodov}, \citenamefont {Blandford}, \citenamefont {Boldyrev}, \citenamefont {Cerutti}, \citenamefont {Fiuza}, \citenamefont {Giannios}, \citenamefont {Grismayer}, \citenamefont {Kunz}, \citenamefont {Loureiro}, \citenamefont {Lyutikov}, \citenamefont {Medvedev}, \citenamefont {Petropoulou}, \citenamefont {Philippov}, \citenamefont {Quataert}, \citenamefont {Schekochihin}, \citenamefont {Schoeffler}, \citenamefont {Silva}, \citenamefont {Sironi}, \citenamefont {Spitkovsky}, \citenamefont {Werner}, \citenamefont {Zhdankin}, \citenamefont {Zrake},\ and\ \citenamefont {Zweibel}}]{Uzdensky2019}%
  \BibitemOpen
  \bibfield  {author} {\bibinfo {author} {\bibfnamefont {D.}~\bibnamefont {Uzdensky}}, \bibinfo {author} {\bibfnamefont {M.}~\bibnamefont {Begelman}}, \bibinfo {author} {\bibfnamefont {A.}~\bibnamefont {Beloborodov}}, \bibinfo {author} {\bibfnamefont {R.}~\bibnamefont {Blandford}}, \bibinfo {author} {\bibfnamefont {S.}~\bibnamefont {Boldyrev}}, \bibinfo {author} {\bibfnamefont {B.}~\bibnamefont {Cerutti}}, \bibinfo {author} {\bibfnamefont {F.}~\bibnamefont {Fiuza}}, \bibinfo {author} {\bibfnamefont {D.}~\bibnamefont {Giannios}}, \bibinfo {author} {\bibfnamefont {T.}~\bibnamefont {Grismayer}}, \bibinfo {author} {\bibfnamefont {M.}~\bibnamefont {Kunz}}, \bibinfo {author} {\bibfnamefont {N.}~\bibnamefont {Loureiro}}, \bibinfo {author} {\bibfnamefont {M.}~\bibnamefont {Lyutikov}}, \bibinfo {author} {\bibfnamefont {M.}~\bibnamefont {Medvedev}}, \bibinfo {author} {\bibfnamefont {M.}~\bibnamefont {Petropoulou}}, \bibinfo {author} {\bibfnamefont {A.}~\bibnamefont {Philippov}}, \bibinfo {author} {\bibfnamefont
  {E.}~\bibnamefont {Quataert}}, \bibinfo {author} {\bibfnamefont {A.}~\bibnamefont {Schekochihin}}, \bibinfo {author} {\bibfnamefont {K.}~\bibnamefont {Schoeffler}}, \bibinfo {author} {\bibfnamefont {L.}~\bibnamefont {Silva}}, \bibinfo {author} {\bibfnamefont {L.}~\bibnamefont {Sironi}}, \bibinfo {author} {\bibfnamefont {A.}~\bibnamefont {Spitkovsky}}, \bibinfo {author} {\bibfnamefont {G.}~\bibnamefont {Werner}}, \bibinfo {author} {\bibfnamefont {V.}~\bibnamefont {Zhdankin}}, \bibinfo {author} {\bibfnamefont {J.}~\bibnamefont {Zrake}},\ and\ \bibinfo {author} {\bibfnamefont {E.}~\bibnamefont {Zweibel}},\ }\href {https://arxiv.org/abs/1903.05328} {\bibinfo {title} {Extreme plasma astrophysics}} (\bibinfo {year} {2019}),\ \Eprint {https://arxiv.org/abs/1903.05328} {arXiv:1903.05328 [astro-ph.HE]} \BibitemShut {NoStop}%
\bibitem [{\citenamefont {Stoneking}\ \emph {et~al.}(2020)\citenamefont {Stoneking}, \citenamefont {Pedersen}, \citenamefont {Helander}, \citenamefont {Chen}, \citenamefont {Hergenhahn}, \citenamefont {Stenson}, \citenamefont {Fiksel}, \citenamefont {von~der Linden}, \citenamefont {Saitoh}, \citenamefont {Surko},\ and\ \citenamefont {et~al.}}]{Stoneking2020}%
  \BibitemOpen
  \bibfield  {author} {\bibinfo {author} {\bibfnamefont {M.~R.}\ \bibnamefont {Stoneking}}, \bibinfo {author} {\bibfnamefont {T.~S.}\ \bibnamefont {Pedersen}}, \bibinfo {author} {\bibfnamefont {P.}~\bibnamefont {Helander}}, \bibinfo {author} {\bibfnamefont {H.}~\bibnamefont {Chen}}, \bibinfo {author} {\bibfnamefont {U.}~\bibnamefont {Hergenhahn}}, \bibinfo {author} {\bibfnamefont {E.~V.}\ \bibnamefont {Stenson}}, \bibinfo {author} {\bibfnamefont {G.}~\bibnamefont {Fiksel}}, \bibinfo {author} {\bibfnamefont {J.}~\bibnamefont {von~der Linden}}, \bibinfo {author} {\bibfnamefont {H.}~\bibnamefont {Saitoh}}, \bibinfo {author} {\bibfnamefont {C.~M.}\ \bibnamefont {Surko}},\ and\ \bibinfo {author} {\bibnamefont {et~al.}},\ }\bibfield  {title} {\bibinfo {title} {A new frontier in laboratory physics: magnetized electron–positron plasmas},\ }\href {https://doi.org/10.1017/S0022377820001385} {\bibfield  {journal} {\bibinfo  {journal} {Journal of Plasma Physics}\ }\textbf {\bibinfo {volume} {86}},\ \bibinfo {pages}
  {155860601} (\bibinfo {year} {2020})}\BibitemShut {NoStop}%
\bibitem [{\citenamefont {Blackett}\ and\ \citenamefont {Occhialini}(1932)}]{blackett1932photography}%
  \BibitemOpen
  \bibfield  {author} {\bibinfo {author} {\bibfnamefont {P.~M.~S.}\ \bibnamefont {Blackett}}\ and\ \bibinfo {author} {\bibfnamefont {G.}~\bibnamefont {Occhialini}},\ }\bibfield  {title} {\bibinfo {title} {Photography of penetrating corpuscular radiation},\ }\href@noop {} {\bibfield  {journal} {\bibinfo  {journal} {Nature}\ }\textbf {\bibinfo {volume} {130}},\ \bibinfo {pages} {363} (\bibinfo {year} {1932})}\BibitemShut {NoStop}%
\bibitem [{\citenamefont {Rossi}(1933)}]{rossi1933eigenschaften}%
  \BibitemOpen
  \bibfield  {author} {\bibinfo {author} {\bibfnamefont {B.}~\bibnamefont {Rossi}},\ }\bibfield  {title} {\bibinfo {title} {{\"U}ber die eigenschaften der durchdringenden korpuskularstrahlung im meeresniveau},\ }\href@noop {} {\bibfield  {journal} {\bibinfo  {journal} {Zeitschrift f{\"u}r Physik}\ }\textbf {\bibinfo {volume} {82}},\ \bibinfo {pages} {151} (\bibinfo {year} {1933})}\BibitemShut {NoStop}%
\bibitem [{\citenamefont {Pfotzer}(1936)}]{pfotzer1936dreifachkoinzidenzen}%
  \BibitemOpen
  \bibfield  {author} {\bibinfo {author} {\bibfnamefont {G.}~\bibnamefont {Pfotzer}},\ }\bibfield  {title} {\bibinfo {title} {Dreifachkoinzidenzen der ultrastrahlung aus vertikaler richtung in der stratosph{\"a}re: I. me$\backslash$methode und ergebnisse},\ }\href@noop {} {\bibfield  {journal} {\bibinfo  {journal} {Zeitschrift f{\"u}r Physik}\ }\textbf {\bibinfo {volume} {102}},\ \bibinfo {pages} {23} (\bibinfo {year} {1936})}\BibitemShut {NoStop}%
\bibitem [{\citenamefont {Auger}\ \emph {et~al.}(1939{\natexlab{a}})\citenamefont {Auger}, \citenamefont {Ehrenfest}, \citenamefont {Maze}, \citenamefont {Daudin},\ and\ \citenamefont {Fréon}}]{Auger1939}%
  \BibitemOpen
  \bibfield  {author} {\bibinfo {author} {\bibfnamefont {P.}~\bibnamefont {Auger}}, \bibinfo {author} {\bibfnamefont {P.}~\bibnamefont {Ehrenfest}}, \bibinfo {author} {\bibfnamefont {R.}~\bibnamefont {Maze}}, \bibinfo {author} {\bibfnamefont {J.}~\bibnamefont {Daudin}},\ and\ \bibinfo {author} {\bibfnamefont {R.~A.}\ \bibnamefont {Fréon}},\ }\bibfield  {title} {\bibinfo {title} {Extensive cosmic-ray showers},\ }\href {https://doi.org/10.1103/revmodphys.11.288} {\bibfield  {journal} {\bibinfo  {journal} {Reviews of Modern Physics}\ }\textbf {\bibinfo {volume} {11}},\ \bibinfo {pages} {288–291} (\bibinfo {year} {1939}{\natexlab{a}})}\BibitemShut {NoStop}%
\bibitem [{\citenamefont {Auger}\ \emph {et~al.}(1939{\natexlab{b}})\citenamefont {Auger}, \citenamefont {Maze}, \citenamefont {Ehrenfest},\ and\ \citenamefont {Freon}}]{auger1939grandes}%
  \BibitemOpen
  \bibfield  {author} {\bibinfo {author} {\bibfnamefont {P.}~\bibnamefont {Auger}}, \bibinfo {author} {\bibfnamefont {R.}~\bibnamefont {Maze}}, \bibinfo {author} {\bibfnamefont {P.}~\bibnamefont {Ehrenfest}},\ and\ \bibinfo {author} {\bibfnamefont {A.}~\bibnamefont {Freon}},\ }\bibfield  {title} {\bibinfo {title} {Les grandes gerbes de rayons cosmiques},\ }\href@noop {} {\bibfield  {journal} {\bibinfo  {journal} {Journal de Physique et le Radium}\ }\textbf {\bibinfo {volume} {10}},\ \bibinfo {pages} {39} (\bibinfo {year} {1939}{\natexlab{b}})}\BibitemShut {NoStop}%
\bibitem [{\citenamefont {Bhabha}\ and\ \citenamefont {Heitler}(1937)}]{bhabha1937passage}%
  \BibitemOpen
  \bibfield  {author} {\bibinfo {author} {\bibfnamefont {H.~J.}\ \bibnamefont {Bhabha}}\ and\ \bibinfo {author} {\bibfnamefont {W.}~\bibnamefont {Heitler}},\ }\bibfield  {title} {\bibinfo {title} {The passage of fast electrons and the theory of cosmic showers},\ }\href@noop {} {\bibfield  {journal} {\bibinfo  {journal} {Proceedings of the Royal Society of London. Series A-Mathematical and Physical Sciences}\ }\textbf {\bibinfo {volume} {159}},\ \bibinfo {pages} {432} (\bibinfo {year} {1937})}\BibitemShut {NoStop}%
\bibitem [{\citenamefont {Carlson}\ and\ \citenamefont {Oppenheimer}(1937)}]{carlson1937multiplicative}%
  \BibitemOpen
  \bibfield  {author} {\bibinfo {author} {\bibfnamefont {J.}~\bibnamefont {Carlson}}\ and\ \bibinfo {author} {\bibfnamefont {J.}~\bibnamefont {Oppenheimer}},\ }\bibfield  {title} {\bibinfo {title} {On multiplicative showers},\ }\href@noop {} {\bibfield  {journal} {\bibinfo  {journal} {Physical Review}\ }\textbf {\bibinfo {volume} {51}},\ \bibinfo {pages} {220} (\bibinfo {year} {1937})}\BibitemShut {NoStop}%
\bibitem [{\citenamefont {Landau}\ and\ \citenamefont {Rumer}(1938)}]{landau1938cascade}%
  \BibitemOpen
  \bibfield  {author} {\bibinfo {author} {\bibfnamefont {L.~D.}\ \bibnamefont {Landau}}\ and\ \bibinfo {author} {\bibfnamefont {G.}~\bibnamefont {Rumer}},\ }\bibfield  {title} {\bibinfo {title} {The cascade theory of electronic showers},\ }\href@noop {} {\bibfield  {journal} {\bibinfo  {journal} {Proceedings of the Royal Society of London. Series A. Mathematical and Physical Sciences}\ }\textbf {\bibinfo {volume} {166}},\ \bibinfo {pages} {213} (\bibinfo {year} {1938})}\BibitemShut {NoStop}%
\bibitem [{\citenamefont {Snyder}(1938)}]{snyder1938transition}%
  \BibitemOpen
  \bibfield  {author} {\bibinfo {author} {\bibfnamefont {H.}~\bibnamefont {Snyder}},\ }\bibfield  {title} {\bibinfo {title} {Transition effects of cosmic rays in the atmosphere},\ }\href@noop {} {\bibfield  {journal} {\bibinfo  {journal} {Physical Review}\ }\textbf {\bibinfo {volume} {53}},\ \bibinfo {pages} {960} (\bibinfo {year} {1938})}\BibitemShut {NoStop}%
\bibitem [{\citenamefont {Tamm}\ and\ \citenamefont {Belenky}(1939)}]{tamm1939soft}%
  \BibitemOpen
  \bibfield  {author} {\bibinfo {author} {\bibfnamefont {I.}~\bibnamefont {Tamm}}\ and\ \bibinfo {author} {\bibfnamefont {S.}~\bibnamefont {Belenky}},\ }\bibfield  {title} {\bibinfo {title} {On the soft component of cosmic rays at sea level},\ }\href@noop {} {\bibfield  {journal} {\bibinfo  {journal} {Journal of Physics USSR}\ }\textbf {\bibinfo {volume} {1}},\ \bibinfo {pages} {177} (\bibinfo {year} {1939})}\BibitemShut {NoStop}%
\bibitem [{\citenamefont {Rossi}\ and\ \citenamefont {Greisen}(1941)}]{rossi1941cosmic}%
  \BibitemOpen
  \bibfield  {author} {\bibinfo {author} {\bibfnamefont {B.}~\bibnamefont {Rossi}}\ and\ \bibinfo {author} {\bibfnamefont {K.}~\bibnamefont {Greisen}},\ }\bibfield  {title} {\bibinfo {title} {Cosmic-ray theory},\ }\href@noop {} {\bibfield  {journal} {\bibinfo  {journal} {Reviews of Modern Physics}\ }\textbf {\bibinfo {volume} {13}},\ \bibinfo {pages} {240} (\bibinfo {year} {1941})}\BibitemShut {NoStop}%
\bibitem [{\citenamefont {Moli{\`e}re}\ and\ \citenamefont {Heisenberg}(1946)}]{moliere1946cosmic}%
  \BibitemOpen
  \bibfield  {author} {\bibinfo {author} {\bibfnamefont {G.}~\bibnamefont {Moli{\`e}re}}\ and\ \bibinfo {author} {\bibfnamefont {W.}~\bibnamefont {Heisenberg}},\ }\href@noop {} {\bibinfo {title} {Cosmic radiation}} (\bibinfo {year} {1946})\BibitemShut {NoStop}%
\bibitem [{\citenamefont {Roberg}\ and\ \citenamefont {Nordheim}(1949)}]{roberg1949angular}%
  \BibitemOpen
  \bibfield  {author} {\bibinfo {author} {\bibfnamefont {J.}~\bibnamefont {Roberg}}\ and\ \bibinfo {author} {\bibfnamefont {L.}~\bibnamefont {Nordheim}},\ }\bibfield  {title} {\bibinfo {title} {The angular and lateral spread of cosmic-ray showers},\ }\href@noop {} {\bibfield  {journal} {\bibinfo  {journal} {Physical Review}\ }\textbf {\bibinfo {volume} {75}},\ \bibinfo {pages} {444} (\bibinfo {year} {1949})}\BibitemShut {NoStop}%
\bibitem [{\citenamefont {Eyges}\ and\ \citenamefont {Fernbach}(1951)}]{eyges1951angular}%
  \BibitemOpen
  \bibfield  {author} {\bibinfo {author} {\bibfnamefont {L.}~\bibnamefont {Eyges}}\ and\ \bibinfo {author} {\bibfnamefont {S.}~\bibnamefont {Fernbach}},\ }\bibfield  {title} {\bibinfo {title} {Angular and radial distributions of particles in cascade showers},\ }\href@noop {} {\bibfield  {journal} {\bibinfo  {journal} {Physical Review}\ }\textbf {\bibinfo {volume} {82}},\ \bibinfo {pages} {23} (\bibinfo {year} {1951})}\BibitemShut {NoStop}%
\bibitem [{\citenamefont {Green}\ and\ \citenamefont {Messel}(1952)}]{green1952spread}%
  \BibitemOpen
  \bibfield  {author} {\bibinfo {author} {\bibfnamefont {H.}~\bibnamefont {Green}}\ and\ \bibinfo {author} {\bibfnamefont {H.}~\bibnamefont {Messel}},\ }\bibfield  {title} {\bibinfo {title} {The spread of the soft component of the cosmic radiation},\ }\href@noop {} {\bibfield  {journal} {\bibinfo  {journal} {Physical Review}\ }\textbf {\bibinfo {volume} {88}},\ \bibinfo {pages} {331} (\bibinfo {year} {1952})}\BibitemShut {NoStop}%
\bibitem [{\citenamefont {Green}\ and\ \citenamefont {Bergmann}(1954)}]{green1954core}%
  \BibitemOpen
  \bibfield  {author} {\bibinfo {author} {\bibfnamefont {H.}~\bibnamefont {Green}}\ and\ \bibinfo {author} {\bibfnamefont {O.}~\bibnamefont {Bergmann}},\ }\bibfield  {title} {\bibinfo {title} {Core structure in soft component showers},\ }\href@noop {} {\bibfield  {journal} {\bibinfo  {journal} {Physical Review}\ }\textbf {\bibinfo {volume} {95}},\ \bibinfo {pages} {516} (\bibinfo {year} {1954})}\BibitemShut {NoStop}%
\bibitem [{\citenamefont {Greisen}(1956)}]{Greisen1956}%
  \BibitemOpen
  \bibfield  {author} {\bibinfo {author} {\bibfnamefont {K.}~\bibnamefont {Greisen}},\ }\bibfield  {title} {\bibinfo {title} {Cosmic ray showers},\ }\href@noop {} {\bibfield  {journal} {\bibinfo  {journal} {Progress in Cosmic Ray Physics}\ }\textbf {\bibinfo {volume} {3}},\ \bibinfo {pages} {1} (\bibinfo {year} {1956})}\BibitemShut {NoStop}%
\bibitem [{\citenamefont {Kamata}\ and\ \citenamefont {Nishimura}(1958)}]{kamata1958lateral}%
  \BibitemOpen
  \bibfield  {author} {\bibinfo {author} {\bibfnamefont {K.}~\bibnamefont {Kamata}}\ and\ \bibinfo {author} {\bibfnamefont {J.}~\bibnamefont {Nishimura}},\ }\bibfield  {title} {\bibinfo {title} {The lateral and the angular structure functions of electron showers},\ }\href@noop {} {\bibfield  {journal} {\bibinfo  {journal} {Progress of Theoretical Physics Supplement}\ }\textbf {\bibinfo {volume} {6}},\ \bibinfo {pages} {93} (\bibinfo {year} {1958})}\BibitemShut {NoStop}%
\bibitem [{\citenamefont {Pouyez}\ \emph {et~al.}(2025{\natexlab{a}})\citenamefont {Pouyez}, \citenamefont {Grismayer}, \citenamefont {Grech},\ and\ \citenamefont {Riconda}}]{pouyez2024kinetic}%
  \BibitemOpen
  \bibfield  {author} {\bibinfo {author} {\bibfnamefont {M.}~\bibnamefont {Pouyez}}, \bibinfo {author} {\bibfnamefont {T.}~\bibnamefont {Grismayer}}, \bibinfo {author} {\bibfnamefont {M.}~\bibnamefont {Grech}},\ and\ \bibinfo {author} {\bibfnamefont {C.}~\bibnamefont {Riconda}},\ }\bibfield  {title} {\bibinfo {title} {Kinetic structure of strong-field qed showers in crossed electromagnetic fields},\ }\href@noop {} {\bibfield  {journal} {\bibinfo  {journal} {Physical Review Letters}\ }\textbf {\bibinfo {volume} {134}},\ \bibinfo {pages} {135001} (\bibinfo {year} {2025}{\natexlab{a}})}\BibitemShut {NoStop}%
\bibitem [{\citenamefont {Agostinelli}\ \emph {et~al.}(2003)\citenamefont {Agostinelli}, \citenamefont {Allison}, \citenamefont {Amako}, \citenamefont {Apostolakis}, \citenamefont {Araujo}, \citenamefont {Arce}, \citenamefont {Asai}, \citenamefont {Axen}, \citenamefont {Banerjee}, \citenamefont {Barrand} \emph {et~al.}}]{agostinelli2003geant4}%
  \BibitemOpen
  \bibfield  {author} {\bibinfo {author} {\bibfnamefont {S.}~\bibnamefont {Agostinelli}}, \bibinfo {author} {\bibfnamefont {J.}~\bibnamefont {Allison}}, \bibinfo {author} {\bibfnamefont {K.~a.}\ \bibnamefont {Amako}}, \bibinfo {author} {\bibfnamefont {J.}~\bibnamefont {Apostolakis}}, \bibinfo {author} {\bibfnamefont {H.}~\bibnamefont {Araujo}}, \bibinfo {author} {\bibfnamefont {P.}~\bibnamefont {Arce}}, \bibinfo {author} {\bibfnamefont {M.}~\bibnamefont {Asai}}, \bibinfo {author} {\bibfnamefont {D.}~\bibnamefont {Axen}}, \bibinfo {author} {\bibfnamefont {S.}~\bibnamefont {Banerjee}}, \bibinfo {author} {\bibfnamefont {G.}~\bibnamefont {Barrand}}, \emph {et~al.},\ }\bibfield  {title} {\bibinfo {title} {Geant4—a simulation toolkit},\ }\href@noop {} {\bibfield  {journal} {\bibinfo  {journal} {Nuclear instruments and methods in physics research section A: Accelerators, Spectrometers, Detectors and Associated Equipment}\ }\textbf {\bibinfo {volume} {506}},\ \bibinfo {pages} {250} (\bibinfo {year}
  {2003})}\BibitemShut {NoStop}%
\bibitem [{\citenamefont {Pouyez}\ \emph {et~al.}(2025{\natexlab{b}})\citenamefont {Pouyez}, \citenamefont {Nicotera}, \citenamefont {Galbiati}, \citenamefont {Grismayer}, \citenamefont {Lancia}, \citenamefont {Riconda},\ and\ \citenamefont {Grech}}]{suppMat}%
  \BibitemOpen
  \bibfield  {author} {\bibinfo {author} {\bibfnamefont {M.}~\bibnamefont {Pouyez}}, \bibinfo {author} {\bibfnamefont {G.}~\bibnamefont {Nicotera}}, \bibinfo {author} {\bibfnamefont {M.}~\bibnamefont {Galbiati}}, \bibinfo {author} {\bibfnamefont {T.}~\bibnamefont {Grismayer}}, \bibinfo {author} {\bibfnamefont {L.}~\bibnamefont {Lancia}}, \bibinfo {author} {\bibfnamefont {C.}~\bibnamefont {Riconda}},\ and\ \bibinfo {author} {\bibfnamefont {M.}~\bibnamefont {Grech}},\ }\bibfield  {title} {\bibinfo {title} {Supplemental material},\ }\href@noop {} {\  (\bibinfo {year} {2025}{\natexlab{b}})}\BibitemShut {NoStop}%
\bibitem [{\citenamefont {Koch}\ and\ \citenamefont {Motz}(1959)}]{koch1959bremsstrahlung}%
  \BibitemOpen
  \bibfield  {author} {\bibinfo {author} {\bibfnamefont {H.}~\bibnamefont {Koch}}\ and\ \bibinfo {author} {\bibfnamefont {J.}~\bibnamefont {Motz}},\ }\bibfield  {title} {\bibinfo {title} {Bremsstrahlung cross-section formulas and related data},\ }\href@noop {} {\bibfield  {journal} {\bibinfo  {journal} {Reviews of modern physics}\ }\textbf {\bibinfo {volume} {31}},\ \bibinfo {pages} {920} (\bibinfo {year} {1959})}\BibitemShut {NoStop}%
\bibitem [{\citenamefont {Motz}\ \emph {et~al.}(1969)\citenamefont {Motz}, \citenamefont {Olsen},\ and\ \citenamefont {Koch}}]{motz1969pair}%
  \BibitemOpen
  \bibfield  {author} {\bibinfo {author} {\bibfnamefont {J.}~\bibnamefont {Motz}}, \bibinfo {author} {\bibfnamefont {H.~A.}\ \bibnamefont {Olsen}},\ and\ \bibinfo {author} {\bibfnamefont {H.}~\bibnamefont {Koch}},\ }\bibfield  {title} {\bibinfo {title} {Pair production by photons},\ }\href@noop {} {\bibfield  {journal} {\bibinfo  {journal} {Reviews of Modern Physics}\ }\textbf {\bibinfo {volume} {41}},\ \bibinfo {pages} {581} (\bibinfo {year} {1969})}\BibitemShut {NoStop}%
\bibitem [{\citenamefont {Martinez}\ \emph {et~al.}(2019)\citenamefont {Martinez}, \citenamefont {Lobet}, \citenamefont {Duclous}, \citenamefont {d'Humi{\`e}res},\ and\ \citenamefont {Gremillet}}]{martinez2019high}%
  \BibitemOpen
  \bibfield  {author} {\bibinfo {author} {\bibfnamefont {B.}~\bibnamefont {Martinez}}, \bibinfo {author} {\bibfnamefont {M.}~\bibnamefont {Lobet}}, \bibinfo {author} {\bibfnamefont {R.}~\bibnamefont {Duclous}}, \bibinfo {author} {\bibfnamefont {E.}~\bibnamefont {d'Humi{\`e}res}},\ and\ \bibinfo {author} {\bibfnamefont {L.}~\bibnamefont {Gremillet}},\ }\bibfield  {title} {\bibinfo {title} {High-energy radiation and pair production by coulomb processes in particle-in-cell simulations},\ }\href@noop {} {\bibfield  {journal} {\bibinfo  {journal} {Physics of Plasmas}\ }\textbf {\bibinfo {volume} {26}} (\bibinfo {year} {2019})}\BibitemShut {NoStop}%
\bibitem [{\citenamefont {Seltzer}\ and\ \citenamefont {Berger}(1985)}]{seltzer1985bremsstrahlung}%
  \BibitemOpen
  \bibfield  {author} {\bibinfo {author} {\bibfnamefont {S.~M.}\ \bibnamefont {Seltzer}}\ and\ \bibinfo {author} {\bibfnamefont {M.~J.}\ \bibnamefont {Berger}},\ }\bibfield  {title} {\bibinfo {title} {Bremsstrahlung spectra from electron interactions with screened atomic nuclei and orbital electrons},\ }\href@noop {} {\bibfield  {journal} {\bibinfo  {journal} {Nuclear Instruments and Methods in Physics Research Section B: Beam Interactions with Materials and Atoms}\ }\textbf {\bibinfo {volume} {12}},\ \bibinfo {pages} {95} (\bibinfo {year} {1985})}\BibitemShut {NoStop}%
\bibitem [{\citenamefont {Seltzer}\ and\ \citenamefont {Berger}(1986)}]{seltzer1986bremsstrahlung}%
  \BibitemOpen
  \bibfield  {author} {\bibinfo {author} {\bibfnamefont {S.~M.}\ \bibnamefont {Seltzer}}\ and\ \bibinfo {author} {\bibfnamefont {M.~J.}\ \bibnamefont {Berger}},\ }\bibfield  {title} {\bibinfo {title} {Bremsstrahlung energy spectra from electrons with kinetic energy 1 kev--10 gev incident on screened nuclei and orbital electrons of neutral atoms with z= 1--100},\ }\href@noop {} {\bibfield  {journal} {\bibinfo  {journal} {Atomic data and nuclear data tables}\ }\textbf {\bibinfo {volume} {35}},\ \bibinfo {pages} {345} (\bibinfo {year} {1986})}\BibitemShut {NoStop}%
\bibitem [{\citenamefont {Heitler}(1954)}]{heitler1984quantum}%
  \BibitemOpen
  \bibfield  {author} {\bibinfo {author} {\bibfnamefont {W.}~\bibnamefont {Heitler}},\ }\bibfield  {title} {\bibinfo {title} {The quantum theory of radiation},\ }\href@noop {} {\bibfield  {journal} {\bibinfo  {journal} {Oxford University Press}\ } (\bibinfo {year} {1954})}\BibitemShut {NoStop}%
\bibitem [{\citenamefont {Kim}\ \emph {et~al.}(2024)\citenamefont {Kim}, \citenamefont {Noh}, \citenamefont {Song}, \citenamefont {Lee}, \citenamefont {Won}, \citenamefont {Song}, \citenamefont {Bae}, \citenamefont {Ryu}, \citenamefont {Nam},\ and\ \citenamefont {Bang}}]{kim2024electron}%
  \BibitemOpen
  \bibfield  {author} {\bibinfo {author} {\bibfnamefont {H.-i.}\ \bibnamefont {Kim}}, \bibinfo {author} {\bibfnamefont {Y.}~\bibnamefont {Noh}}, \bibinfo {author} {\bibfnamefont {J.}~\bibnamefont {Song}}, \bibinfo {author} {\bibfnamefont {S.}~\bibnamefont {Lee}}, \bibinfo {author} {\bibfnamefont {J.}~\bibnamefont {Won}}, \bibinfo {author} {\bibfnamefont {C.}~\bibnamefont {Song}}, \bibinfo {author} {\bibfnamefont {L.}~\bibnamefont {Bae}}, \bibinfo {author} {\bibfnamefont {C.-M.}\ \bibnamefont {Ryu}}, \bibinfo {author} {\bibfnamefont {C.~H.}\ \bibnamefont {Nam}},\ and\ \bibinfo {author} {\bibfnamefont {W.}~\bibnamefont {Bang}},\ }\bibfield  {title} {\bibinfo {title} {Electron-positron generation by irradiating various metallic materials with laser-accelerated electrons},\ }\href@noop {} {\bibfield  {journal} {\bibinfo  {journal} {Results in Physics}\ }\textbf {\bibinfo {volume} {57}},\ \bibinfo {pages} {107310} (\bibinfo {year} {2024})}\BibitemShut {NoStop}%
\bibitem [{\citenamefont {NIST}()}]{NistWeb}%
  \BibitemOpen
  \bibfield  {author} {\bibinfo {author} {\bibnamefont {NIST}},\ }\href@noop {} {\bibinfo {title} {Photon attenuation table}},\ \bibinfo {note} {\url{https://physics.nist.gov/cgi-bin/Xcom/xcom2?Method=Elem&Output2=Hand}}\BibitemShut {NoStop}%
\bibitem [{\citenamefont {Klein}(1999)}]{klein1999suppression}%
  \BibitemOpen
  \bibfield  {author} {\bibinfo {author} {\bibfnamefont {S.}~\bibnamefont {Klein}},\ }\bibfield  {title} {\bibinfo {title} {Suppression of bremsstrahlung and pair production due to environmental factors},\ }\href@noop {} {\bibfield  {journal} {\bibinfo  {journal} {Reviews of Modern Physics}\ }\textbf {\bibinfo {volume} {71}},\ \bibinfo {pages} {1501} (\bibinfo {year} {1999})}\BibitemShut {NoStop}%
\bibitem [{\citenamefont {Migdal}(1956)}]{Migdal1956}%
  \BibitemOpen
  \bibfield  {author} {\bibinfo {author} {\bibfnamefont {A.~B.}\ \bibnamefont {Migdal}},\ }\bibfield  {title} {\bibinfo {title} {Bremsstrahlung and pair production in condensed media at high energies},\ }\href {https://doi.org/10.1103/physrev.103.1811} {\bibfield  {journal} {\bibinfo  {journal} {Physical Review}\ }\textbf {\bibinfo {volume} {103}},\ \bibinfo {pages} {1811–1820} (\bibinfo {year} {1956})}\BibitemShut {NoStop}%
\bibitem [{\citenamefont {Ter-Mikael{\u ia}n}(1972)}]{Terikaelian1972}%
  \BibitemOpen
  \bibfield  {author} {\bibinfo {author} {\bibfnamefont {M.~L.}\ \bibnamefont {Ter-Mikael{\u ia}n}},\ }\href@noop {} {\bibinfo {title} {High-energy electromagnetic processes in condensed media}} (\bibinfo {year} {1972})\BibitemShut {NoStop}%
\bibitem [{\citenamefont {Akhiezer}\ \emph {et~al.}(1994)\citenamefont {Akhiezer}, \citenamefont {Merenkov},\ and\ \citenamefont {Rekalo}}]{akhiezer1994kinetic}%
  \BibitemOpen
  \bibfield  {author} {\bibinfo {author} {\bibfnamefont {A.}~\bibnamefont {Akhiezer}}, \bibinfo {author} {\bibfnamefont {N.}~\bibnamefont {Merenkov}},\ and\ \bibinfo {author} {\bibfnamefont {A.}~\bibnamefont {Rekalo}},\ }\bibfield  {title} {\bibinfo {title} {On a kinetic theory of electromagnetic showers in strong magnetic fields},\ }\href@noop {} {\bibfield  {journal} {\bibinfo  {journal} {Journal of Physics G: Nuclear and Particle Physics}\ }\textbf {\bibinfo {volume} {20}},\ \bibinfo {pages} {1499} (\bibinfo {year} {1994})}\BibitemShut {NoStop}%
\bibitem [{\citenamefont {Matthews}(2005)}]{matthews2005heitler}%
  \BibitemOpen
  \bibfield  {author} {\bibinfo {author} {\bibfnamefont {J.}~\bibnamefont {Matthews}},\ }\bibfield  {title} {\bibinfo {title} {A heitler model of extensive air showers},\ }\href@noop {} {\bibfield  {journal} {\bibinfo  {journal} {Astroparticle Physics}\ }\textbf {\bibinfo {volume} {22}},\ \bibinfo {pages} {387} (\bibinfo {year} {2005})}\BibitemShut {NoStop}%
\bibitem [{\citenamefont {Selivanov}\ and\ \citenamefont {Fedotov}(2024)}]{selivanov2024final}%
  \BibitemOpen
  \bibfield  {author} {\bibinfo {author} {\bibfnamefont {Y.}~\bibnamefont {Selivanov}}\ and\ \bibinfo {author} {\bibfnamefont {A.}~\bibnamefont {Fedotov}},\ }\bibfield  {title} {\bibinfo {title} {Final multiplicity of a qed cascade in generalized heitler model},\ }\href@noop {} {\bibfield  {journal} {\bibinfo  {journal} {Physical Review D}\ }\textbf {\bibinfo {volume} {110}},\ \bibinfo {pages} {096022} (\bibinfo {year} {2024})}\BibitemShut {NoStop}%
\bibitem [{\citenamefont {Capdevielle}\ and\ \citenamefont {Cohen}(2005)}]{capdevielle2005relation}%
  \BibitemOpen
  \bibfield  {author} {\bibinfo {author} {\bibfnamefont {J.-N.}\ \bibnamefont {Capdevielle}}\ and\ \bibinfo {author} {\bibfnamefont {F.}~\bibnamefont {Cohen}},\ }\bibfield  {title} {\bibinfo {title} {The relation between the lateral profile of giant extensive air showers and the age parameter},\ }\href@noop {} {\bibfield  {journal} {\bibinfo  {journal} {Journal of Physics G: Nuclear and Particle Physics}\ }\textbf {\bibinfo {volume} {31}},\ \bibinfo {pages} {507} (\bibinfo {year} {2005})}\BibitemShut {NoStop}%
\bibitem [{\citenamefont {Apel}\ \emph {et~al.}(2006)\citenamefont {Apel}, \citenamefont {Badea}, \citenamefont {Bekk}, \citenamefont {Bercuci}, \citenamefont {Bl{\"u}mer}, \citenamefont {Bozdog}, \citenamefont {Brancus}, \citenamefont {Chilingarian}, \citenamefont {Daumiller}, \citenamefont {Doll} \emph {et~al.}}]{apel2006comparison}%
  \BibitemOpen
  \bibfield  {author} {\bibinfo {author} {\bibfnamefont {W.}~\bibnamefont {Apel}}, \bibinfo {author} {\bibfnamefont {A.}~\bibnamefont {Badea}}, \bibinfo {author} {\bibfnamefont {K.}~\bibnamefont {Bekk}}, \bibinfo {author} {\bibfnamefont {A.}~\bibnamefont {Bercuci}}, \bibinfo {author} {\bibfnamefont {J.}~\bibnamefont {Bl{\"u}mer}}, \bibinfo {author} {\bibfnamefont {H.}~\bibnamefont {Bozdog}}, \bibinfo {author} {\bibfnamefont {I.}~\bibnamefont {Brancus}}, \bibinfo {author} {\bibfnamefont {A.}~\bibnamefont {Chilingarian}}, \bibinfo {author} {\bibfnamefont {K.}~\bibnamefont {Daumiller}}, \bibinfo {author} {\bibfnamefont {P.}~\bibnamefont {Doll}}, \emph {et~al.},\ }\bibfield  {title} {\bibinfo {title} {Comparison of measured and simulated lateral distributions for electrons and muons with kascade},\ }\href@noop {} {\bibfield  {journal} {\bibinfo  {journal} {Astroparticle Physics}\ }\textbf {\bibinfo {volume} {24}},\ \bibinfo {pages} {467} (\bibinfo {year} {2006})}\BibitemShut {NoStop}%
\bibitem [{\citenamefont {Dey}\ and\ \citenamefont {Dam}(2016)}]{dey2016slope}%
  \BibitemOpen
  \bibfield  {author} {\bibinfo {author} {\bibfnamefont {R.~K.}\ \bibnamefont {Dey}}\ and\ \bibinfo {author} {\bibfnamefont {S.}~\bibnamefont {Dam}},\ }\bibfield  {title} {\bibinfo {title} {Slope of the lateral density function of extensive air showers around the knee region as an indicator of shower age},\ }\href@noop {} {\bibfield  {journal} {\bibinfo  {journal} {The European Physical Journal Plus}\ }\textbf {\bibinfo {volume} {131}},\ \bibinfo {pages} {402} (\bibinfo {year} {2016})}\BibitemShut {NoStop}%
\bibitem [{\citenamefont {Moliere}(1948)}]{moliere1948theorie}%
  \BibitemOpen
  \bibfield  {author} {\bibinfo {author} {\bibfnamefont {G.}~\bibnamefont {Moliere}},\ }\bibfield  {title} {\bibinfo {title} {Theorie der streuung schneller geladener teilchen ii mehrfach-und vielfachstreuung},\ }\href@noop {} {\bibfield  {journal} {\bibinfo  {journal} {Zeitschrift f{\"u}r Naturforschung A}\ }\textbf {\bibinfo {volume} {3}},\ \bibinfo {pages} {78} (\bibinfo {year} {1948})}\BibitemShut {NoStop}%
\bibitem [{\citenamefont {Bethe}(1953)}]{bethe1953moliere}%
  \BibitemOpen
  \bibfield  {author} {\bibinfo {author} {\bibfnamefont {H.~A.}\ \bibnamefont {Bethe}},\ }\bibfield  {title} {\bibinfo {title} {Moliere's theory of multiple scattering},\ }\href@noop {} {\bibfield  {journal} {\bibinfo  {journal} {Physical review}\ }\textbf {\bibinfo {volume} {89}},\ \bibinfo {pages} {1256} (\bibinfo {year} {1953})}\BibitemShut {NoStop}%
\bibitem [{\citenamefont {Scott}(1963)}]{scott1963theory}%
  \BibitemOpen
  \bibfield  {author} {\bibinfo {author} {\bibfnamefont {W.~T.}\ \bibnamefont {Scott}},\ }\bibfield  {title} {\bibinfo {title} {The theory of small-angle multiple scattering of fast charged particles},\ }\href@noop {} {\bibfield  {journal} {\bibinfo  {journal} {Reviews of modern physics}\ }\textbf {\bibinfo {volume} {35}},\ \bibinfo {pages} {231} (\bibinfo {year} {1963})}\BibitemShut {NoStop}%
\bibitem [{\citenamefont {Highland}(1975)}]{highland1975some}%
  \BibitemOpen
  \bibfield  {author} {\bibinfo {author} {\bibfnamefont {V.~L.}\ \bibnamefont {Highland}},\ }\bibfield  {title} {\bibinfo {title} {Some practical remarks on multiple scattering},\ }\href@noop {} {\bibfield  {journal} {\bibinfo  {journal} {Nuclear Instruments and Methods}\ }\textbf {\bibinfo {volume} {129}},\ \bibinfo {pages} {497} (\bibinfo {year} {1975})}\BibitemShut {NoStop}%
\bibitem [{\citenamefont {Lynch}\ and\ \citenamefont {Dahl}(1991)}]{lynch1991approximations}%
  \BibitemOpen
  \bibfield  {author} {\bibinfo {author} {\bibfnamefont {G.~R.}\ \bibnamefont {Lynch}}\ and\ \bibinfo {author} {\bibfnamefont {O.~I.}\ \bibnamefont {Dahl}},\ }\bibfield  {title} {\bibinfo {title} {Approximations to multiple coulomb scattering},\ }\href@noop {} {\bibfield  {journal} {\bibinfo  {journal} {Nuclear Instruments and Methods in Physics Research Section B: Beam Interactions with Materials and Atoms}\ }\textbf {\bibinfo {volume} {58}},\ \bibinfo {pages} {6} (\bibinfo {year} {1991})}\BibitemShut {NoStop}%
\bibitem [{\citenamefont {Capdevielle}\ and\ \citenamefont {Gawin}(1982)}]{capdevielle1982radial}%
  \BibitemOpen
  \bibfield  {author} {\bibinfo {author} {\bibfnamefont {J.}~\bibnamefont {Capdevielle}}\ and\ \bibinfo {author} {\bibfnamefont {J.}~\bibnamefont {Gawin}},\ }\bibfield  {title} {\bibinfo {title} {The radial electron distribution in extensive air showers},\ }\href@noop {} {\bibfield  {journal} {\bibinfo  {journal} {Journal of Physics G: Nuclear Physics}\ }\textbf {\bibinfo {volume} {8}},\ \bibinfo {pages} {1317} (\bibinfo {year} {1982})}\BibitemShut {NoStop}%
\bibitem [{\citenamefont {Song}\ \emph {et~al.}(2023)\citenamefont {Song}, \citenamefont {Kim}, \citenamefont {Won}, \citenamefont {Song}, \citenamefont {Lee}, \citenamefont {Ryu}, \citenamefont {Bang},\ and\ \citenamefont {Nam}}]{song2023characterization}%
  \BibitemOpen
  \bibfield  {author} {\bibinfo {author} {\bibfnamefont {H.}~\bibnamefont {Song}}, \bibinfo {author} {\bibfnamefont {C.~M.}\ \bibnamefont {Kim}}, \bibinfo {author} {\bibfnamefont {J.}~\bibnamefont {Won}}, \bibinfo {author} {\bibfnamefont {J.}~\bibnamefont {Song}}, \bibinfo {author} {\bibfnamefont {S.}~\bibnamefont {Lee}}, \bibinfo {author} {\bibfnamefont {C.-M.}\ \bibnamefont {Ryu}}, \bibinfo {author} {\bibfnamefont {W.}~\bibnamefont {Bang}},\ and\ \bibinfo {author} {\bibfnamefont {C.~H.}\ \bibnamefont {Nam}},\ }\bibfield  {title} {\bibinfo {title} {Characterization of relativistic electron--positron beams produced with laser-accelerated gev electrons},\ }\href@noop {} {\bibfield  {journal} {\bibinfo  {journal} {Scientific Reports}\ }\textbf {\bibinfo {volume} {13}},\ \bibinfo {pages} {310} (\bibinfo {year} {2023})}\BibitemShut {NoStop}%
\bibitem [{\citenamefont {Williams}\ \emph {et~al.}(2020)\citenamefont {Williams}, \citenamefont {Chen}, \citenamefont {Kim}, \citenamefont {Kerr},\ and\ \citenamefont {Khater}}]{williams2020comment}%
  \BibitemOpen
  \bibfield  {author} {\bibinfo {author} {\bibfnamefont {G.}~\bibnamefont {Williams}}, \bibinfo {author} {\bibfnamefont {H.}~\bibnamefont {Chen}}, \bibinfo {author} {\bibfnamefont {J.}~\bibnamefont {Kim}}, \bibinfo {author} {\bibfnamefont {S.}~\bibnamefont {Kerr}},\ and\ \bibinfo {author} {\bibfnamefont {H.}~\bibnamefont {Khater}},\ }\bibfield  {title} {\bibinfo {title} {Comment on “table-top laser-based source of femtosecond, collimated, ultrarelativistic positron beams”},\ }\href@noop {} {\bibfield  {journal} {\bibinfo  {journal} {Physical Review Letters}\ }\textbf {\bibinfo {volume} {124}},\ \bibinfo {pages} {179501} (\bibinfo {year} {2020})}\BibitemShut {NoStop}%
\bibitem [{\citenamefont {Ludwig}\ \emph {et~al.}(2025)\citenamefont {Ludwig}, \citenamefont {Wilks}, \citenamefont {Kemp}, \citenamefont {Williams}, \citenamefont {Lemos}, \citenamefont {Rockafellow}, \citenamefont {Miao}, \citenamefont {Shrock}, \citenamefont {Milchberg}, \citenamefont {Vay} \emph {et~al.}}]{ludwig2025laser}%
  \BibitemOpen
  \bibfield  {author} {\bibinfo {author} {\bibfnamefont {J.}~\bibnamefont {Ludwig}}, \bibinfo {author} {\bibfnamefont {S.}~\bibnamefont {Wilks}}, \bibinfo {author} {\bibfnamefont {A.}~\bibnamefont {Kemp}}, \bibinfo {author} {\bibfnamefont {G.}~\bibnamefont {Williams}}, \bibinfo {author} {\bibfnamefont {N.}~\bibnamefont {Lemos}}, \bibinfo {author} {\bibfnamefont {E.}~\bibnamefont {Rockafellow}}, \bibinfo {author} {\bibfnamefont {B.}~\bibnamefont {Miao}}, \bibinfo {author} {\bibfnamefont {J.}~\bibnamefont {Shrock}}, \bibinfo {author} {\bibfnamefont {H.}~\bibnamefont {Milchberg}}, \bibinfo {author} {\bibfnamefont {J.-L.}\ \bibnamefont {Vay}}, \emph {et~al.},\ }\bibfield  {title} {\bibinfo {title} {Laser based 100 gev electron acceleration scheme for muon production},\ }\href@noop {} {\bibfield  {journal} {\bibinfo  {journal} {Scientific Reports}\ }\textbf {\bibinfo {volume} {15}},\ \bibinfo {pages} {1} (\bibinfo {year} {2025})}\BibitemShut {NoStop}%
\bibitem [{\citenamefont {Shen}\ \emph {et~al.}(2021)\citenamefont {Shen}, \citenamefont {Pukhov},\ and\ \citenamefont {Qiao}}]{ShenPRX2021}%
  \BibitemOpen
  \bibfield  {author} {\bibinfo {author} {\bibfnamefont {X.~F.}\ \bibnamefont {Shen}}, \bibinfo {author} {\bibfnamefont {A.}~\bibnamefont {Pukhov}},\ and\ \bibinfo {author} {\bibfnamefont {B.}~\bibnamefont {Qiao}},\ }\bibfield  {title} {\bibinfo {title} {Monoenergetic high-energy ion source via femtosecond laser interacting with a microtape},\ }\href {https://doi.org/10.1103/PhysRevX.11.041002} {\bibfield  {journal} {\bibinfo  {journal} {Phys. Rev. X}\ }\textbf {\bibinfo {volume} {11}},\ \bibinfo {pages} {041002} (\bibinfo {year} {2021})}\BibitemShut {NoStop}%
\bibitem [{\citenamefont {Marini}\ \emph {et~al.}(2023)\citenamefont {Marini}, \citenamefont {Grech}, \citenamefont {Kleij}, \citenamefont {Raynaud},\ and\ \citenamefont {Riconda}}]{MariniPRR2023}%
  \BibitemOpen
  \bibfield  {author} {\bibinfo {author} {\bibfnamefont {S.}~\bibnamefont {Marini}}, \bibinfo {author} {\bibfnamefont {M.}~\bibnamefont {Grech}}, \bibinfo {author} {\bibfnamefont {P.~S.}\ \bibnamefont {Kleij}}, \bibinfo {author} {\bibfnamefont {M.}~\bibnamefont {Raynaud}},\ and\ \bibinfo {author} {\bibfnamefont {C.}~\bibnamefont {Riconda}},\ }\bibfield  {title} {\bibinfo {title} {Electron acceleration by laser plasma wedge interaction},\ }\href {https://doi.org/10.1103/PhysRevResearch.5.013115} {\bibfield  {journal} {\bibinfo  {journal} {Phys. Rev. Res.}\ }\textbf {\bibinfo {volume} {5}},\ \bibinfo {pages} {013115} (\bibinfo {year} {2023})}\BibitemShut {NoStop}%
\end{thebibliography}
\end{document}


\title{Supplementary material for \\ \tys{Properties of Pair Plasmas Emerging from Electromagnetic Showers in Matter}}

\author{M. Pouyez}
\affiliation{LULI, Sorbonne Université, CNRS, CEA, École Polytechnique, Institut Polytechnique de Paris, F-75255 Paris, France}
\author{G. Nicotera}
\affiliation{LULI, CNRS, CEA, Sorbonne Universit\'{e}, École Polytechnique, Institut Polytechnique de Paris, F-91128 Palaiseau, France}
\author{M. Galbiati}
\affiliation{LULI, CNRS, CEA, Sorbonne Universit\'{e}, École Polytechnique, Institut Polytechnique de Paris, F-91128 Palaiseau, France}
\author{T. Grismayer}
\affiliation{GoLP/Instituto de Plasmas e Fusão Nuclear, Instituto Superior Técnico, Universidade de Lisboa, 1049-001 Lisboa, Portugal}
\author{L. Lancia}
\affiliation{LULI, CNRS, CEA, Sorbonne Universit\'{e}, École Polytechnique, Institut Polytechnique de Paris, F-91128 Palaiseau, France}
\author{C. Riconda}
\affiliation{LULI, Sorbonne Université, CNRS, CEA, École Polytechnique, Institut Polytechnique de Paris, F-75255 Paris, France}
\author{M. Grech}
\affiliation{LULI, CNRS, CEA, Sorbonne Universit\'{e}, École Polytechnique, Institut Polytechnique de Paris, F-91128 Palaiseau, France}

\maketitle

\section{Units and definition}

SI units and standard notations for physical constants will be used throughout the paper:
$c$ stands for the speed of light in vacuum, $m$ for the electron mass, $e$ for the elementary charge, $\alpha = e^2/(4\pi\varepsilon_0\hbar c)$ for the fine structure constant, $\varepsilon_0$ for the permittivity of vacuum and $r_e=e^2/(4\pi\varepsilon_0m_ec^2)$ and $r_c=r_e/\alpha$ for the electron and Compton radius.

\section{Rates definitions}\label{appendixA}

As discussed in the manuscript, multiple definitions of the Bremsstrahlung and Bethe–Heitler cross-sections \cite{bethe1934stopping} exist in the literature. They are discussed in \cite{koch1959bremsstrahlung, motz1969pair}. In this work, we adopt the same formulations as those used in \cite{martinez2019high}, which provide a mathematical model for the interaction rates and allow us to account for particle production in both neutral and ionised targets.

\subsection{Useful functions}

The Thomas Fermi length $L_{\rm{TF}}$ and Debye length $L_D$ are defined by:

\begin{eqnarray}
    \Ltf&=&4\pi\varepsilon_0\frac{\hbar^2}{m_ee^2}Z^{-1/3} \\
    \Ld&=&\sqrt{\frac{\varepsilon_0k_BT}{e^2n_iZ^{*}(Z^{*}+1)}}
\end{eqnarray}
with $n_i$ and $T$ the density and temperature of ions and $Z^*$ the ionisation degree. In the following, we used the normalised Thomas Fermi $\Ltfbarr =\Ltf/r_c$ and Debye $\Ldbarr =\Ld/r_c$ length.

The cross-sections for both Bremsstrahlung and Bethe-Heitler processes depend on the functions $I$ and $J$:

\begin{eqnarray}
    I(\delta)&=&\int_\delta^1\frac{du}{u^3}(u-\delta)^2[1-F_e(u)]^2 \\
    J(\delta)&=&\int_\delta^1\frac{du}{u^4}[u^3-6\delta^2u\ln(u/\delta)+3\delta^2u-4\delta^3][1-F_e(u)]^2 
\end{eqnarray}
where $1-F_e$ is the form factor:
\begin{eqnarray}
    1-F_e(u)&=& \etf \frac{\Ltfbar{2}u^2}{1+\Ltfbar{2}u^2}+\ed\frac{\Ldbar{2}u^2}{1+\Ldbar{2}u^2}
\end{eqnarray}
with $\etf=1-Z^*/Z$ and  $\ed=Z^*/Z$. Injecting this in the definition of $I$ and $J$ we have:

\begin{eqnarray}
    I(\delta,\Ltfbar{},\Ldbar{},\etf,\ed)&=&I_F(\delta,\Ltfbar{},\etf)+I_F(\delta,\Ldbar{},\ed)+I_\Gamma(\delta,\Ltfbar{},\Ldbar{},\etf,\ed)\label{eq:Idelta}\\
    J(\delta,\Ltfbar{},\Ldbar{},\etf,\ed)&=&J_F(\delta,\Ltfbar{},\etf)+J_F(\delta,\Ldbar{},\ed)+J_\Gamma(\delta,\Ltfbar{},\Ldbar{},\etf,\ed)\label{eq:Jdelta}
\end{eqnarray}
with:
\begin{eqnarray}
    I_F(\delta,L,\eta)&=&\eta^2\left[L\delta(\arctan(\delta L ) -\arctan(L))-\frac{L^2}{2}\frac{(1-\delta)^2}{1+L^2}+\frac{1}{2}\ln\left(\frac{1+L^2}{1+L^2 \delta^2 }\right)\right], \\
    J_F(\delta,L,\eta)&=&\frac{\eta^2}{2}\left[4L^3\delta^3(\arctan(L\delta)-\arctan(L))+(1+3L^2\delta^2)\ln\left(\frac{1+L^2}{1+L^2\delta^2}\right)\right] \nonumber \\
    &+&\frac{\eta^2}{2}\left[\frac{6L^4\delta^2}{1+L^2}\ln(\delta)+\frac{L^2(\delta-1)(\delta+1-4L^2\delta^2)}{1+L^2}\right],
\end{eqnarray}
and 

\begin{eqnarray}
I_\Gamma(\delta,L_1,L_2,\eta_1,\eta_2)&=& \frac{\eta_1\eta_2}{L_1^2-L_2^2}4L_1L_2\delta\left[L_2(\arctan(L_1)-\arctan(L_1\delta))-L_1(\arctan(L_2)-\arctan(L_2\delta))\right] \nonumber \\
&+&\frac{\eta_1\eta_2}{L_1^2-L_2^2}\left[L_1^2\ln\left(\frac{1+L_2^2}{1+L_2^2\delta^2}\right)-L_2^2\ln\left(\frac{1+L_1^2}{1+L_1^2\delta^2}\right)\right]\nonumber \\
&+&\frac{\eta_1\eta_2}{L_1^2-L_2^2}L_1^2L_2^2\delta^2\ln\left(\frac{(1+L_1^2)(1+L_2^2\delta^2)}{(1+L_2^2)(1+L_1^2\delta^2)}\right), \\
J_\Gamma(\delta,L_1,L_2,\eta_1,\eta_2)&=& 
    \frac{\eta_1\eta_2}{L_1^2-  L_2^2}8L_1^2L_2^2\delta^3\left[L_2(\arctan(L_2)-\arctan(L_2\delta))-L_1(\arctan(L_1)-\arctan(L_1\delta))\right] \nonumber \\
    &+&\frac{\eta_1\eta_2}{L_1^2-L_2^2}\left[L_1^2\ln\left(\frac{1+L_2^2}{1+L_2^2\delta^2}\right)-L_2^2\ln\left(\frac{1+L_1^2}{1+L_1^2\delta^2}\right)\right] \nonumber \\
    &+& \frac{\eta_1\eta_2}{L_1^2-L_2^2}3L_1^2L_2^2\delta^2\ln\left(\frac{(1+L_1^2)(1+L_2^2\delta^2)}{(1+L_2^2)(1+L_1^2\delta^2)}\right) \nonumber \\
    &+& \frac{\eta_1\eta_2}{L_1^2-L_2^2}6L_1^2L_2^2\delta^2\ln(\delta)\ln\left(\frac{1+L_1^2}{1+L_2^2}\right) \nonumber \\
    &+& \frac{\eta_1\eta_2}{L_1^2-L_2^2}3L_1^2L_2^2\delta^2\left[\Li{2}(-L_2^2)-\Li{2}(-L_2^2\delta^2)-\Li{2}(-L_1^2)+\Li{2}(-L_1^2\delta^2)\right].
\end{eqnarray}
with $\Li{2}(z)=\sum_{k=1}^{\infty}z^k/k^2$ the polylogarithm function of second order.

In the case of a neutral target, $\eta_D=0$ and the functions simply read $I(\delta,\Ltfbar{},\Ldbar{},\etf,\ed)=I_F(\delta,\Ltfbar{},\etf)$ and $J(\delta,\Ltfbar{},\Ldbar{},\etf,\ed)=J_F(\delta,\Ldbar{},\etf)$. For brevity, we did not write all the dependencies of the $I$ and $J$ functions in the following.

\subsection{Bremsstrahlung cross section}

\subsubsection{Moderately relativistic regime}
For $2\leq \gamma \leq 100$, the differential cross-section of the Bremsstrahlung emission reads:
\begin{eqnarray}\label{eq:br:mr:dsigdg}
    \frac{d\sigma}{d\gamma_\gamma}=\frac{4Z^2r_e^2\alpha}{\gamma_\gamma}\left[1+\left(\frac{\gamma-\gamma_\gamma}{\gamma}\right)^2\right][I(\delta)+1]
    -\frac{4Z^2r_e^2\alpha}{\gamma_\gamma}\frac{2}{3}\frac{\gamma-\gamma_\gamma}{\gamma}\left[J(\delta)+\frac{5}{6}\right],
\end{eqnarray}
with $\delta=\gamma_\gamma/(2\gamma(\gamma-\gamma_\gamma))$.

\subsubsection{Ultrarelativistic regime}

For ultra-relativistic electrons $\gamma>100$, the differential cross section reads:
\begin{eqnarray}\label{eq:br:ur:dsigdg}
    \frac{d\sigma}{d\gamma_\gamma}=\frac{4Z^2r_e^2\alpha}{\gamma_\gamma}\left[\left[1+\left(\frac{\gamma-\gamma_\gamma}{\gamma}\right)^2\right][I(\delta)+1-f_C(Z)]-\frac{2}{3}\frac{\gamma-\gamma_\gamma}{\gamma}\left[J(\delta)+\frac{5}{6}-f_C(Z)\right]\right]
\end{eqnarray}
where $f_C$ is the Coulomb correction defined as:
\begin{eqnarray}\label{eq:ur:fc}
   f_C(Z)=\frac{\alpha^2Z^2}{1+\alpha^2Z^2}\sum_{n=1}^\infty (-\alpha^2Z^2)^n[\zeta(2n+1)-1] 
\end{eqnarray}
\tys{Note that in the majority of cases, it is sufficient to stop the sum after the first five terms when computing Eq.~\eqref{eq:ur:fc}.}

\subsection{Bethe-Heitler cross sections}

\subsubsection{Moderately relativistic regime}

For $2\leq \gamma_\gamma\leq 100$, the cross differential cross-section of the Bethe-Heitler pair creation is:
\begin{eqnarray}\label{eq:bh:mr:dsigdg}
    \frac{d\overline{\sigma}}{d\gamma}=\frac{4Z^2r_e^2\alpha}{\gamma_\gamma^3}\left[\left[\gamma^2+(\gamma_\gamma-\gamma)^2\right][I(\delta)+1]+\frac{2}{3}\gamma(\gamma_\gamma-\gamma)\left[J(\delta)+\frac{5}{6}\right]\right]
\end{eqnarray}
with $\delta=\gamma_\gamma/(2\gamma(\gamma_\gamma-\gamma))$.

\subsubsection{Utlrarelativistic regime}
In the ultra-energetics limits $\gamma_\gamma\geq 100$, the differential cross-section reads:
\begin{eqnarray}\label{eq:bh:ur:dsigdg}
    \frac{d\overline{\sigma}}{d\gamma}=\frac{4Z^2r_e^2\alpha}{\gamma_\gamma^3}\left[\left[\gamma^2+(\gamma_\gamma-\gamma)^2\right][I(\delta)+1-f_C(Z)]+\frac{2}{3}\gamma(\gamma_\gamma-\gamma)\left[J(\delta)+\frac{5}{6}-f_C(Z)\right]\right]
\end{eqnarray}
with $f_C$ the Coulomb correction given by Eq. \eqref{eq:ur:fc}.

\subsection{Rates and limits}

\noindent We denote:

\begin{eqnarray}
    w(\gamma,\gamma_\gamma)=n_i c\frac{d\sigma}{d\gamma_\gamma} \\
    \overline{w}(\gamma_\gamma,\gamma)=n_i c\frac{d\overline{\sigma}}{d\gamma}
\end{eqnarray}
the energy differential rate of Bremsstrahlung ($w$) and Bethe Heitler ($\overline{w}$) processes. In the limit $\gamma\gg \gamma_\gamma \gg 1$:
\begin{eqnarray}\label{eq:w:asymptotic}
    w(\gamma,\gamma_\gamma)\rightarrow K(n_i,Z,Z^*,T)\frac{1}{\gamma_\gamma}\Theta(\gamma-\gamma_\gamma)
\end{eqnarray}
with $\Theta$ the Heaviside function and
\begin{eqnarray}\label{eq:K}
    K(n_i,Z,Z^*,T)=4n_icZ^2r_e^2\alpha\left[\frac{4}{3}I(0)+\frac{13}{9}-\frac{4}{3}f_C(Z)\right].
\end{eqnarray}
Integrating these differential rates over all possible daughter particle energies leads to the photon emission and pair production rates:
\begin{eqnarray}
    W(\gamma)=n_i c\int_\epsilon^{\gamma-1} d\gamma_\gamma\frac{d\sigma}{d\gamma_\gamma} \\
    \overline{W}(\gamma_\gamma)=n_i c\int_1^{\gamma_\gamma-1} d\gamma \frac{d\overline{\sigma}}{d\gamma}.
\end{eqnarray}
\tys{We have integrated the photon emission spectrum starting from $\epsilon$ since the integral diverge when $\gamma_\gamma\rightarrow 0$. Thus, the quantity $W(\gamma)dt$ corresponds to the number of photons of energy larger than $\epsilon mc^2$ emitted between time $t$ and $t+dt$. In practice, we choose $\epsilon$ such that the energy in the sub-$\epsilon$ photons is negligible with respect to that carried by higher-energy photons. As discussed by Martinez et al. \cite{martinez2019high}, choosing $\epsilon=10^{-7}\gamma$ is sufficient.}
In the limit $\gamma\gg 1$, we have:
\begin{eqnarray}\label{eq:Wbr:limit}
    W(\gamma)\longrightarrow K(n_i,Z,Z^*,T)[\ln(\gamma-1)-\ln(\epsilon)].
\end{eqnarray}
Similarly for $\gamma_\gamma\gg 1$,
\begin{eqnarray}\label{eq:Wbh:limit}
    \overline{W}(\gamma_\gamma)\longrightarrow R(n_i,Z,Z^*,T)
\end{eqnarray}
with 
\begin{eqnarray}\label{eq:R}
    R(n_i,Z,Z^*,T)=4n_icZ^2r_e^2\alpha\left[\frac{7}{9}I(0)+\frac{41}{54}-\frac{7}{9}f_C(Z)\right].
\end{eqnarray}

\tys{\section{Assumptions behind Eqs. (1) and (2) of the manuscript.}
Equations (1) and (2) of the manuscript account only for the Bremsstrahlung and Bethe–Heitler processes. We detail below the reasons for neglecting other processes:
\begin{itemize}
    \item Collisional and ionisation processes are neglected since they are not dominant for the range of lepton and photon energies that contribute to the shower (larger than $\gamma_c mc^2$, the threshold photon energy defined in the manuscript). For thick targets (equivalently, long-time), the effect of these processes is included by introducing the cut-off energy $\gamma_cmc^2$, below which pair production is suppressed.
    \item Production of heavier leptons like muons is neglected as the cross-section for lepton production scales with the inverse of the square of the lepton mass. For example, the cross-section for muon production is 5 orders of magnitude below that of electron-positron production.
    \item Because it is a direct process, Trident pair production can be important for very thin targets. Geant4 simulations (not shown) indeed show that Trident can dominate for very thin targets $L\sim 10^{-3}\Lcol$. For thicker targets, however, it can be neglected since the Bremsstrahlung and Bethe-Heitler processes largely dominate pair production.
    \item Linear Breit-Wheeler pair production is neglected since in this geometry all photons propagate in the quasi-same direction.
\end{itemize}
}

\section{Shower time}\label{sec:radiativetime}

\tys{To compute the shower time, it is interesting to quantify how an incident lepton loses its energy. From the kinetic equation [Eq. (1) of the manuscript] that describes the evolution of an initial population of lepton $f_-^{(0)}$, we compute the average energy of a lepton \mica{subject} to radiation reaction only. It reads:
\begin{eqnarray}
    \frac{d}{dt} \langle \gamma \rangle= \int_0^\infty  d\gamma_\gamma \gamma_\gamma w(\langle \gamma \rangle,\gamma_\gamma),
\end{eqnarray}
where $\langle \gamma \rangle$ represent the average energy of a population $f_-$. We then define the shower time, similarly to \cite{akhiezer1994kinetic}, as the the time required for an electron to transfer all its energy to photons:
}
\begin{eqnarray}
    \Tcol(\gamma_0)= \int_1^{\gamma_0}\frac{d\gamma}{\int_0^\infty  d\gamma_\gamma \gamma_\gamma w(\gamma,\gamma_\gamma)}.
\end{eqnarray}
In the limit of $\gamma \gg 1$, we have:
\begin{eqnarray}
    \int_0^\infty  d\gamma_\gamma \gamma_\gamma w(\gamma,\gamma_\gamma)= \int_0^{\gamma-1}  d\gamma_\gamma \gamma_\gamma w(\gamma,\gamma_\gamma) 
    = n_i c \int_0^{\gamma-1} d\gamma_\gamma \gamma_\gamma \frac{d\sigma}{d\gamma_\gamma}.
\end{eqnarray}
Assuming that the integral is dominated by the low energetic part $\gamma_\gamma\ll \gamma$, we have $\delta\simeq 0$ and $(\gamma-\gamma_\gamma)/\gamma\simeq 1$. It follows that:

\begin{eqnarray}
    \int_0^\infty  d\gamma_\gamma \gamma_\gamma w(\gamma,\gamma_\gamma)&\simeq& 4n_icZ^2r_e^2\alpha\left[\frac{4}{3}I(0)+\frac{13}{9}-\frac{4}{3}f_C(Z)\right]\gamma \nonumber \\
    &=& K(n_i,Z,Z^*,T) \gamma
\end{eqnarray}
where we use $I(0)=J(0)$. Injecting into the definition of $\Tcol$ we obtain:
\begin{eqnarray}
    \Tcol \equiv \frac{1}{K(n_i,Z,Z^*,T)}\ln(\gamma_0) \,,
\end{eqnarray}
or equivalently, $\Lcol=c/K(n_i,Z,Z^*,T)\times \ln(\gamma_0)$.

\section{Calculation for the long time scale: Eqs. (14)-(16) of the manuscript.}

By considering the Breit–Wheeler process for pair production and non-linear Compton scattering for photon emission, we derive a recurrence relation linking the total photon spectrum emitted by generation $n$ to that of generation $n-1$. The detailed proof is provided in the supplementary material of \cite{pouyez2024kinetic}. A similar derivation can be carried out for the Bethe–Heitler process (pair production) and Bremsstrahlung (photon emission), since the corresponding kinetic equations differ only in the definition of the interaction rates.

In the limit where $1 - \exp[-(t - t') \Wbw(\gamma_\gamma, t)] \simeq \Theta(\gamma_\gamma - \gamma_c)$, with $\gamma_c$ representing the minimum photon energy required for pair creation (as discussed in the manuscript), and under the approximation that each converted photon transfers half of its energy to each lepton, i.e., $\overline{w}(\gamma_\gamma, \gamma) = \Wbw(\gamma_\gamma) \delta(\gamma - \gamma_\gamma/2)$, we obtain:

\begin{eqnarray}\label{eq:Fn:recurrence}
    F_\gamma^{(n)}(\gamma_\gamma,t) = 2 \int_{\gamma_c}^\infty\!\!\!d\gamma_\gamma'\,\,
    F_\gamma^{(n-1)}(\gamma_\gamma',t)\,\,
    I_\gamma\!\left(\gamma_\gamma'/2,\gamma_\gamma\right)\,\,.
\end{eqnarray}
where $F_\gamma^{(n)}$ denotes the total photon spectrum emitted by the leptons of the $nth$ generation, and where we have introduced the transfer function $I_\gamma(\gamma_i, \gamma_\gamma)$, which represents the total photon spectrum emitted by a single charged particle initially at energy $\gamma_i$. It is given by:

\begin{eqnarray}
I_\gamma(\gamma_i,\gamma_\gamma)=\int_0^\infty dt w(\gamma_\gamma,\gamma(t))
\end{eqnarray}
where $\gamma(t)$ is the energy of the electron as a function of time. It can be estimated by the equation of evolution for the average energy:
\begin{eqnarray}
    \frac{d}{dt}\gamma(t)\simeq- K(n_i,Z,Z^*,T)\gamma(t).
\end{eqnarray}
Finally, the function $I_\gamma$ can be simplified as:

\begin{eqnarray}\label{eq:I}
    I_\gamma(\gamma_i,\gamma_\gamma)=\frac{1}{K(n_i,Z,Z^*,T)}\int_1^{\gamma_i} d\gamma \frac{w(\gamma,\gamma_\gamma)}{\gamma}
    \simeq \frac{1}{\gamma_\gamma}\ln\left(\frac{\gamma_i}{\gamma_\gamma}\right)\Theta(\gamma_i-\gamma_\gamma)
\end{eqnarray}
where we used the Eq. \eqref{eq:w:asymptotic} for the rate. Injecting Eq. \eqref{eq:I} in the recurrence relation Eq. \eqref{eq:Fn:recurrence} and noting that for the first generation:

\begin{eqnarray}
    F^{(0)}(\gamma_\gamma)=I_\gamma(\gamma_0,\gamma_\gamma),
\end{eqnarray}
we finally obtained:

\begin{eqnarray}
    F_\gamma^{(n)}(\gamma_\gamma)=\frac{2^n}{(2n+1)!}\frac{1}{\gamma_\gamma}\ln\left(\frac{\gamma_0}{2^n \gamma_\gamma}\right)^{2n+1}\Theta(\gamma_0 -2^n \gamma_\gamma).\label{eq:lt:sol:Fn}
\end{eqnarray}
In these limits, the number of pairs of generation $(n)$ reads \cite{pouyez2024kinetic}:
\begin{eqnarray}\label{eq:Npm-longtime}
   N_{\pm}^{(n)}(t) = \int_{\gamma_c}^{\infty}\!\!d\gamma_\gamma\,F_{\gamma}^{(n-1)}(\gamma_\gamma,t).
\end{eqnarray}
After integration, we obtain:

\begin{eqnarray}
    N_\pm^{(n)}/N_0&=&\frac{2^{n-1}}{(2n)!}\ln\left(\frac{\gamma_0}{2^{n-1}\gamma_c}\right)^{2n} \Theta(\gamma_0-2^{n-1}\gamma_c).\label{eq:lt:sol:Nn}
\end{eqnarray}
This equation provides the maximum number of generations, $n_{\rm\scriptscriptstyle{max}}$, in the shower. The Heaviside function leads that $N_\pm^{(n)}(t)$ is non zero only if $\gamma_0>2^{n-1}\gamma_c$, thus:

\begin{eqnarray}
    n_{\rm\scriptscriptstyle{max}}=1+\ln\left(\frac{\gamma_0}{\gamma_c}\right)/\ln(2)
\end{eqnarray}
similarly to the analysis of Heitler \cite{heitler1984quantum}. The total number of pairs is given by the sum of Eq. \eqref{eq:lt:sol:Nn} over each generation, it reads:

\begin{eqnarray}\label{eq:lt:Ntot:sumgen}
N_\pm(t)/N_0&=&\sum_{i=1}^{n_{\rm\scriptscriptstyle{max}}}N_\pm^{(n)}(t)/N_0.
\end{eqnarray}
In the limit $\gamma_0\rightarrow \infty$, 

\begin{eqnarray}
N_\pm(t)/N_0&\simeq&\int_1^{n_{\rm\scriptscriptstyle{max}}}dn\frac{2^{n-1}}{\Gamma(2n+1)}\ln\left(\frac{\gamma_0}{2^{n-1}\gamma_c}\right)^{2n}
\end{eqnarray}
where $\Gamma$ is the gamma function. Using Stirling's approximation for $\Gamma$ and using a saddle-node method for the final integration, we obtain:
\begin{eqnarray}
N_\pm(t)/N_0&\simeq&\frac{1}{2+\ln(2)}\frac{\gamma_0}{\gamma_c}.
\end{eqnarray}

\tys{
\section{Angular aperture of the outgoing pairs: Eq. (19) of the manuscript.}

When a lepton passes through matter, it accumulates an angle due to Multiple Coulomb scattering (MCS). The root-mean square (RMS) angle of the resulting angular distribution as been first studied by Rossi and Greisen \cite{rossi1941cosmic} (Part I, Sec E. 22). Since then, the angular distribution due to  MCS has been extensively studied \cite{moliere1948theorie,bethe1953moliere,scott1963theory,highland1975some,lynch1991approximations} and the RMS angle due to MCS is commonly refereed as Molière's angle. Today, one of the most common representations of this angle is given by Lynch \cite{lynch1991approximations} and reads:
\begin{eqnarray}\label{eq:angle:lynch}
    \theta_{\rm rms}\sim\frac{13.6\, \mbox{MeV}}{\gamma_0mc^2}\sqrt{\frac{L}{L_r}}\left[1+0.088\log_{10}\left(\frac{L}{L_r}\right)\right]
\end{eqnarray}
where $L_r$ is the radiation length [$L_r\equiv \Lcol/\ln(\gamma_0)$]. Neglecting the logarithmic contribution in Eq. \eqref{eq:angle:lynch} and replacing $L_r$ by $\Lcol/\ln(\gamma_0)$, we obtain:
\begin{eqnarray}
    \theta_{\rm rms} \sim \frac{13.6 \, \mbox{MeV}}{\gamma_0mc^2}\sqrt{\ln(\gamma_0)}\sqrt{\frac{L}{\Lcol}}.
\end{eqnarray}
For $\gamma_0 mc^2 \in [0.1, 50]$ GeV, the factor $\sqrt{\ln(\gamma_0)}$ varies within $[2.3, 3.4]$. This term can therefore be approximated by $\sim 2.8$, yielding:
\begin{eqnarray}
    \theta_{\rm rms} \sim \frac{74.5}{\gamma_0}\sqrt{\frac{L}{\Lcol}}.
\end{eqnarray}
This is the dominant term of the angle at which the distribution is maximised ($\theta_{\rm{peak}}$) from Eq. (19) of the manuscript. Hence, we conclude that the primary contribution to the angular spectrum of the outgoing positrons arises from Multiple Coulomb Scattering.

Molière's theory was not developed to model the angular aperture of an electromagnetic shower. Hence, it does not account for the fact that Bremsstrahlung and Bethe-Heitler lead to an additional angular spread (accounted for in Geant4 simulations), nor does it account for the fact that several successive generations of particles emerge from the shower. The second term of Eq. (19) originates from the successive shower generations and additional angular broadening mechanisms. 
}

\mica{
\section{Useful quantities for computing the skin-depth}

\subsection{Center-of-mass frame}

In order to compare the skin depth to the size of the escaping pair jet, it is interesting to place ourselves in the centre of mass frame of those escaping pairs. This frame moves with respect to the laboratory frame, in the direction of the incident electron beam, at a velocity:
\begin{eqnarray}\label{eq:Vb:exact}
    V_b = \frac{c^2{\vert\bf P}_{\rm tot}\vert}{\mathcal{E}_{\rm tot}} \,.
\end{eqnarray}
The associated Lorentz boost reads $\Gamma_b=(1-(V_b/c)^2)^{-1/2}$.
While this velocity can be computed directly from the simulation output, we are now looking for an estimate directly from the derived pair distribution and characteristic angle. As discussed in the manuscript, the angular distribution of the pair beam is difficult to estimate, thus for simplicity, we start from the ansatz that the energy-angular distribution can be casted in the form: $f_\pm(\gamma,\theta,t)=f_\pm^{(1)}(\gamma,t)\delta(\theta-\theta_{\rm{peak}}\gamma_0/(2\gamma))$. This general form is motivated by Geant4 simulations. The factor $1/2$ in the Dirac distribution ensures that the maxima of the total angular distribution $\int d\gamma f_\pm(\gamma,\theta,t)$ is at $\theta_{\rm{peak}}$. The distribution $f_\pm^{(1)}(\gamma,t)$ is the energy distribution of the first generation of pairs and reads [Eq.~(8) of the manuscript]:
\begin{eqnarray}\label{eq:sh:fpm:1}
    f_\pm^{(1)}(\gamma,t)=\frac{t^2}{2}\int_0^{\gamma_0} d\gamma_\gamma \overline{w}(\gamma_\gamma,\gamma)(\gamma_\gamma)w(\gamma_0,\gamma_\gamma).
\end{eqnarray} 
This way, the velocity $V_b$ is estimated as:
\begin{eqnarray}\label{eq:Vb:approx}
    V_b/c \simeq \frac{\int_1^{\gamma_0} d\gamma \int_0^{\pi/2} d\theta \sqrt{\gamma^2-1}\cos(\theta) f_\pm(\gamma,\theta) }{\int_1^{\gamma_0} d\gamma \gamma f_\pm^{(1)}(\gamma)} \,.
\end{eqnarray}
Considering the asymptotic forms for the Bremsstrahlung rate [Eq. \eqref{eq:w:asymptotic}] and Bethe-Heitler differential rate [Eq.~\eqref{eq:Wbh:limit}], one obtains:
\begin{eqnarray}
    V_b/c = \int_0^1 d\nu\cos\left(\frac{\theta_{\rm{peak}}}{\nu}\right) \xrightarrow[\theta_{\rm{peak}} \ll 1]{} 1-\frac{\pi}{2}\theta_{\rm{peak}}\,,\\
\end{eqnarray}
and (for $\theta_{peak}\pi<1$)
\begin{eqnarray}\label{eq:LorentzBoost:asymtptotic}
    \Gamma_b \simeq \frac{1}{\sqrt{\pi \theta_{\rm{peak}}}}.
\end{eqnarray}
The Lorentz boost Eq.~\eqref{eq:LorentzBoost:asymtptotic} is compared in Fig. \ref{fig:figSupp1} (a) to the one extracted from Geant4 simulation [using the exact form of $V_b$ Eq. \eqref{eq:Vb:exact}]. Two simulations are shown corresponding to an incident electron beam of $3$ GeV (blue) and $10$ GeV (red). As illustrated, Eq.~\eqref{eq:LorentzBoost:asymtptotic} shows similar dependencies to the Geant4 results, but, due to the different approximations, especially regarding the angular description, it underestimates (although by a factor less than $2$) the simulated Lorentz boost.

\begin{figure*}[htpb]
    \centering
    \includegraphics[width=\linewidth]{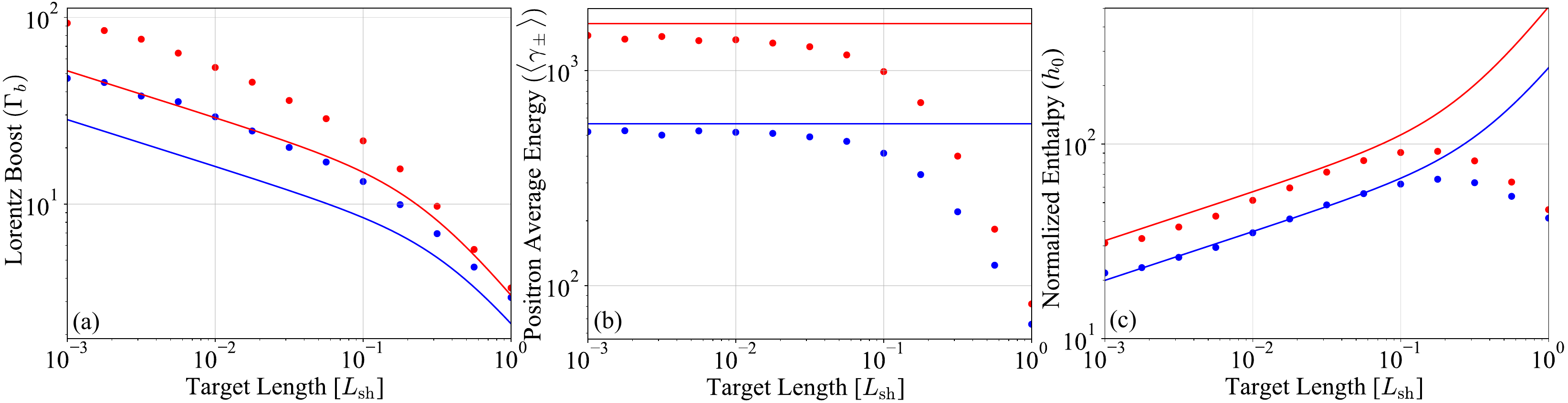}
    \caption{\mica{(a) Lorentz boost factor $\Gamma_b$, (b)  positron average energy $\langle \gamma_\pm \rangle$ in the laboratory-frame, and (c) normalized enthalpy $h_0^R$, as functions of the target thickness $L/\Lcol$. Blue and red  are associated to an incident electron beam of $3$ and $10$ GeV, respectively. Dots are results extracted from Geant4 simulation [for panels (c) using Eq. \eqref{eq:enthalpyexact}]. Solid lines are theoretical estimates from Eqs. \eqref{eq:Vb:approx} [panel (a)], \eqref{eq:average_energy} [panel (b)] and \eqref{eq:enthalpyHotMJ} [panel (c)].}}
    \label{fig:figSupp1}
\end{figure*}

\subsection{Enthalpy and its link to the average energy in the laboratory-frame}

When considering a relativistically hot plasma moving in the lab frame with a preferential direction (denoted by the subscript $\parallel$), the skin-depth, as given by Eq.~(28) of the manuscript, is a function of the normalised enthalpy
\begin{eqnarray}\label{eq:enthalpyexact}
    h_0^R= \langle \gamma_\pm \rangle_R + \langle v_{\parallel}p_{\parallel} \rangle_R/(mc^2)
\end{eqnarray}
here written in the centre-of-mass, and where $\langle . \rangle_R$ denotes the average over the rest frame distribution. 
The first term in Eq. \eqref{eq:enthalpyexact} corresponds to the average energy of the pairs, while the second term is related to the thermal pressure. 

For the particular case of a relativistically hot Maxwell-Jüttner distribution, the normalised enthalpy can be rewritten in terms of the average positron energy $\langle \gamma_\pm \rangle$ measured in the laboratory frame ~\cite{melzani2013apar}:
\begin{eqnarray}\label{eq:enthalpyHotMJ}
    h_0^R\xrightarrow[\mbox{hot plasma}]{} \frac{\langle \gamma_\pm \rangle}{\Gamma_b}.
\end{eqnarray}
The average energy of the pairs in the laboratory frame is estimated considering the short time limit solution for the first pairs distribution Eq. \eqref{eq:sh:fpm:1}, as:
\begin{eqnarray}\label{eq:average_energy}
    \langle \gamma_\pm \rangle &&= \frac{\int_0^{\gamma_0}d\gamma_\gamma \Wbw(\gamma_\gamma)w(\gamma_0,\gamma_\gamma)\gamma_\gamma/2}{\int_0^{\gamma_0}d\gamma_\gamma \Wbw(\gamma_\gamma)w(\gamma_0,\gamma_\gamma)} \sim 0.25\gamma_0^{0.89}\,.
\end{eqnarray}
This prediction is compared to Geant4 simulations in Fig. \ref{fig:figSupp1} (b) as a function of the normalised target length $L/L_{\rm sh}$. Solid lines correspond to Eq.~\eqref{eq:average_energy}, and dots to the simulation results (blue for a $3$-GeV electron beam, red for a $10$-GeV one). A very good agreement is found for $L/\Lcol\lesssim 0.1$, but for $L/\Lcol \simeq 1$, our estimates no longer capture the good scaling since the effect of cooling, neglected in our estimate, sets in. 

Although Eq. \eqref{eq:enthalpyHotMJ} can be derived rigorously only for a relativistically hot Maxwell-Jüttner distribution, it also provides a very good approximation for the enthalpy of the escaping jet of pairs obtained in our geometry. To prove this, we plot in Fig. \ref{fig:figSupp1}(c) the normalised enthalpy $h_0^R$ as a function of the target length for two incident electron beams of $3$ (blue) and $10$ GeV (red). Dots are computed from the Geant4 results, boosted to the centre-of-mass frame, using the definition Eq.~\eqref{eq:enthalpyexact}. Solid lines are computed using the relativistically-hot-limit enthalpy Eq. \eqref{eq:enthalpyHotMJ} with our estimate for the Lorentz boost Eq.~\eqref{eq:LorentzBoost:asymtptotic} and average energy Eq.~\eqref{eq:average_energy}. A very good agreement is obtained in the important limit $L/L_{\rm sh} \lesssim 0.1$. 


}